\documentclass[aip,jmp,author-numerical,11pt]{revtex4-2}

\usepackage[english,french]{babel}
\usepackage[colorlinks=true,linkcolor=blue]{hyperref}
\usepackage{amssymb,amsthm}
\usepackage{amsmath}
\usepackage[all]{xy}
\usepackage{pstricks}
\usepackage{fancyhdr}
\usepackage{graphicx}\graphicspath{./}
\usepackage{enumerate,paralist}
\usepackage[T1]{fontenc}
\usepackage{url}
\usepackage{tikz,tkz-tab}
\usepackage{float}
\usepackage{graphics}
\usepackage{setspace}

\usepackage{caption}
\usepackage{dsfont}
\usepackage[utf8]{inputenc}
\usepackage[autostyle]{csquotes}

\usepackage{latexsym,amscd,amsfonts}

\def \Z {\mathbb Z}
\def \R {\mathbb R}
\def \C {\mathbb C}
\def \Q {\mathbb Q}
\def \N {\mathbb N}
\def \T {\mathbb T}

\def \E {\mathbb E}

\def \SM {\mathbb S}

\def \WW {\mathbb W}
\def \UU {\mathbb U}
\def \VV {\mathbb V}
\providecommand{\dd}{\,\mathrm{d}}
\providecommand{\ee}{\mathrm{e}}
\providecommand{\ii}{\mathrm{i}}
\providecommand{\DD}{\mathrm{D}}

\newcommand{\supp}{{\rm supp}\; }

\newcommand{\SLD}{\mathrm{SL}_{2}(\R)\,}

\newcommand{\SpD}{\mathrm{Sp}_{\mathrm{D}}(\R)\,}
\newcommand{\spDD}{\mathfrak{sp}_{2}(\R)\,}
\newcommand{\spD}{\mathfrak{sp}_{\mathrm{D}}(\R)\,}

\newcommand{\MDR}{\mathcal{M}_{\mathrm{D}}(\R)}

\newcommand{\UL}{\textrm{U}(D)}

\newcommand{\Hof}{\{H_{\omega} \}_{\omega \in \Omega} }

\newcommand{\omO}{\omega^{(0)}}

\newcommand{\SD}{\mathrm{S}_{\mathrm{D}}(\R)\,}

\DeclareMathOperator{\dlO}{d_{\log\, \mathcal{O}}}

\newcommand{\one}{\mathrm{I}_{\mathrm{D}}}
\newcommand{\Ll}{\mathcal{L}}
\newcommand{\inv}{{\mbox{\rm\tiny inv}}}
\newcommand{\CMV}{{\mathbb U}}
\newcommand{\CMVL}{{\mathbb V}}
\newcommand{\CMVR}{{\mathbb W}}

\theoremstyle{plain}

{\vspace{3mm}}

\newtheorem{thm}{Theorem}
\newtheorem{defi}{Definition}
\newtheorem{cor}{Corollary}
\newtheorem{prop}{Proposition}
\newtheorem{lem}{Lemme}

\newtheorem{openq}{Open question}

\theoremstyle{definition}

\begin{document}
\title{Localization for random quasi-one-dimensional models}
\author{H. Boumaza}
\affiliation{H. Boumaza, Universit\'e Sorbonne Paris Nord, LAGA, CNRS, UMR 7539,  F-93430, Villetaneuse, France}
\email{boumaza@math.univ-paris13.fr}
\thanks{Supported by ANR JCJC RAW}

\date{\today}

\begin{abstract}
In this paper we review results of Anderson localization for different random families of operators which enter in the framework of random quasi-one-dimensional models. We first recall what is Anderson localization from both physical and mathematical point of views. From the Anderson-Bernoulli conjecture in dimension 2 we justify the introduction of quasi-one-dimensional models. Then we present different types of these models : the Schr\"odinger type in the discrete and continuous cases, the unitary type, the Dirac type and the point interactions type. In a second part we present tools coming from the study of dynamical systems in dimension one : the transfer matrices formalism, the Lyapunov exponents and the Furstenberg group. We then prove a criterion of localization for quasi-one-dimensional models of Schr\"odinger type involving only geometric and algebraic properties of the Furstenberg group. Then, in the last two sections, we review results of localization, first for Schr\"odinger-type models and then for unitary type models. Each time, we reduce the question of localization to the study of the Furstenberg group and show how to use more and more refined algebraic criterions to prove the needed properties of this group. All the presented results for quasi-one-dimensional models of Schr\"odinger type include the case of Bernoulli randomness.
\end{abstract}

\maketitle

\section{Anderson localization and quasi-1D models}\label{sec_loc_Anderson}

The physics of condensed matter, which studies solids with a crystalline structure, teaches us that the electrons are distributed among all the atoms of the solid. At equilibrium, the Pauli principle states that two electrons, these being fermions, cannot share the same quantum state, which implies that all the electrons of the solid cannot be at the fundamental level. The lowest energy levels therefore fill up until they reach the Fermi energy level. When the solid is subjected to a change of temperature or to an electric potential, some electrons are excited and their energy increases beyond the Fermi level. These electrons are scattered in the solid and produce an electronic transport.

This description of the electronic transport is valid for a solid whose crystal structure is periodic and thus does not contain impurities. However, in the nature, the ideal crystals do not exist, they always contain impurities. These can be of different natures. For example, one can observe the presence of ionized atoms in the crystalline network or, in the case where the crystal is not constituted of all identical atoms but is an alloy between several materials, it is possible that the network is no longer perfectly periodic, since there is here and there an atom which is not in the right place. Finally, some atoms are sometimes slightly out of place with respect to their ideal position on the periodic lattice. In all these cases, the physical properties of the crystal are modified. 

How can we model these impurities in a crystal and their impact on electronic transport? The first to propose a model explaining the effects of disorder on the quantum behavior of electrons in a crystal lattice containing impurities was the American physicist Philip Warren Anderson\cite{A58}. By introducing random terms into the Schr\"odinger equation, two new phenomena were demonstrated: the Anderson localization and the existence of a phase transition in dimension $3$ and beyond.

The Anderson localization phenomenon can be stated as follows: at a fixed energy, beyond a certain amount of disorder in the crystal, the diffusion of electrons ceases and any excited electron will remain confined in a localized region instead of diffusing in the crystal. The crystal stops being a conductor and becomes an insulator.

A possible explanation of the Anderson localization is given by the following wave interpretation: any excited electron in the crystal has an associated wave, and with each collision of the electron with an impurity in the crystal, its associated wave scatters. One would expect that, as the disorder increases, the mean free path (the average distance traveled by the electron between two collisions) would decrease continuously. But this is not what happens. After a certain critical quantity of impurities, the diffusion of the electron stops suddenly. This sudden stop takes place when the mean free path becomes shorter than the wavelength of the electron: if the wave is scattered before even a first period, we cannot really consider it as a wave anymore... Let us take note that the localization phenomenon goes beyond the framework of quantum mechanics. It can be observed in other situations where a wave propagates in a disordered medium. This can be the case of a light wave, microwaves or acoustic waves. For a complete introduction to condensed matter physics one can read the classical reference by Anderson\cite{And18} or the more recent book by Girvin and Yang \cite{GY19}. For further readings about physical aspects of Anderson localization and how it appears in several domains of physics, we refer to \cite{H00,BAA06,BrKe03}.


Anderson's paper  \cite{A58} also predicts the existence of an insulator/conductor phase transition as soon as the dimension of the crystal lattice is greater than or equal to $3$. Whatever the amount of disorder in the crystal, there is an energy below which there is Anderson localization and above which there is scattering of excited electrons. For one-dimensional crystals this transition does not exist and for two-dimensional crystals the absence of transition is conjectured : at any energy, the Anderson localization phenomenon appears when there is disorder. For the Anderson model, it is therefore conjectured that there is a critical dimension, in this case $3$, for which the behavior of the system changes drastically. Let us take note that the non-existence of a phase transition is well demonstrated mathematically in dimension 1 but that it remains an open conjecture in dimension 2. We will come back to this point after having presented the Anderson model from the mathematical point of view.

Anderson's original idea is to consider that the charge of the atoms of the crystal is a random variable. More precisely, let $(\tilde{\Omega},\tilde{\mathcal{A}},\tilde{\mathsf{P}})$ be a complete probability space and let us pose
\begin{equation}\label{eq_def_espace_proba_tensor_1}
(\Omega,\mathcal{A},\mathsf{P})=\left(\bigotimes_{n\in \Z^d} \tilde{\Omega},\bigotimes_{n\in \Z^d} \tilde{\mathcal{A}}, \bigotimes_{n\in \Z^d} \tilde{\mathsf{P}}\right). 
\end{equation}
We identify the coordinates $\omega_n$ of $\omega \in \Omega$ to random variables $\omega^{(n)}$ on $(\tilde{\Omega}, \tilde{\mathcal{A}},\tilde{\mathsf{P}})$, $\omega^{(n)}$ representing for example the charge of the atom at the site $n\in \Z^d$.

Anderson's idea leads to consider a potential felt at the point $x$ of the form
$$\forall \omega \in \Omega,\forall x \in \R^d,\ V_{\omega}(x)=\sum_{n\in \Z^d} \omega^{(n)} f(x-n).$$
The $ \omega^{(n)}$ can take a priori only a finite number of values, but we can also consider the case where the $ \omega^{(n)}$ have a continuous law, which is simpler from the mathematical point of view. We also assume that $\{ \omega^{(n)} \}_{n\in \Z^d}$ is a family of independent and identically distributed (\emph{i.i.d.} for short) random variables. 
\vskip 3mm

This idea leads to the introduction of a random family of Schr\"odinger operators:
\begin{equation}\label{eq_def_Anderson_continu}
\forall \omega \in \Omega,\ H_{\omega} = -\Delta_d + V_{\mbox{per}}+\lambda V_{\omega},
\end{equation}
acting on the space $L^2(\R^d)$ and self-adjoint on the Sobolev space $H^2(\R^d)$, where $\Delta_d$ is the usual Laplacian in dimension $d$, $V_{\mbox{per}}$ is an operator of multiplication by a $\Z^d$-periodic function, $V_{\omega}$ is the multiplication operator by the function $V_{\omega}$ introduced above and $\lambda$ is a positive real number which measures the intensity of the disorder. The family of operators $\{ H_{\omega} \}_{\omega \in \Omega}$ is called the \textbf{continuous Anderson model}.
\vskip 3mm

The discrete analog of this model is given by : 
\begin{equation}\label{eq_def_Anderson_discret}
\forall \omega \in \Omega,\forall u\in \ell^2(\Z^d),\forall n \in \Z^d,\ (h_{\omega} u)_n= -\sum_{||m-n||_1=1} u_m + \lambda \omega^{(n)} u_n,
\end{equation}
acting on $\ell^2(\Z^d)$ and where for all $n=(n_1,\ldots, n_d)\in \Z^d$, $||n||_1=|n_1|+\cdots+|n_d|$. Again, $\lambda$ is a positive real number that measures the intensity of the disorder. The family of operators $\{ h_{\omega} \}_{\omega \in \Omega}$ is called the \textbf{discrete Anderson model}.
\vskip 3mm

We already notice that these two families of random operators are $\Z^d$-ergodic. For $\{ H_{\omega} \}_{\omega \in \Omega}$, this is related to the particular form of the potential $V_{\omega}$. More precisely, the $\Z^d$-ergodicity of $\{ H_{\omega} \}_{\omega \in \Omega}$ is a consequence of the \emph{i.i.d.} hypothesis made on the family of random variables  $\{ \omega^{(n)}\}_{n\in \Z^d}$ and of the fact that the supports of the translates of the one-site potential $f$ do not superpose.
\vskip 3mm

Let us briefly recall the definition of this property of families of random operators. Let $\Gamma$ be a lattice in $\R^d$ (often $\Z^d$). Let $(\Omega, \mathcal{A}, \mathsf{P})$ be a complete probabilistic space and  $\{ \tau_{i} \}_{i \in \Gamma}$  a group of measurable transformations which preserve the measure $\mathsf{P}$. A set $A\in \mathcal{A}$ is said to be invariant under the action of $\{ \tau_{i}\}_{i \in \Gamma}$ when $\tau_{i}^{-1}A=A$ for all $i\in \Gamma$. Then $\{ \tau_{i}\}_{i \in \Gamma}$ is said to be $\Gamma$-ergodic if any invariant set is of measure zero or one.

\begin{defi}
 A measurable family of self-adjoint $\{ H_{\omega} \}_{\omega \in \Omega}$ is $\Gamma$-ergodic when there exists a group of ergodic transformations  $\{ \tau_{i} \}_{i \in \Gamma}$ and a family of unitary operators  $\{U_{i}\}_{i\in \Gamma}$  such that: 
$$\forall i \in \Gamma,\ \forall \omega \in \Omega,\ H_{\tau_{i}\omega}=U_{i}H_{\omega}U_{i}^{*}.$$
\end{defi}

\noindent The interesting thing about the ergodicity is that then, for any $\omega \in \Omega$ and any $i\in \Gamma$, $\sigma(H_{\tau_{i}\omega})= \sigma(H_{\omega})$, which leads, with the help of Pastur's theorem\cite[Corollary 1]{P80}, to the existence of a deterministic set equal $\mathsf{P}$-almost surely to the spectrum of $H_{\omega}$ . This deterministic set is called the almost-sure spectrum of the family  $\{ H_{\omega} \}_{\omega \in \Omega}$ and is denoted by $\Sigma$. More precisely, a result by Kunz and Souillard\cite[Th\'eor\`eme IV.2]{KS80} assures us that for an ergodic family of self-adjoint operators, their pure point, singular continuous and absolutely continuous spectra are independent of $\omega$ almost surely (see also \cite[Theorem 1]{KM82}). These almost-sure spectra are denoted respectively by $\Sigma_{\mathrm{pp}}$, $\Sigma_{\mathrm{sc}}$ and $\Sigma_{\mathrm{ac}}$ . 

The existence of the different types of almost sure spectra is essential to be able to demonstrate  spectral properties which are almost-surely independent of $\omega$ for the families $\{ H_{\omega} \}_{\omega \in \Omega}$ and $\{ h_{\omega} \}_{\omega \in \Omega}$ and leads to the first definition of Anderson localization.

There are several mathematical definitions to translate the Anderson localization phenomenon for a family of random operators. Recall that $(\Omega, \mathcal{A}, \mathsf{P})$ denotes a complete probability space and that we consider a family of self-adjoint random operators $\{ H_{\omega} \}_{\omega \in \Omega}$  on a Hilbert space which will be the space $L^2(\R^d)$ in the continuous case or the space $\ell^2(\Z^d)$ in the discrete case.

\begin{defi}
 Let $I$ be an interval of $\R$. We say that the family $\{ H_{\omega} \}_{\omega \in \Omega}$ is \textbf{spectrally localized in $I$} when the spectrum of $H_{\omega}$ in $I$ is nonempty and pure point for $\mathsf{P}$-almost every $\omega \in \Omega$.
\end{defi}

This first definition expresses that for almost any $\omega$ in $\Omega$, $H_{\omega}$ has no continuous spectrum in $I$. The R.A.G.E. theorem (see \cite{RS3,AW15}) ensures then that there are no diffusive states for $H_{\omega}$ and this, almost surely in $\omega$. But this does not give a clear idea of the behavior of the eigenfunctions associated to the eigenvalues in $I$. For this we will give a second definition which is more precise and better reflects the idea of a localized state and not only a non-diffusive one.

\begin{defi}\label{def_anderson_loc}
Let $I$ be an interval of $\R$. We say that the family $\{ H_{\omega} \}_{\omega \in \Omega}$ of almost-sure spectrum $\Sigma$ has the property of \textbf{Anderson localization in $I$} when:
\begin{enumerate}
\item $\Sigma \cap I \neq \emptyset$ and $\Sigma \cap I = \Sigma_{\mathrm{pp}} \cap I$,
\item the eigenfunctions associated to the eigenvalues in $\Sigma \cap I$ decay exponentially to $0$ at infinity. 
\end{enumerate}
\end{defi}

Note that if $\{ H_{\omega} \}_{\omega \in \Omega}$ exhibits Anderson localization in $I$, $\mathsf{P}$-almost surely the point spectrum of $H_{\omega}$ is dense in $\Sigma \cap I$. This phenomenon is quite surprising in comparison with spectra observed at the level of atoms or molecules or in periodic media without disorder. These spectra, as for example that of hydrogen, generally show isolated eigenvalues and eventually an absolutely continuous component. The periodic Schr\"odinger operators whose periodic potential verifies reasonable hypotheses (see \cite{RS4}) have a purely absolutely continuous band spectrum and no eigenvalue. The fact of obtaining a dense set of eigenvalues is mainly found in models where there is a disorder, either of a random nature as here, or for example in quasi-periodic operators.  
\vskip 3mm

The definition just given of Anderson localization is a stationary definition, involving only the Hamiltonian $H_{\omega}$ and not the associated one-parameter group. The following definition takes into account the dynamics in time of the wave packets. Let us denote by $\Sigma$ either the almost-sure spectrum of $\{ H_{\omega} \}_{\omega \in \Omega}$ in the continuous case or the almost-sure spectrum of  $\{ h_{\omega} \}_{\omega \in \Omega}$ in the discrete one.

\begin{defi}\label{def_dyn_loc}
  Let $I$ be an interval of $\R$. We say that the family $\{ H_{\omega} \}_{\omega \in \Omega}$ (respectively $\{ h_{\omega} \}_{\omega \in \Omega}$) is \textbf{dynamically localized} in $I$ when
\begin{enumerate}
 \item   $\Sigma \cap I = \emptyset$,
\item for every compact interval $I_0\subset I$, every $\psi\in L^2(\R^d)$ and every $p\geq 0$,
\begin{equation}\label{eq_def_loc_dynamique_continu}
\E\left( \sup_{t\in \R} ||(1+|\cdot|^2)^{\frac{p}{2}} \ee^{-\mathrm{i}tH_{\omega}}  \mathbf{1}_{I_0}(H_{\omega}) \psi ||_{L^2(\R^d)}^2 \right) < +\infty
\end{equation}
where $\mathbf{1}_{I_0}(H_{\omega})$ denotes the spectral projector on $I_0$ associated with $H_{\omega}$ and $\E$ denotes the expectation taken with respect to the probability measure $\mathsf{P}$;

respectively, for every $u\in \ell^2(\Z^d)$ and every $ p\geq 0$,
\begin{equation}\label{eq_def_loc_dynamique_discret}
\E\left( \sup_{t\in \R} ||\ (1+||\cdot||_1)^{p} \ee^{-\mathrm{i}th_{\omega}}  \mathbf{1}_{I}(h_{\omega}) u ||_{\ell^2(\Z^d)}^2 \right) < +\infty
\end{equation}
\end{enumerate}
\end{defi}

\noindent  The definition \ref{def_dyn_loc} is dynamic in nature and follows the evolution of wave packets over time. It tells us that the solutions of the Schr\"odinger equation are localized in space in the vicinity of their initial position and this, uniformly over time. This reflects the absence of quantum transport.

\noindent Let us point out that the dynamical localization implies the Anderson localization\cite{DS01,S01}. It is also possible to define even stronger notions of localization. For an exhaustive presentation of these notions we refer to the third part of \cite{Ki08} which is written by Abel Klein.

Recall that the Anderson model is actively studied because it is the subject of two important conjectures: that of the existence of an insulator/conductor phase transition in dimension greater than or equal to $3$ and that this transition does not exist in dimension $1$ or $2$.

From the mathematical point of view, quite few answers are brought to these conjectures, in spite of a sustained effort of the community since the end of the 70s. The question of the nature of the spectrum for the Anderson model is not easy. Indeed, if the Laplacian has a purely absolutely continuous spectrum, the multiplication operator by $V_{\omega}$ is (in the discrete case) a diagonal random matrix and its spectrum is therefore discrete. The two effects counterbalance each other when we look at the spectrum of the sum of these two operators. Many mathematical results are perturbative in nature and include a coefficient measuring the size of the order in front of the term $V_{\omega}$. When this parameter is large, the random potential prevails over the Laplacian and a priori there will be localization. When this parameter is small, it is the opposite. Intuitively, the larger the order, the more likely it is that localized states will appear.

The case of dimension $d=1$ with scalar-valued operators is the only one that has been completely solved so far: whatever the common law of the random variables $\omega^{(n)}$ which appear in \eqref{eq_def_Anderson_continu} and \eqref{eq_def_Anderson_discret} and whatever the value of the disorder parameter $\lambda >0$, there is localization to all energies in the discrete case\cite{KS80, CKM87} and to all energies outside a discrete set in the continuous case\cite{DSS02}.  We will present these results with more details in Section \ref{sec_localization_results}.

The case of dimension $d=2$ is particular. We restrict our discussion to the case of Anderson-Bernoulli operators. It is conjectured that localization takes place at all energies independently of the value of the disorder parameter, as in dimension $d=1$. However, it is not impossible that the spectrum remaining pure point at all energies, the eigenfunctions are no longer exponentially decreasing at large energies and that this introduces a transition at the level of quantum transport as in the case of the Landau Hamiltonian\cite{GKS07}. The only thing known is the localization at the bottom of the spectrum in the continuous case for $d\geq 1$ arbitrary\cite{BK05} and in the discrete case for $d=2$ \cite{DS20} and $d=3$ \cite{LZ22}.

The question of localization in dimension $2$ for the Anderson-Bernoulli model at all energies is a question that has turned out to be far too difficult to be tackled head on and the first possible simplification is to consider not the Anderson model on the whole plan $\R^2$ but only on a continuous band  $\R\times [0,1]$.

With the notations introduced at \eqref{eq_def_espace_proba_tensor_1}, let us consider the operator acting on $L^2(\R \times [0,1])\otimes \C$ and defined, for every $\omega \in \Omega$, by
\begin{equation}\label{eq_def_Hcs}
H_{cs,\omega}=-\Delta_2 + \sum_{n\in \Z} \omega^{(n)} V(x-n,y),
\end{equation}
with Dirichlet boundary conditions on $\R\times \{0\}$ and $\R\times \{1\}$ and with $V$ supported in $[0,1]^2$. 

The question of localization at all energies for $\{ H_{\mathrm{cs},\omega}\}_{\omega \in \Omega}$ presents difficulties comparable to those encountered in the study of the Anderson-Bernoulli operator on $\R^2$, due to the fact that both problems are related to the theory of partial differential equations and therefore cannot be approached with tools specific to the dimension $1$.

However, with $\{ H_{\mathrm{cs},\omega}\}_{\omega \in \Omega}$, it is possible to operate a discretization in the bounded direction $[0,1]$ of the band. For that we make a Fourier transform in the second variable, which leads to look at a continuous model in one direction and a discrete one in the second one, acting therefore on a space $L^2(\R \times \Z)$ instead of $L^2(\R\times [0,1])$. Formally, we have therefore reduced ourselves to a continuous Anderson operator in dimension $1$ whose potential is a matrix of infinite size. Of course, this model is still essentially a two-dimensional model and to pass to a one-dimensional model, we restrict ourselves to a finite size for the matrix potential by keeping only a bounded interval of Fourier frequencies in the second variable. We then obtain an Anderson model in dimension $1$ whose potential is a matrix of any size but finite. It acts on $L^2(\R)\otimes \C^D$, where $D\geq 1$ is an integer. This transforms the initial partial differential equation problem into a differential system problem for which we will be able to use the techniques of dynamical systems in dimension $1$ such as transfer matrices and Lyapunov exponents. The hope is then to obtain the localization for this matrix-valued one-dimensional model with localization intervals and localization lengths at best independent of $D$, at worst with a good control on these quantities when $D$ tends to infinity. However, if one looks at the behavior of the integrated density of states at the bottom of the spectrum, it is shown in \cite{BN15JSP} that it has Lifschitz tails behavior with a Lifschitz exponent independent of $D$. Hence from this point of view, it remains unclear if this approximation approach has a chance to work or not. 

The study of this Anderson-Bernoulli model in dimension $1$ whose potential is matrix-valued enters into a more general framework. 

\begin{defi}\label{def_quasi1d}
Let $D\geq 1$ be an integer and $(\Omega,\mathcal{A},\mathsf{P})$ a complete probability space. We call \textbf{random quasi-one-dimensional model} any measurable family of operators acting on $\ell^2(\Z)\otimes \C^D$ (discrete model) or $L^2(\R)\otimes \C^D$ (continuous model) and indexed by $(\Omega,\mathcal{A},\mathsf{P})$.
\end{defi}

This kind of definition could also be transposed to any operator acting on $\ell^2(\Z)\otimes \C^D$ or $L^2(\R)\otimes \C^D$ or to families of quasi-periodic operators acting on these spaces and depending on a frequency parameter for example. We choose to consider only the random case because we will only deal with this case in the following.
\bigskip

%

A random quasi-one-dimensional model can be seen as acting on $D$ copies of $\Z$ (discrete case) or $D$ copies of $\R$ (continuous case). In the following, the models studied will couple these $D$ copies through non-diagonal matrix-valued potentials. This means that these quasi-one-dimensional models will not trivially reduce to a finite direct sum of one-dimensional models with scalar values.
\bigskip

The definition adopted for the notion of quasi-one-dimensional model covers a large number of possible situations. Among these, let us quote four types of models.

\begin{enumerate}
 \item \textbf{The Schr\"odinger type} : in the discrete case, the operators act on  $\ell^2(\Z)\otimes \C^D$ and are of the form
 $$\forall u\in \ell^2(\Z)\otimes \C^D,\ \forall n\in \Z,\  (h_{\omega}u)_{n} =-(u_{n+1}+u_{n-1}) + \lambda V_{\omega^{(n)}}u_n$$
where $\lambda$ is a positive real number, $(V_{\omega^{(n)}})_{n\in \Z}$ is a sequence of \emph{i.i.d.} random variables on $(\Omega,\mathcal{A},\mathsf{P})$, taking its values in the space of symmetric matrices of size $D\times D$. 
 
In the continuous case, the operators act on $L^2(\R)\otimes \C^D$ and have the following form
$$H_{\omega}=-\frac{\dd^{2}}{\dd x^{2}}\otimes I_{\mathrm{D}}+ V_{\mathrm{per}} + \lambda V_{\omega}$$
where  $\lambda$ is a positive real number, $V_{\mathrm{per}}$ is a periodic interaction potential and $V_{\omega}$ is a random potential, both taking their values in the space of regular symmetric matrices of size $D\times D$. The Schr\"odinger type includes discrete and continuous Anderson models.

\item \textbf{The unitary type} :  It includes the case of unitary random matrices acting on $\ell^2(\Z)\otimes \C^D$, in particular the cases of CMV matrices, the  unitary Anderson model or the Chalker-Coddington model on a cylinder. It also includes the random scattering zipper model. This model is defined as follows : let $\UU_{\omega}\;=\;\VV_{\omega}\, \WW_{\omega}$, where
$$\VV_{\omega}\;=\;\left( \begin{smallmatrix}
\ddots & & & \\
 & S_{\omega^{(0)}} & &  \\
& & S_{\omega^{(2)}} &  \\
& & & \ddots 
\end{smallmatrix}  \right)  \circ s_g^{D}                                                                                        
\;,
\qquad
\WW_{\omega}\;=\;\left(\begin{smallmatrix}
\ddots & & &  \\
& S_{\omega^{(-1)}}  & &  \\
 & & S_{\omega^{(1)}}  &  \\
& & & \ddots   
\end{smallmatrix}   \right),$$
$s_g$ is the shift operator to the left ($(v_n)_{n\in \Z} \mapsto  (v_{n+1})_{n\in \Z}$) and the $S_{\omega^{(n)}}$ are unitary matrices in the unitary group $U(2D)$ of the particular form
{\small $$S_{\omega^{(n)}} = \left( \begin{smallmatrix} \alpha & \rho(\alpha) U_{\omega^{(n)}} 
\\ 
V_{\omega^{(n)}}\tilde{\rho}(\alpha) &\ \ -V_{\omega^{(n)}} \alpha^* U_{\omega^{(n)}} 
\end{smallmatrix}  \right),\ \rho(\alpha)=(\mathrm{Id}_{\mathrm{D}}-\alpha\alpha^*)^{\frac{1}{2}},\ \tilde{\rho}(\alpha)=(\mathrm{Id}_{\mathrm{D}}-\alpha^*\alpha)^{\frac{1}{2}},\ ||\alpha \alpha^*||<1. $$}
The exact assumptions on $(U_{\omega^{(n)}},V_{\omega^{(n)}})_{n\in \Z}$ and $\alpha$ will be presented at Section \ref{sec_unitary}.

\item \textbf{The Dirac type} : the operators act on $L^2(\R)\otimes \C^D$ and have the following form 
$$D_{\omega} = \left( \begin{smallmatrix}
                      0 &  -\frac{\dd}{\dd x}\otimes I_{\mathrm{D}} \\
\frac{\dd}{\dd x}\otimes I_{\mathrm{D}} & 0
                      \end{smallmatrix}
 \right)+ V_{\mathrm{per}} + \lambda V_{\omega}$$
where $\lambda>0$, $V_{\mathrm{per}}$ is a periodic potential, linear combination of tensorized Pauli matrices of the form $V_{\mathrm{per}} = \left( \alpha_1  \left( \begin{smallmatrix}
0 & 1 \\
1 & 0                                                                           \end{smallmatrix} \right) + \alpha_2 \left( \begin{smallmatrix}
1 & 0 \\
0 & -1                                                                           \end{smallmatrix} \right) + \alpha_3 \left( \begin{smallmatrix}
0 & -\ii \\
\ii & 0                                                                           \end{smallmatrix} \right) + \alpha_4 \mathrm{I}_2 \right) \otimes  \hat{V}_{\mathrm{per}},$
with $ \hat{V}_{\mathrm{per}}$ a periodic function taking values in real symmetric matrices.

The random potential is of the form 
$V_{\omega} =\left( \beta_1  \left( \begin{smallmatrix}
0 & 1 \\
1 & 0                                                                           \end{smallmatrix} \right) + \beta_2 \left( \begin{smallmatrix}
1 & 0 \\
0 & -1                                                                           \end{smallmatrix} \right) + \beta_3 \left( \begin{smallmatrix}
0 & -\ii \\
\ii & 0                                                                           \end{smallmatrix} \right) + \beta_4 \mathrm{I}_2 \right) \otimes \hat{V}_{\omega}$ where $\hat{V}_{\omega} = \mathrm{diag} \left(
 \sum_{n\in \Z}\omega_{1}^{(n)} f_1(\cdot -n),\ldots,  \sum_{n\in \Z} \omega_{D}^{(n)} f_D(\cdot - n) \right)$
and $f_1,\ldots ,f_D$ are supported in $[0,1]$.  The sequences $(\omega_1^{(n)})_{n\in \Z},\ldots,(\omega_D^{(n)})_{n\in \Z}$ are sequences of random variables defined on a complete probability space $(\Omega,\mathcal{A},\mathsf{P})$. These sequences are assumed to be independent of each other.

\item \textbf{The point interactions type} : Formally, such a model is represented by the following family of random Schr\"odinger operators: 
 
\begin{equation}\label{def_modele_int_pt_formel_N}
\forall \omega \in \Omega,\ H_{P,\omega}=-\frac{\dd^{2}}{\dd x^{2}}\otimes I_{\DD}+ V_{\mathrm{per}} +\sum_{n\in \Z} \left(
\begin{smallmatrix}
c_1\omega_{1}^{(n)} \delta_{n} & & 0 \\
& \ddots & \\
0 & & c_D \omega_{D}^{(n)} \delta_{n}
\end{smallmatrix}\right)
\end{equation}
acting on $L^2(\R)\otimes \C^D$. The numbers $c_1,\ldots,c_D$ are non-zero real numbers, $\delta_n$ is the Dirac distribution at $n\in \Z$ and $V_{\mathrm{per}}$ is the multiplication operator by a periodic function with values in the real symmetric matrices.  The sequences $(\omega_1^{(n)})_{n\in \Z},\ldots,(\omega_D^{(n)})_{n\in \Z}$ are sequences of random variables defined on a complete probability space $(\Omega,\mathcal{A},\mathsf{P})$. These sequences are assumed to be independent of each other. This is a matrix-valued version of the point interaction model introduced in the scalar-valued case in\cite{BF61}.

The definition of the $H_{P,\omega}$ operators in \eqref{def_modele_int_pt_formel_N} is formal because of the presence of Dirac distributions. It is therefore necessary to give a precise definition as in \cite{AGHKH05}. For all $\omega \in \Omega$, let's define,
\begin{equation}\label{def_modele_int_pt_rig_N}
H_{P,\omega} = \bigoplus_{i=1}^{D} H_{\omega_i} + V_{\mathrm{per}}
\end{equation}
acting on $L^2(\R)\otimes \C^D = L^2(\R,\C) \oplus \cdots \oplus L^2(\R,\C)$. For every $i\in \{1,\ldots ,D\}$,  $H_{\omega_i}$ acts on $L^2(\R,\C)$ by $H_{\omega_i}f=-f''$ for $f$ in the domain : 
\begin{align*} \label{def_modele_int_pt_rig_N_domaine}
D(H_{\omega_i}) 
& =  \{ f\in L^2(\R,\C)\mid f,f'\text{ are absolutely continuous on }\R\setminus \Z, \: f'' \in L^2(\R,\C), \notag \\
&f \text{ is continuous on }\R, \;f'(n^+)=f'(n^-) + c_i \omega_i^{(n)} f(n)\text{ for every }n\in\Z\},
\end{align*}
where the left and right derivatives $f'(n^-)$ and $f'(n^+)$ at any integer point $n$ are assumed to exist.  
\end{enumerate}

The Schr\"odinger type models, discrete or continuous, describe the electronic transport in $D$ one-dimensional layers, which are a priori in interaction. An electron can jump from one layer to another and on each layer it may or may not encounter an impurity at each integer point.

In order to study the Anderson localization for quasi-one-dimensional random models, one can use dynamical systems techniques specific to the dimension one: the formalism of transfer matrices and the Lyapunov exponents. In the next Section we present these tools in the framework of quasi-one-dimensional random models of Schr\"odinger type.

\section{Lyapunov exponents and Furstenberg group}\label{sec_lyap}

To address the question of Anderson localization for quasi-one-dimensional Schr\"odinger-type models, we can start by looking at the conditions for the exponential growth of the eigenfunctions for these models. We are then led to study the asymptotic behavior of a linear equation of order $2$, either of finite differences in the discrete case, or differential in the continuous case. 

Let $D\geq 1$ and $E\in \R$.  In the discrete Schr\"odinger case we study the exponential asymptotic behavior of the sequences $u\in \ell^2(\Z)\otimes \C^D$ which satisfy
\begin{equation}\label{eq_valeurs_propres_discret}
h_{\omega}u=Eu \quad \Leftrightarrow \quad \forall n\in \Z,\ -(u_{n+1}+u_{n-1}) + \lambda V_{\omega^{(n)}} u_n = E u_n.
\end{equation}

In the continuous Schr\"odinger case we study the asymptotic exponential behavior of the functions $u\in L^2(\R)\otimes \C^D$ which satisfy

\begin{equation}\label{eq_valeurs_propres_continu}
H_{\omega}u=Eu \quad \Leftrightarrow \quad \forall x\in \R,\ -u''(x)+(V_{\mbox{per}} + \lambda V_{\omega}) (x) u(x) = Eu(x).
\end{equation}

We are not looking for a precise equivalent of $u$ at infinity but simply to know if $u$ behaves asymptotically like an exponential. This remark allows us to reduce the study of the exponential asymptotic behavior in $+\infty$ (resp. $-\infty$) of $u$ to the study of the successive jumps from site $n$ to site $n+1$ (resp. $-n$ to $-n-1$) for any natural number $n$. This leads to introduce the notion of transfer matrix and to apply the idea of transforming a second order linear equation into a first order system.

First, in the discrete case, the equation \eqref{eq_valeurs_propres_discret} is equivalent to the system
\begin{equation}\label{eq_mat_trans_discret}
\forall n\in \Z,\ \left(\begin{smallmatrix} u_{n+1}\\ u_n  \end{smallmatrix} \right) = \left( \begin{smallmatrix}
\lambda V_{\omega^{(n)}} -E & -\mathrm{I}_{\mathrm{D}} \\ \mathrm{I}_{\mathrm{D}} & 0 \end{smallmatrix} \right) \left( \begin{smallmatrix} u_{n}\\ u_{n-1}  \end{smallmatrix} \right). 
\end{equation}
For $n\in \Z$, let us set :
\begin{equation}\label{eq_def_transfer_mat_discret}
T_{\omega^{(n)}}(E) = \left( \begin{smallmatrix}
        \lambda V_{\omega^{(n)}} -E & -\mathrm{I}_{\mathrm{D}} \\ 
		\mathrm{I}_{\mathrm{D}} & 0 
                \end{smallmatrix}\right).
\end{equation}

\noindent The matrix $T_{\omega^{(n)}}(E)$ is called \textbf{transfer matrix} from $n$ to $n+1$. The sequence $(T_{\omega^{(n)}}(E))_{n\in \Z}$ is a sequence of random matrices in $\SpD$, the symplectic group of order $D$. The symplectic group $\SpD$ is the subgroup of $\mathrm{GL}_{\mathrm{2D}}(\R)$ constituted by the matrices $M$ satisfying
${^tM}JM=J$ where $J$ is the matrix of size $2D\times 2D$ defined by 
$J=\left(\begin{smallmatrix}
0 & -I_{\mathrm{D}} \\
I_{\mathrm{D}} & 0
\end{smallmatrix}\right)$. 
Moreover, if we assume that $(V_{\omega^{(n)}})_{n\in \Z}$ is a sequence of independent and \emph{i.i.d.} random variables, then the sequence $(T_{\omega^{(n)}}(E))_{n\in \Z}$ is also. 

Iterating (\ref{eq_mat_trans_discret}), the asymptotic behavior in $+\infty$ of $(u_n)_{n\in \Z}$ reduces to that of the product $T_{\omega^{(n)}}(E)\cdots T_{\omega^{(0)}}(E)$, and  in $-\infty$ to that of the product $(T_{\omega^{(-n+1)}}(E))^{-1}\cdots (T_{\omega^{(-1)}}(E))^{-1}$ by 
\begin{equation}\label{eq_transfert_matrices_iteration_positif}
 \forall n\geq 0,\ \left( \begin{smallmatrix} u_{n+1}\\ u_n  \end{smallmatrix}\right) =T_{\omega^{(n)}}(E)\cdots T_{\omega^{(0)}}(E) \left(\begin{smallmatrix} u_{0}\\ u_{-1}  \end{smallmatrix} \right)
\end{equation}
and
\begin{equation}\label{eq_transfert_matrices_iteration_negatif}
 \forall n<0 ,\ \left( \begin{smallmatrix} u_{n+1}\\ u_n  \end{smallmatrix} \right) = (T_{\omega^{(-n+1)}}(E))^{-1}\cdots (T_{\omega^{(-1)}}(E))^{-1}\left(\begin{smallmatrix} u_{0}\\ u_{-1}  \end{smallmatrix} \right).
\end{equation}
\vskip 3mm

In the continuous case, the equation \eqref{eq_valeurs_propres_continu} is equivalent to the differential system
\begin{equation}\label{eq_mat_trans_continu}
\left( \begin{smallmatrix} u\\ u'  \end{smallmatrix} \right)' = \left( \begin{smallmatrix}
0 & \mathrm{I}_{\mathrm{D}} \\ V_{\mbox{per}} + \lambda V_{\omega}-E & 0 \end{smallmatrix} \right) \left( \begin{smallmatrix} u\\ u'  \end{smallmatrix} \right). 
\end{equation}
Then, for any $n \in \Z$, we introduce the \textbf{transfer matrix} from $n$ to $n+1$, which is again denoted by $T_{\omega^{(n)}}(E)$, as the application which sends a solution $\left( \begin{smallmatrix} u\\ u'  \end{smallmatrix} \right)$ of the system \eqref{eq_mat_trans_continu} at time $n$ onto the solution at time $n+1$. The transfer matrix $T_{\omega^{(n)}}(E)$ is thus defined, for every $n\in \Z$, by the relation
\begin{equation}\label{eq_def_transfer_mat_continu}
\left( \begin{smallmatrix}
u(n+1) \\
u'(n+1) 
\end{smallmatrix} \right) = T_{\omega^{(n)}}(E) \left( \begin{smallmatrix}
u(n) \\
u'(n) 
\end{smallmatrix} \right).
\end{equation}
Since $T_{\omega^{(n)}}(E)$ is the solution of a Hamiltonian system of order $1$ at time $1$, $T_{\omega^{(n)}}(E)$ belongs to $\SpD$. Moreover, we will assume that the process $(V_{\omega^{(n)}})_{n\in \Z}$ is such that $(T_{\omega^{(n)}}(E))_{n\in \Z}$ is a sequence of \emph{i.i.d.} random matrices  in $\SpD$.
\bigskip

We are thus led, in the discrete and continuous cases, to study the asymptotic behavior of a sequence of \emph{i.i.d.} random matrices in $\SpD$. More precisely, we introduce the cocycle $\Phi_E : \Z\times \Omega \to \SpD$ defined for every $E\in \R$ by, 
\begin{equation} \label{eq_def_cocycle}
\forall n\in \Z,\ \forall \omega \in \Omega,\  \Phi_E(n,\omega) \;=\; \left\lbrace 
\begin{array}{lcl}
T_{\omega^{(n-1)}}(E)\cdots T_{\omega^{(0)}}(E) & \mbox{ if } & n >0 \\
I_{D} & \mbox{ if } & n=0 \\
(T_{\omega^{(n)}}(E))^{-1} \cdots (T_{\omega^{(-1)}}(E))^{-1} & \mbox{ if } & n <0
\end{array} \right.
\end{equation}

We now define the dominant Lyapunov exponents in plus and minus infinity associated with the cocycle $\Phi_E$.

\begin{defi}\label{def_upper_lyap}
The following limits exist and belong to $\R \cup \{ -\infty \}$:
\begin{equation}\label{eq_def_upper_lyap}
 \gamma^+(E)= \lim_{n\to +\infty} \frac{1}{n}\E( \log{||\Phi_E(n,\cdot)|| }) \ \ \mbox{ and }\ \ \gamma^-(E)= \lim_{n\to -\infty} \frac{1}{|n|}\E( \log||\Phi_E(n,\cdot) ||)
\end{equation}
We call them the dominant Lyapunov exponents in $\pm\infty$ associated to the cocycle $\Phi_E$.
\end{defi}

By equivalence of norms in finite dimension, $\gamma^+(E)$ and $\gamma^-(E)$ do not depend on the choice of the norm on $\MDR$. The existence of limits in \eqref{eq_def_upper_lyap} comes from the subadditivity of the sequence $( \E( \log ||\Phi_E(n,\cdot)||))_{n\in \Z}$.

It turns out that the dominant Lyapunov exponents can also be obtained as an almost sure limit and not only as a limit in expectation. This is the object of the Furstenberg-Kesten theorem\cite{FK60} which can be obtained by a direct proof\cite{BL85}, or which can be seen as a consequence of the subadditive ergodic Kingman theorem\cite{Kg73}.

The dominant Lyapunov exponents, when it is non-zero, allow us to understand the global exponential asymptotic behavior of the sequence of random matrices. However, it does not allow us to study more finely the asymptotic behavior of the sequence $||\Phi_E(n,\omega) v ||$ for any vector $v\in \R^D$. Indeed, the space $\R^D$ decomposed in several subspaces in which the exponential dynamics is given by a family of Lyapunov exponents.

\begin{defi}\label{def_lyapounov_GLN}
The Lyapunov exponents $\gamma_{1}^{\pm}(E),\ldots,\gamma_{D}^{\pm}(E)$ associated to the cocycle $\Phi_E$ are defined inductively by $\gamma_{1}^{\pm}(E)=\gamma^{\pm}(E)$ (the dominant Lyapunov exponents) and for $p\in \{2,\ldots, D\}$,
$$\sum_{i=1}^{p} \gamma_{i}^{\pm}(E) = \lim_{n \to \pm \infty} \frac{1}{|n|} \E(\log ||\wedge^{p} \Phi_E(n,\cdot) ||)$$
where $\wedge^p$ denotes the $p$-th exterior power of the matrix $\Phi_E(n,\cdot)$ (see\cite{BL85}).
\end{defi}
\bigskip

We see that the sums $\sum_{i=1}^{p} \gamma_{i}^{\pm}(E)$ are in fact the dominant Lyapunov exponents associated to the cocycle $\wedge^{p}\Phi_E $ when $p$ varies. Thus, the limits exist and these sums belong to $\R \cup \{ -\infty \}$. We give a characterization of these Lyapunov exponents as a function of the sequence of singular values of the matrices $\Phi_E(n,\omega) $, i.e. the square roots of the eigenvalues of the matrices ${^t\Phi_E(n,\omega)}\Phi_E(n,\omega)$.

\begin{prop}\label{prop_lyapvps}
If for $n\in \Z$, $\lambda_{1}(n,E,\omega)\geq \ldots \geq \lambda_{D}(n,E,\omega)>0$ are the singular values of $\Phi_E(n,\omega)$, then, for $\mathsf{P}$-almost every $\omega \in \Omega$,
$$\forall p\in \{1,\ldots, D\},\ \gamma_{p}^{\pm}(E)= \lim_{n \to \pm \infty} \frac{1}{|n|} \E(\log \lambda_{p}(n,E,\cdot)) = \lim_{n \to \pm \infty} \frac{1}{|n|} \log \lambda_{p}(n,E,\omega)$$
\end{prop}

In particular, this proposition justifies the numeration of Lyapunov exponents and the terminology of dominant Lyapunov exponent since it implies that $\gamma_1^{\pm}(E) \geq \cdots \geq \gamma_D^{\pm}(E)$. We find the demonstration in \cite{BL85} or in \cite{Ar98}.

Since the cocycle $\Phi_E$ takes values in $\SpD$, we have two additional properties. First, all the Lyapunov exponents are finite and moreover,
$$\forall n<0,\ \E(||\Phi_E(n,\cdot)||) = \E(||(\Phi_E(-n,\cdot))^{-1}||)= \E(||\Phi_E(-n,\cdot)||) $$
and $\gamma^{-}(E)=\gamma^{+}(E)$. Similarly, for any $i\in \{1,\ldots, 2D\}$, $\gamma_i^{-}(E)=\gamma_i^{+}(E)$. 
Indeed, the first equality comes from the fact that the shift on $\Omega$ preserves the product measure (which is also implied by the stronger assumption of the \emph{i.i.d.} character of the transfer matrices and would be false without the expectancy), and the second equality comes from the fact that for a symplectic matrix, its norm is equal to the norm of its inverse. In the following, we therefore omit the exponents $+$ and $-$ without ambiguity.

The second property is a property of symmetry. If $\gamma_{1}(E) \geq \ldots \geq \gamma_{2D}(E)$ are the Lyapunov exponents associated to $\Phi_E$, then, for every $i\in \{1,\ldots, D\}$, $\gamma_{2D-i+1}=-\gamma_{i}.$

This property of symmetry tells us that the Lyapunov exponents associated with an \emph{i.i.d.} sequence of symplectic random matrices can be grouped in pairs of opposite exponents and so it is sufficient to study the first $D$ exponents, $\gamma_1(E),\ldots, \gamma_D(E)$. In particular, if the Lyapunov exponents are distincts, because of the symmetry relation, the first $D$ are strictly positive and the following $D$ strictly negative: 
$$\gamma_{1}(E) > \cdots > \gamma_{D}(E) >0 > \gamma_{D+1}(E) > \cdots > \gamma_{2D}(E).$$ 

By Oseledets' theorem\cite{Os68}, the dominant Lyapunov exponent gives, when it is different from zero, the exponential rate of decay of the solution of \eqref{eq_valeurs_propres_discret} or \eqref{eq_valeurs_propres_continu} in some direction. Hence, in order to prove exponential decay of an eigenfunction of $h_{\omega}$ or $H_{\omega}$, one should first look at the non-vanishing of the Lyapunov exponents. Before discussing the question of positivity of the Lyapunov exponents, let us first recall what this positivity implies.

To simplify the discussion, we place ourselves in the case $D=1$ of scalar valued operators. In this case, we state a simple deterministic form of Oseledets' theorem due to Ruelle\cite{R79}.

\begin{thm}[Oseledets in $\mathrm{SL}_2(\R)$]\label{thm_oseledets_SL2}
Let $(T_n)_{n\in \N}$ a sequence in $\mathrm{SL}_2(\R)$ such that  
$$\lim_{n\to +\infty} \frac{1}{n} \ln ||T_n|| =0 \mbox{ and } \gamma=\lim_{n\to +\infty} \frac{1}{n} \ln (||T_n \cdots T_1||)>0.$$
Then there exists a subspace $V_{-}\subset \R^2$ of dimension $1$ such that
\begin{enumerate}
 \item $\forall v\in V_{-},\ v\neq 0,\ \displaystyle\lim_{n\to +\infty} \frac{1}{n} \ln (||T_n \cdots T_1 v||) =-\gamma$.
\item   $\forall v\notin V_{-},\ \displaystyle\lim_{n\to +\infty} \frac{1}{n} \ln (||T_n \cdots T_1 v||) =\gamma$.
\end{enumerate}
\end{thm}

We apply Theorem \ref{thm_oseledets_SL2} to the sequence of transfer matrices for $D=1$ and for $n>0$. Assume that $E\in \R$ is such that $\gamma(E)>0$. By Oseledets' theorem in $\mathrm{SL}_2(\R)$, there exist $\mathsf{P}$-almost surely only exponentially increasing or exponentially decreasing solutions to the equations $h_{\omega}u=Eu$ or $H_{\omega}u=Eu$. An exponentially decreasing solution (in $+\infty$), is obtained only for an initial condition $v_{+\infty}\in V_{-}^{+}$. Any other initial condition leads to an exponentially increasing solution in $+\infty$. 

Still by Oseledets' theorem, an exponentially increasing solution in $-\infty$  is obtained only for an initial condition $v_{-\infty}\in V_{-}^{-}$. Thus, to obtain an eigenvector in $\ell^2(\Z)$ or in $L^2(\R)$, we must have $\mathrm{Span}(v_{+\infty})=\mathrm{Span}(v_{-\infty})$ which is not a priori satisfied.

It is also necessary to pay attention to the fact that what we have just stated depends on $E$. If $E$ varies in a non-countable set (for example an interval of $\R$), the set of $\omega\in \Omega$ for which for any $E$ the Lyapounov exponent is not strictly positive, could be of non-zero measure. For example, one cannot say that $\mathsf{P}$-almost surely, for any $E$, any solution of $h_{\omega}u=Eu$ or $H_{\omega}u=Eu$ is exponentially increasing or decreasing. This requires further analysis. However, it already implies the absence of almost-sure absolutely continuous spectrum, using Kotani theory. For $S\subset \R$, denote by $\overline{S}^{\mathrm{ess}}$ its essential closure and for $j\in \{1,\ldots,D\}$, set 
$$Z_j=\{E\in \R\ |\ \exists \ l_1,\ldots, l_{2j}\in \{1,\ldots,2D \},\ \gamma_{l_1}(E)=\cdots=\gamma_{l_{2j}}(E)=0 \}.$$
Then we have the following generalization of Ishii-Pastur's theorem to quasi-one-dimensionnal Schr\"odinger operators due to Kotani and Simon.

\begin{thm}[Kotani-Simon\cite{KS88}]\label{thm_ishii_pastur_kotani}
\begin{enumerate}
 \item The set $Z_j$ is the essential support of the absolutely continuous spectrum of multiplicity $2j$. 
\item There is no absolutely continuous spectrum of odd multiplicity.
\item We have 
$$\Sigma_{\mathrm{ac}}=\overline{Z_D}^{\mathrm{ess}}=\overline{ \{E\in \R\ |\ \gamma_{1}(E)=\cdots=\gamma_{2D}(E)=0 \} }^{\mathrm{ess}}.$$
\end{enumerate}
\end{thm}

In particular, this implies that if for Lebesgue-almost every $E\in \R$, $\gamma_1(E)\geq \cdots \geq \gamma_D(E) >0$, then $\Sigma_{\mathrm{ac}}=\emptyset$. The R.A.G.E. theorem\cite{RS3,AW15} assures us then that there are no diffusive states for $H_{\omega}$ (respectively $h_{\omega}$) and this almost surely in $\omega$.

Let us return to the question of positivity of Lyapunov exponents. Intuitively, in order to hope to obtain a strictly positive limit in the formula \eqref{eq_def_upper_lyap}, one must make sure that when the random parameter varies, a large number of products and inverse of transfer matrices have a sufficiently large norm so that on average and after division by $n$, the limit is not zero. For example, if all products of transfer matrices are bounded, the limit in \eqref{eq_def_upper_lyap} will automatically be zero. We must therefore study the set that contains all the products of transfer matrices when the random parameter varies.

We have assumed above that the transfer matrices are \emph{i.i.d.}. Let us note $\mu_E$ the common law of these transfer matrices. The support of $\mu_E$ gives us the set of values that a transfer matrix can take when we vary the random parameter. If we then want to make products and inverse of such matrices as in $\Phi_E$, the natural algebraic object to consider is the group generated by the support of $\mu_E$.

\begin{defi}\label{def_groupe_furstenberg_intro}
The \textbf{Furstenberg group} $G_{\mu_E}$ associated to the sequence of transfer matrices $(T_{\omega^{(n)}}(E))_{n\in \Z}$ is the closure of the group generated by the support of the common law of $T_{\omega^{(n)}}(E)$ : $G_{\mu_E} = \overline{\langle\supp \mu_E\rangle}.$
\end{defi}

We choose to include the closure for the topology induced by the usual topology on the space of matrices in the definition of the Furstenberg group because this makes it a closed subgroup of the linear real group $\mathrm{GL}_{2\mathrm{D}}(\R)$, hence a Lie group. In the following it is essential that the Furstenberg group be provided with a structure of a linear real Lie group, this allowing to use all the techniques proper to these groups, in particular to use their Lie algebra. 
\bigskip

The strict positivity of the Lyapunov exponents comes down to the study of the "size" of the Furstenberg group. If this group is large enough in some sense to be specified, there will be strict positivity of the Lyapunov exponents and they will be distincts. Let us now define two properties of the Furstenberg group which, when they are both satisfied, imply that the Lyapunov exponents are distincts and positive.

\begin{defi}[\textbf{p-contractivity}]\label{def_p_contractant}
Let $T\subset \mathrm{GL}_{2\mathrm{D}}(\R)$ and $p\in \{ 1,\ldots,D \}$. We say that $T$ is \textbf{$p$-contracting} if there exists a sequence $(M_{n})_{n\in \N}$ in $T$ such that the following limit exists:
\[
\lim_{n\to \infty} \frac{\wedge^{p}M_n}{||\wedge^{p}M_n||} = M
\]
and is a matrix of rank $1$.
\end{defi}

We introduce the $p$-Lagrangian submanifold of $\R^{2D}$. Let $(e_{1},\ldots ,e_{2D})$ be the canonical basis of $\R^{2D}$.

For all $p\in \{1,\ldots ,D\}$, let $L_{p}=\mathrm{Span}(\{Me_1 \wedge \ldots \wedge Me_p \ | \ M\in \SpD \})$. It is called the $p$-Lagrangian subspace of $\R^{2D}$.
 
\begin{defi}[\textbf{$L_p$-strong irreducibility}]\label{def_lpstrong}
Let $T$ be a part of $\SpD$ and $p$ an integer in $\{ 1,\ldots,D \}$. We say that $T$ is \textbf{$L_p$-strongly irreducible} if there is no finite union $W$ of proper subspaces of $L_p$ such that $\wedge^{p}M(W)=W$ for all $M\in T$.
\end{defi}

By proper subspace, we mean a subspace of $L_{p}$ different from $L_{p}$ and $\{ 0 \}$. We now state a theorem of separation of Lyapunov exponents in $\SpD$. This theorem is taken from \cite{BL85} and was initially proven by Guivarch and Raugi\cite{GR85}.

\begin{thm}[\cite{BL85,GR85}]\label{thm_lyapmainSPN}
Let $E\in \R$ and $p\in \{1,\ldots ,D\}$. Assume that $G_{\mu_E}$ is $p$-contracting and $L_p$-strongly irreducible. Then $\gamma_{p}(E) > \gamma_{p+1}(E)$ and for any non-zero $x\in L_p$ and for $\mathsf{P}$-almost any $\omega$, 
\[
\lim_{n\to \infty} \frac{1}{n} \log ||\wedge^{p}(\Phi_E(n,\omega))x||=\sum_{i=1}^{p} \gamma_i(E).
\]
\end{thm}

This result will especially interest us in the form of the following corollary.

\begin{cor}\label{cor_lyapmain}
Let $E\in \R$. Assume that $G_{\mu_E}$ is $p$-contracting and $L_p$-strongly irreducible for all $p \in \{1,\ldots ,D\}$. Then, the Lyapunov exponents associated to $\Phi_E$ are distinct and in particular: 
\[
\gamma_{1}(E) > \gamma_{2}(E) > \ldots > \gamma_{D}(E) > 0.
\]
\end{cor}
\vskip 5mm

The link is made between a dynamical property associated to a cocycle with values in the symplectic group, in this case the separation of its Lyapunov exponents, and more geometrical properties of an algebraic object associated to this cocycle, the Furstenberg group. This is a very useful criterion in practice, since it allows us not to have to study directly limits of products of transfer matrices, but simply to reduce ourselves to finite products of such matrices in the Furstenberg group.

\section{Localization criterion for quasi-1D models of Schr\"odinger type}\label{sec_localization_criterion}

In this Section we present an algebraic localization criterion for quasi-one-dimensional models of Schr\"odinger type, either discrete or continuous.  

Let $D\geq 1$ be an integer. Let $(\Omega,\mathcal{A},\mathsf{P})$ be a complete probability space and let $\omega \in \Omega$. 

In the discrete case, we consider random operators acting on $\ell^2(\Z)\otimes \C^D$ by 
\begin{equation}\label{model_H_disc}
\forall u\in \ell^2(\Z)\otimes \C^D,\ \forall n\in \Z,\  (h_{\omega}u)_{n} =-(u_{n+1}+u_{n-1}) + V_{\omega^{(n)}}u_n 
\end{equation}
where $(V_{\omega^{(n)}})_{n\in \Z}$ is a sequence of \emph{i.i.d.} random variables taking values in the real symmetric matrices. 

In the continuous case, we consider random operators acting on $L^2(\R)\otimes \C^D$ by
\begin{equation}\label{model_H_cont}
H_{\omega} = -\frac{\dd^2}{\dd x^2}\otimes I_{\mathrm{D}} + \sum_{n\in \Z} V_{\omega}^{(n)} (x-\ell n),
\end{equation}
where $I_{\mathrm{D}}$ is the identity matrix of order $D$ and $\ell>0$ is a real number.

It is assumed that for any $n\in \Z$, the functions $x\mapsto V_{\omega}^{(n)} (x)$ have values in the space of symmetric real matrices, are  supported in $[0,\ell]$ and are uniformly bounded in $x$, $n$ and $\omega$. 

We also assume that the sequence $(V_{\omega}^{(n)})_{n\in \Z}$ is a sequence of \emph{i.i.d.} random variables on $\Omega$. Finally, we assume that the potential $x\mapsto \sum_{n\in \Z} V_{\omega}^{(n)} (x-\ell n)$ is such that the family  $\Hof$ of random operators is $\ell \Z$-ergodic. As bounded perturbations of $-\frac{\dd^2}{\dd x^2}\otimes I_{\mathrm{D}}$, the operators $H_{\omega}$ are self-adjoint on the Sobolev space $H^2(\R)\otimes \C^D$ and thus, for any $\omega\in \Omega$, the spectrum of $H_{\omega}$ is included in $\R$.

Let us denote by $\Sigma \subset \R$ the almost-sure spectrum of either the family $\Hof$ or the family $\{ h_{\omega} \}_{\omega \in \Omega}$. For $E\in \R$, let $G_{\mu_{E}}$ be the Furstenberg group of either $\Hof$ or  $\{ h_{\omega} \}_{\omega \in \Omega}$. 

\begin{thm}[Localization criterion\cite{KLS90,B09MPAG}]\label{thm_localization}
Let $I\subset \R$ be a compact interval such that $\Sigma \cap I\neq \emptyset$ and let $\tilde{I}$ be an open interval containing $I$ and such that for any $E\in \tilde{I}$, $G_{\mu_{E}}$ is $p$-contracting and $L_p$-strongly irreducible for all $p$ in $\{1,\ldots, D\}$. Then, $\Hof$ (resp. $\{ h_{\omega} \}_{\omega \in \Omega}$) has both the dynamical localization and Anderson localization properties on $\Sigma \cap I$.
\end{thm}

In the discrete case, this criterion was obtained by Klein, Lacroix and Speis\cite{KLS90}. In the continuous case, this result was obtained in \cite{B09MPAG}.
\bigskip

The demonstration of this theorem, as detailed in \cite{KLS90,B09MPAG}, follows the plan: 
\begin{enumerate}
\item The fact that the Furstenberg group is $p$-contracting and $L_p$-strongly irreducible for all $p$ in $\{1,\ldots, D\}$ implies not only the separability of the Lyapunov exponents but also the existence of an integral formula for these exponents. This integral formula allows us to demonstrate the H\"older regularity of the Lyapunov exponents.

\item With the help of a Thouless formula, we deduce the same H\"older regularity for the integrated density of states associated with $\Hof$.

\item The H\"older regularity of the integrated density of states implies a weak Wegner estimate, adapted to the case of the Bernoulli randomness.

\item With the Wegner estimate, one can then apply a multi-scale analysis scheme. This requires in particular to demonstrate an initial length scale estimate.
\end{enumerate}

This is the approach adopted in the continuous scalar case by Damanik, Sims and Stolz\cite{DSS02} and in the discrete quasi-one-dimensional case by Klein, Lacroix and Speis\cite{KLS90}. In \cite{B09MPAG}, we mix the results of these two references to prove Theorem \ref{thm_localization}.
\bigskip

\textbf{Step 1.} As explained in Section \ref{sec_lyap} the positivity of the Lyapunov exponents is not a sufficient condition to get the existence of eigenfunctions exponentially decaying to $0$ at infinity. Actually, Theorem \ref{thm_lyapmainSPN} can be precised with an integral formula for the sum of Lyapunov exponents.

\begin{thm}[\cite{BL85}]\label{thm_lyap_int_rep}
Let $E$ be a real number and $p$ an integer in $\{1,\ldots ,D\}$. Let us assume that $G_{\mu_E}$ is $p$-contracting and $L_p$-strongly irreducible. Then, there exists a unique $\mu_{E}$-invariant probability measure on $\mathbb{P}(L_{p})= \{ \bar{x} \in \mathbb{P}(\wedge^{p} \R^{2D})\ | \ x \in \ L_{p} \}$, denoted by $\nu_{p,E}$, such that: 
\begin{equation}\label{eq_lyap_formule_integrale}
\int_{\SpD\times \mathbb{P}(L_{p})} \log \frac{||\wedge^{p} M x||}{||x||}\dd\mu_{E}(M)\dd\nu_{p,E} (\bar{x}) = \sum_{i=1}^{p} \gamma_i(E)
\end{equation}
\end{thm}
\bigskip
The fact that we have an integral representation of the Lyapunov exponents involving the measure $\nu_{p,E}$, implies that to study the regularity of these exponents as a function of $E$  it is sufficient to study the regularity of this measure seen as a function of $E$. Applying this approach, one proves the following general theorem of H\"older regularity of Lyapunov exponents\cite{B08RMP}.

\begin{thm}[\cite{B08RMP}]\label{thm_holder_lyap}
Let $I$ be a compact interval in $\R$ such that for every $E\in I$ :
\begin{enumerate}
\item  $G_{\mu_E}$ is $p$-contracting and $L_p$-strongly irreducible for every $p\in \{1,\ldots ,D\}$.
\item There exist $C_{1}>0$, $C_{2}>0$ independent of $n,\omega,E$ such that
\begin{equation}\label{estim1_holder}
\forall p\in \{1,\ldots ,D\},\ ||\wedge^{p} T_{\omega^{(n)}}(E)||^{2} \leq \exp( pC_{1} + p|E| +p) \leq C_{2}.
\end{equation}
\item There exist $C_{3}>0$ independent of $n,\omega,E$ such that 
\begin{equation}\label{estim2_holder}
\forall p\in \{1,\ldots ,D\},\ \forall E,E'\in I,\ ||\wedge^{p} T_{\omega^{(n)}}(E)-\wedge^{p} T_{\omega^{(n)}}(E')||\leq C_{3} |E-E'|.
\end{equation}
\end{enumerate}
\vskip 2mm

\noindent Then there exist two real numbers $\alpha >0$ and $C>0$ such that  :
$$\forall p\in \{1,\ldots D\},\ \forall E,E' \in I,\ |\gamma_{p}(E)-\gamma_{p}(E')| \leq C |E-E'|^{\alpha}.$$
\end{thm}

To prove this result, we use a result on negative cocycles as stated in \cite[Proposition IV 3.5]{CL90}. It is on this precise point that we use the separability of Lyapunov exponents induced by the first hypothesis of Theorem \ref{thm_holder_lyap}. We also need estimates on the Laplace operators on the H\"older spaces as in \cite[Proposition V 4.13]{CL90}, which use the estimates (\ref{estim1_holder}) and (\ref{estim2_holder}). Finally by using the decomposition of these Laplace operators given in \cite[Proposition IV 3.12]{CL90} and involving the invariant measure, we show the H\"older continuity of $E\mapsto \nu_{p,E}$ on $I$. At this stage of the demonstration of Theorem \ref{thm_localization}, an important part of the work is to show that the transfer matrices associated to the  family $\Hof$ verify the estimates \eqref{estim1_holder} and \eqref{estim2_holder}. If this is relatively obvious in the case of discrete quasi-one-dimensional Schr\"odinger-type operators, because of the very explicit character of the transfer matrices, the estimates are less obvious in the continuous case and are related to a priori estimates on solutions of differential equations. For this we refer to \cite[Lemma 2 and Lemma 3]{B08RMP}.

\textbf{Step 2.} The integrated density of states is a counting function of the energy levels located under a fixed value of energy $E$. For operators whose spectrum is continuous or dense, the question of the existence of such a function arises because a naive definition inevitably leads to the definition of a function equal to infinity at any point beyond the bottom of the spectrum. To get around this difficulty, we define the integrated density of states using a thermodynamical limit. 

Let us fix $\omega \in \Omega$. We start by restricting $H_{\omega}$ to intervals of finite length in $\R$. Let $L$ be an integer larger than $1$ and let $\Lambda=[-\ell L,\ell L]\subset \R$ the interval centered in $0$ and of length $2\ell L$. We set:
\begin{equation}\label{model_IDS_cube}
H_{\omega}^{(\Lambda)}=\left(-\frac{\dd^2}{\dd x^2}\right)^{(\Lambda)}\otimes I_{\DD}+\sum_{n\in \Z} V_{\omega}^{(n)}(x-\ell n) 
\end{equation}
the restriction of $H_{\omega}$ acting on $L^{2}(\Lambda)\otimes \C^{D}$ with Dirichlet (or Neumann) boundary conditions.

\begin{defi}\label{def_IDS}
  The integrated density of state of $\Hof$ is the increasing function from $\R$ to $\R_{+}$, $E \mapsto N(E)$ where for any real number $E$, $N(E)$ is defined as : 
\begin{equation}\label{IDSdef}
N(E)=\lim_{L\to +\infty} \frac{1}{|\Lambda|} \# \left\{ \lambda \leq E |\ \lambda \in \sigma (H_{\omega}^{(\Lambda)}) \right\}
\end{equation}
where $|\Lambda|=2\ell L$ is the length of $\Lambda$.
\end{defi}

There is a double existence problem in the expression (\ref{IDSdef}). We must  prove that the cardinal $\# \{ \lambda \leq E |\ \lambda \in \sigma (H_{\omega}^{(\Lambda)})\}$ is finite for all $E$ fixed and then prove the existence of the limit. The answer to these two problems is given by the existence of an   $L^{2}$ kernel for the one-parameter group $(\ee^{-tH_{\omega}^{(\Lambda)}})_{t>0}$. This kernel is obtained using a Feynman-Kac formula involving an ordered exponential\cite[Proposition 1]{B08RMP}.

Combining the results of \cite{KS88} and \cite{DSS02}, we show in \cite{B08RMP} a Thouless formula which makes the link between the density of states and the sum of the positive Lyapunov exponents. This link is made through a function $w$ called Kotani function\cite{Ko84} and by using tools of harmonic analysis. We introduce the space of Herglotz functions: $\mathcal{H}=\{ h : \C_{+}\to \C_{+} \ |\ h\ \mathrm{is}\ \mathrm{holomorphic}\ \mathrm{on}\ \C_{+}  \}$ and consider the subspace of $\mathcal{H}$, $\mathcal{W}= \{ w\in \mathcal{H}\ |\ w,\ w',\ -\mathrm{i}w\in \mathcal{H} \}.$

One proves in \cite[Proposition 8]{B08RMP} that the Kotani function $w$ is in this space $\mathcal{W}$. Since $-(\gamma_{1}+\ldots +\gamma_{D})$ and $\pi N$ are respectively, in the tangential limit, the real and imaginary parts of the function $w$ which is in the space $\mathcal{W}$, the harmonic analysis developed for this space by Kotani in \cite{Ko84} gives us that these two functions are linked by an integral relation and one deduces from it the following Thouless formula.

\begin{thm}[Thouless formula\cite{B08RMP}]\label{thm_thouless}
For Lebesgue-almost any $E\in \R$, we have : 
\begin{equation}\label{eq_thouless_formula}
(\gamma_{1}+\ldots+\gamma_{\DD})(E)= -\alpha+\int_{\R} \log \left( \left| \frac{E'-E}{E'-\mathrm{i}}\right| \right)~\dd \mathfrak{n}(E')
\end{equation}
where $\alpha$ is a real number independent of $E$ and $\mathfrak{n}$ is the density of states. Moreover, if $I\subset \R$ is an interval on which $E\mapsto (\gamma_{1}+\ldots +\gamma_{\DD})(E)$ is continuous, then (\ref{eq_thouless_formula}) is true for any $E\in I$.
\end{thm}

As for the discrete or continuous Anderson operators with scalar values\cite{CKM87,DSS02}, or discrete matrix-valued ones\cite{KS88}, one can use the Thouless formula to show that the integrated density of states has the same H\"older continuity as the Lyapunov exponents. This is the object of the main result of \cite{B08RMP}.

\begin{thm}[\cite{B08RMP}]\label{thm_holderids}
Let $I$ be a compact interval of $\R$ and let $\tilde{I}$ be an open interval containing $I$. If the Furstenberg group $G_{\mu_{E}}$ of $\Hof$ is $p$-contracting and $L_p$-strongly irreducible for any $p\in \{1, \ldots ,\DD\}$ and any $E\in \tilde{I}$, then the integrated density of states associated to $\Hof$ is H\"older continuous on $I$.
\end{thm}

The proof of Theorem \ref{thm_holderids} is based on the Thouless formula and on the properties of the Hilbert transform.
 
Also note that it is possible to recover H\"older continuity of the integrated density of states in the Bernoulli case by using general functional analysis results of\cite{SVW98} which are based upon the supersymmetric formalism.

\textbf{Step 3.}  For a continuous  Anderson model like  \eqref{eq_def_Anderson_continu} or a discrete one like \eqref{eq_def_Anderson_discret} in any dimension, a Wegner estimate is obtained when there exists a constant $C_W>0$ and real exponents $\alpha,\beta>0$ such that for any interval $I\subset \R$ and any cube $\Lambda \subset \R^d$ (or $\Lambda \subset \Z^d$), 
\begin{equation}\label{eq_estimee_wegner_generale}
    \mathsf{P} \left( \left\{ H_{\omega}^{(\Lambda)} \mbox{ has an eigenvalue in } I \right\} \right) \leq C_W |I|^{\alpha}\cdot |\Lambda|^{\beta}   
\end{equation}
where for any $\omega$, $H_{\omega}^{(\Lambda)}$ is the restriction of $H_{\omega}$ to the cube $\Lambda$ with Dirichlet (or Neumann) boundary conditions and $|\cdot|$ denotes the Lebesgue measure, either in $\R$ for $I$ or in $\R^d$ for $\Lambda$ (or the cardinal of $\Lambda$ if $\Lambda \subset \Z^d$).

Such an estimate is an essential ingredient in the application of a multi-scale analysis scheme. The Wegner estimate also plays a central role in the results of spectral statistics, and in this context, it is necessary to have linear estimates in the energy and in the volume of $\Lambda$. In this case we speak about optimal Wegner's estimate. For example, we can prove an optimal Wegner estimate for a discrete and scalar Anderson $h_{\omega}$ operator with random variables whose common law has a density:

\begin{equation}\label{eq_wegner_optimale}
 \E\left(\mathrm{Tr}\left( \mathbf{1}_{I} \left(h_{\omega}^{(\Lambda)}\right) \right)\right) \leq C_{W}\cdot |I| \cdot |\Lambda|
\end{equation}
where $\mathbf{1}_{I} \left(h_{\omega}^{(\Lambda)}\right)$ is the spectral projector of $h_{\omega}^{(\Lambda)}$ on the interval $I$. Let's mention that \eqref{eq_wegner_optimale} is related to \eqref{eq_estimee_wegner_generale} via the Markov inequality which implies that in general, 
$$ \mathsf{P} \left( \left\{ h_{\omega}^{(\Lambda)} \mbox{ has an eigenvalue in } I \right\} \right)\leq \mathrm{Tr}\left( \mathbf{1}_{I} \left( h_{\omega}^{\Lambda} \right)\right).$$

The optimal Wegner estimate is also obtained under additional assumptions on the randomness in the continuous case. For these results, we refer to \cite{His08,Ves08}. In general, the proof of a Wegner estimate relies on the use of spectral averaging\cite{AW15}, which explains the assumption of regularity on the algebra in the sense that the law of the random variables that define the Anderson potential is required to have a density. However, in the case of Anderson-Bernoulli operators, a Wegner estimate such as \eqref{eq_estimee_wegner_generale} cannot be true for any interval $I$. Indeed, if the random variables in the Anderson's potential follow a Bernoulli distribution of parameter $\frac12$, then the probability of the  event $\left\{ H_{\omega}^{(\Lambda)} \mbox{ has an eigenvalue in } I \right\}$ is lower bounded by $2^{-|\Lambda|}$ and for $I$ of too small length, \eqref{eq_estimee_wegner_generale} cannot be valid. In the case of the Bernoulli randomness, it is still possible to show a weaker form of the Wegner estimate for the Anderson model. 

\begin{thm}[\cite{B09MPAG}]\label{thm_wegner_bernoulli}
    Let $I\subset \R$ be a compact interval and $\tilde{I}$ an open interval, $I \subset \tilde{I}$, such that, for any $E\in \tilde{I}$, $G_{\mu_E}$ is $p$-contracting and $L_p$-strongly irreducible for any $p\in \{1,\ldots,D\}$. Then, for any $\beta \in (0,1)$ and any $\kappa >0$, there exist $L_0\in \N$ and $\xi>0$ such that,
\begin{equation}\label{eq_thm_wegner}
\mathsf{P}\left( d\left(E, \sigma(H_{\omega}^{(\Lambda)}) \right) \leq \ee^{-\kappa(\ell L)^{\beta}} \right) \leq \ee^{-\xi(\ell L)^{\beta}},
\end{equation}
for all $E\in I$ and all $L\geq L_0$.
\end{thm}

The proof of Theorem \ref{thm_wegner_bernoulli} is based upon specific one-dimensional tools. Contrary to the case of random variables whose law have a  density, here the regularity of the integrated density of states is not a consequence of Wegner's estimate, but a key ingredient of its demonstration. In dimension 1, the H\"older regularity of the integrated density of states is obtained directly from that of the Lyapunov exponents and from the Thouless formula as explained in Step 2.

We begin by showing, under the hypotheses of Theorem \ref{thm_wegner_bernoulli}, that the probability that there exists an eigenvalue of $H_{\omega}^{(\Lambda)}$ in an interval $[E-\varepsilon, E+\varepsilon]$ associated to a normalized eigenfunction and whose eigenvalues at the edges of $[-\ell L, \ell L]$ are controlled by $\varepsilon$, is smaller than a multiple of $L$ and $\varepsilon^{\alpha}$ where $\alpha$ is the exponent of H\"older continuity of the integrated density of states. This first result is thus based on the H\"older continuity of the integrated density of states but also on a priori estimates of the solutions of the differential deterministic system $-u''+Vu=0$ where $V$ is a locally integrable function with matrix values. Also used in this first demonstration are the fact that the Anderson potential is invariant in law by translation, and the law of large numbers to relate a given sequence of events to the integrated density of states. 

The rest of the proof of Theorem \ref{thm_wegner_bernoulli} then relies on the estimate of the probabilities of the intersections of the event in \eqref{eq_thm_wegner} with a family of events for which the norms of the products of transfer matrices allowing to pass from one edge to the other of $[-\ell L, \ell L]$ are strictly minorized by a strictly positive constant of the form $\ee^{\theta (\ell L)^{\beta}}$. We also use the Lipschitzian character of the transfer matrices seen as functions of the energy, which is the case in the considered model. Again, the use of transfer matrices makes the proof of the inequality \eqref{eq_thm_wegner} quite specific to dimension 1.

\textbf{Step 4.}  The last step of the proof of Theorem \ref{thm_localization} is the implementation of the multi-scale analysis. It is a procedure of proof by induction on the sizes of the cubes to which we restrict the studied random family of operators.

The idea is the following: the exponential growth of the eigenfunctions is implied by that of the norm of the resolvent of the restricted operators at larger and larger cubes. To show that the norm of the local resolvent decreases exponentially with the size of the cubes, we start by showing an upper bound  of this norm by an exponentially small term in a cube of initial size. From there we extend this exponential domination of the norm of the resolvent to all the cubes of a larger size, which is a positive power of the initial size. For this we have to show that we almost surely keep an exponential estimate of the resolvent. In particular, the estimates that we want for the resolvent must be uniform in energy when this one is in a compact interval. It is thus necessary to make sure when one passes from a scale of cubes to a larger scale that one avoids the possible energies which would make explode the norm of the resolvent: those in the spectrum of the restricted operator. This is where the Wegner estimate comes in : it ensures that with a probability exponentially close to 1, the energies in the interval considered are at least at a strictly positive distance from the spectrum of the restricted operator.

For more details about multi-scale analysis in general, we refer to the articles of Abel Klein\cite{Kl08} and Germinet and Klein\cite{GK01}. For details of the use of multi-scale analysis in the proof of Theorem \ref{thm_localization}, we refer to\cite{B09MPAG}.

In the case of one-dimensional operators, it is legitimate to ask whether a more direct approach, based solely on studying the sequence of transfer matrices and proving large deviations inequalities for this sequence, can lead to localization. This approach has been successfully adopted in \cite{BDFGVW19,JZ19} for the discrete case and in \cite{BDFGVW19b} for the continuous case and gives a more elementary proof of localization in dimension $d=1$. Extending the methods of \cite{JZ19}, dynamical localization was obtained in\cite{LX20}. Also noteworthy is the purely dynamical proof of Anderson localization in \cite{GoKl21}, based on a parametric version of Furstenberg's theorem. Finally, we mention a recent proof of localization for a generalization of the discrete quasi-one-dimensional model in \cite{MS21}, based on the same techniques as \cite{BDFGVW19, JZ19}. For more general results about large deviations theorems, see also\cite{DuSi20}.

\section{Localization results for quasi-1D models}\label{sec_localization_results}

In this Section, we review localization results for different type of quasi-one-dimensional random models for which the randomness appears through random variables which can be Bernoulli variables. We do not present other localization results which do not include this case. We classify the results by order of algebraic complexity of their associated Furstenberg group.

In all the studied models, the randomness appear in the following way. For $i\in \{1,\ldots ,D\}$, let $(\tilde{\Omega}_i,\tilde{\mathcal{A}}_i,\tilde{\mathsf{P}}_i)$ a complete probability space and let us pose 
\begin{equation}\label{eq_def_espace_proba_tensor}
(\Omega,\mathcal{A},\mathsf{P})=\left(\bigotimes_{n\in \Z} \tilde{\Omega}_1\otimes \cdots \otimes \tilde{\Omega}_D,\bigotimes_{n\in \Z} \tilde{\mathcal{A}}_1\otimes \cdots \otimes\tilde{\mathcal{A}}_D, \bigotimes_{n\in \Z}\tilde{\mathsf{P}}_1\otimes \cdots \otimes\tilde{\mathsf{P}}_D \right) 
\end{equation}

Let $(\omega_{1}^{(n)})_{n\in \Z}, \ldots, (\omega_{D}^{(n)})_{n\in \Z}$ be  sequences of \emph{i.i.d.} real random variables respectively on $(\tilde{\Omega}_1,\tilde{\mathcal{A}}_1, \tilde{\mathsf{P}}_1)$,$\ldots$, $(\tilde{\Omega}_D,\tilde{\mathcal{A}}_D, \tilde{\mathsf{P}}_D)$ and of respective common laws  $\nu_1,\ldots, \nu_D$ whose supports  $\supp \nu_i$ contains at least two different points $a_i$ and $b_i$ (for example $0$ and $1$ if they are Bernoulli variables) and are bounded. We also set $\omega^{(n)}=(\omega_{1}^{(n)}, \ldots, \omega_{D}^{(n)})$ of law $\nu_1 \otimes \cdots \otimes \nu_D$.

\subsection{The Furstenberg criterion}

We start reviewing localization results for scalar-valued operators, corresponding to the case $D=1$ in the definition of quasi-one-dimensional random models.

Due to Theorem \ref{thm_localization}, to obtain a localization result on some interval $I$, it suffices to show that the Furstenberg group associated to the studied model is  $p$-contracting and $L_p$-strongly irreducible for all $p$ in $\{1,\ldots, D\}$. In the case $D=1$, these properties reduces to the assumptions of the following result due to Furstenberg.

\begin{thm}[Furstenberg \cite{F63}]\label{thm_furstenberg_GLN}
Let $E\in \R$, $(T_{\omega^{(n)}}(E))_{n\in \Z}$ a sequence of \emph{i.i.d.} random matrices in $\SLD$ of common law $\mu_E$ and let $G_{\mu_E}$ be its Furstenberg group. Assume that $G_{\mu_E}$ is not compact and $G_{\mu_E}$ is strongly irreducible \emph{i.e.}, there is no finite family $V_1,\ldots, V_k$ of strict subspaces of $\R^2$ such that : $\forall M\in G_{\mu_E},\ M(V_1 \cup \cdots \cup V_k)=V_1 \cup \cdots \cup V_k.$

Then there exists $\gamma(E) >0$ such that for every $x\in \R^2$, $x\neq 0$, $\mathsf{P}$-almost surely,
$$\lim_{n\to +\infty} \frac{1}{n} \log(||\Phi_E(n,\omega) x||)=\lim_{n\to +\infty} \frac{1}{n} \log(||\Phi_E(n,\omega)||)= \gamma(E).$$
where $\Phi_E$ is defined as in \eqref{eq_def_cocycle}. Moreover,  there exists a unique $\mu_{E}$-invariant probability measure on $\mathbb{P}(\R^2)$, denoted by $\nu_{E}$, such that: 
$$\gamma(E) = \int_{\SLD \times \mathbb{P}(\R^2)} \log \frac{|| M x||}{||x||}\dd\mu_{E}(M)\dd\nu_{E} (\bar{x}) $$
\end{thm}
Here $P(\R^2)$ is the projective space of $\R^2$  and for $x\in \R^2\setminus \{0 \}$, $\bar{x}$ denotes its class in $P(\R^2)$.

The Furstenberg theorem is sufficient to deal with the case of models for which the transfer matrices are in $\SLD$ and for which the question of strict positiveness arises only for one Lyapunov exponent, the dominant exponent.

\subsubsection{The discrete one-dimensional Anderson model}

We are going to apply Furstenberg's theorem to a first example of a sequence of random matrices in $\SLD$  coming from the discrete Anderson model in dimension $1$ and with scalar values. We use the notations introduced in \eqref{eq_def_espace_proba_tensor} and since here $D=1$ we omit the index $1$ in the notations of $(\omega_{1}^{(n)})_{n\in \Z}$ or $\nu_1$.

Consider $\{ h_{\omega} \}_{\omega\in \Omega}$ the ergodic family of random operators defined by
\begin{equation}\label{eq_def_Anderson_discret_1d_N1}
 \forall \omega \in \Omega,\ h_{\omega} \ :\ \begin{array}{ccl}
                   \ell^2(\Z) & \to &\ell^2(\Z) \\
(u_n)_{n\in \Z} & \mapsto & (-(u_{n+1}+u_{n-1}) + \omega^{(n)} u_n)_{n\in \Z}
                  \end{array}
\end{equation}

Its of almost-sure spectrum is $[-2,2]+\supp \nu$ (see \cite[Theorem 3.9]{Ki08}). The transfer matrices associated with  $\{ h_{\omega} \}_{\omega\in \Omega}$ are given by 
$$\forall n\in \Z,\ \forall E\in \R,\ T_{\omega^{(n)}}(E) = \left( \begin{smallmatrix}
\omega^{(n)} -E & -1 \\ 
1 & 0                                                                           \end{smallmatrix}
 \right) $$
and the sequence $(T_{\omega^{(n)}}(E))_{n\in \Z}$ is a sequence of random matrices in $\SLD$, \emph{i.i.d.} and of common law $\mu$ induced by the law $\nu$ of the random variables $\omega^{(n)}$. Since the transfer matrices are \emph{i.i.d.}, the  Furstenberg group is given by  
$$\forall E\in \R,\ G_{\mu_E}=\overline{\langle \{ T_{\omega^{(0)}}(E)\ |\ \omega^{(0)}\in \supp \nu \} \rangle}.$$
We  first remark that all the matrices in $G_{\mu_E}$ are of determinant $1$. To show that $G_{\mu_E}$ is not compact, we will exhibit a non-bounded sequence in $G_{\mu_E}$. By hypothesis on the law of $\omega^{(0)}$, we have 
$$\left\langle  \left( \begin{smallmatrix}                                                                       a -E & -1 \\ 
1 & 0  \end{smallmatrix} \right),\ \left( \begin{smallmatrix}                                                                       b -E & -1 \\ 
1 & 0  \end{smallmatrix} \right) \right\rangle \subset G_{\mu_E}.$$
Then,
$$ \left( \begin{smallmatrix}                                                                       a -E & -1 \\ 
1 & 0  \end{smallmatrix} \right) \left( \begin{smallmatrix}                                                                       b -E & -1 \\ 
1 & 0  \end{smallmatrix} \right)^{-1} = \left( \begin{smallmatrix}                                                                       1 & a-b \\ 
0 & 1  \end{smallmatrix} \right) \in G_{\mu_E}$$
and for every $n\in \N$, 
$$ \left( \begin{smallmatrix}                                                                       1 & a-b \\ 
0 & 1  \end{smallmatrix} \right)^n =  \left( \begin{smallmatrix}                                                                       1 & n(a-b) \\ 
0 & 1  \end{smallmatrix} \right) \in G_{\mu_E}$$
which contains an unbounded sequence since $a-b \neq 0$.
\bigskip

It can be shown\cite{BL85} that the strong irreducibility of $G_{\mu_E}$ is equivalent (under the hypothesis of non-compactness of the Furstenberg group) to :
$$\forall \bar{x} \in P(\R^2),\  \# \{ M \bar{x}\ |\ M\in G_{\mu_E} \} \geq 3.$$
 Let $x=(x_1,x_2)\in \R^2$, $x\neq 0$. Assume that $x_2\neq 0$. Then, we show by simply solving linear systems that for $A=\left(\begin{smallmatrix} 1 & a-b \\ 0 & 1 \end{smallmatrix}\right)$, the vectors $Ax$, $A^2 x$ and $A^3 x$ are non-colinear two by two which allows to build $3$ distinct elements in $\{ M \bar{x}\ |\ M\in G_{\mu_E} \}$. If $x_2=0$ then $x_1\neq 0$ and if $B=\left(\begin{smallmatrix} 1 & 0 \\ a-b & 1 \end{smallmatrix}\right)=\left(\begin{smallmatrix} a-E & -1 \\ 1 & 0 \end{smallmatrix}\right)^{-1}\left(\begin{smallmatrix} b-E & -1 \\ 1 & 0 \end{smallmatrix}\right)\in G_{\mu_E}$, then the vectors  $Bx$, $B^2 x$ and $B^3 x$ are non-colinear two by two.

Hence, Furstenberg theorem applies for every $E\in \R$ and using Theorem \ref{thm_localization} one gets the following localization result initially due to Carmona, Klein and Martinelli\cite{CKM87}.

\begin{thm}[Carmona, Klein, Martinelli\cite{CKM87}]
The family  $\{h_{\omega} \}_{\omega \in \Omega}$ exhibits dynamical localization in every interval $I\subset \R$.
\end{thm}

There is a huge literature on localization results for random Jacobi matrices which extend the definition of the one-dimensional discrete Anderson model. Still, very few of them handle Bernoulli randomness. Among recent papers we quote the results of Rangamani about singular random Jacobi matrices\cite{R19} and about localization for random word models which include several models as the Anderson-Bernoulli model or dimer models\cite{R22}. We also mentionned at the end of Section \ref{sec_localization_criterion} recents papers\cite{BDFGVW19,JZ19,GoKl21} which gives elementary proofs of localization in dimension $d=1$ using large deviations theorems or a parametric version of Furstenberg's theorem.

\subsubsection{The continuous one-dimensional Anderson model}

Consider the ergodic family of operators $\{ H_{\omega} \}_{\omega\in \Omega}$ defined by 
\begin{equation}\label{eq_def_Anderson_continu_1d_N1}
\forall \omega \in \Omega,\ H_{\omega}  \ :\ \begin{array}{ccl}
                   L^2(\R) & \to &L^2(\R) \\
u & \mapsto & -u'' + \left( V_{\mbox{per}} + \displaystyle\sum_{n\in \Z^d} \omega^{(n)} f(\cdot-n)\right) u
                  \end{array} 
\end{equation}

where $V_{\mbox{per}}$ is a multiplication operator by a locally integrable $\Z$-periodic function and where $f$ is an integrable non-zero function with support in $[-\frac12,\frac12]$. 

For this ergodic family $\Hof$, Damanik, Sims and Stolz\cite{DSS02} proved the following localization result.

\begin{thm}[Damanik, Sims and Stolz\cite{DSS02}]\label{thm_DSS02}
There exists a discrete set $M\subset \R$ such that for every $E\in \R \setminus M$,  $G_{\mu_E}$ is non compact and strongly irreducible. Hence, for every $E\in  \R \setminus M$, $\gamma(E)>0$ and on every interval $I\subset \R \setminus M$, $\{ H_{\omega} \}_{\omega\in \Omega}$ exhibits dynamical localization.
\end{thm}

The demonstration of the strict positivity result of the dominant Lyapunov exponent associated to $\{ H_{\omega} \}_{\omega\in \Omega}$ is clearly more delicate than in the discrete case. For the details we refer directly to \cite{DSS02}. The main difficulty that appears with continuous models is that the transfer matrices have no longer an explicit form as in the discrete case, but require to solve a differential system on the intervals $[n,n+1]$ to compute them. It is then a question of trying to find an expression of the solutions involved in the transfer matrices which allows to verify the hypotheses of the Furstenberg theorem. In \cite{DSS02}, the idea is to consider first the Floquet solutions of $-u''+ V_{\mbox{per}}u =Eu$ and then to use them to construct Jost solutions of $-u''+ (V_{\mbox{per}}+f)u =Eu$. We will have previously reduced ourselves to the case where $a=0$ and $b=1$ and we are therefore led to solve on an interval $[n,n+1]$ either $-u''+ V_{\mbox{per}}u =Eu$, or $-u''+ (V_{\mbox{per}}+f)u =Eu$ in order to determine the two transfer matrices which are contained in the Furstenberg group. 

In the first case, the transfer matrix obtained will be that of a rotation, in the second it will be a linear combination of rotations. The idea is then to start with a unit vector in the plane and to apply the second transfer matrix to it. By choosing well the direction $\theta_0$ of the initial unit vector, we obtain a new vector of norm strictly greater than 1 and of direction $\theta_1$. By applying the second transfer matrix again we obtain a new vector whose norm depends on $\theta_1$. By turning it with the help of the first transfer matrix, we can modify $\theta_1$ in a new direction so that when we apply the second transfer matrix we obtain a new vector of norm strictly greater than the second vector obtained. In this way we manage to construct a non-bounded sequence of vectors from a unit vector. We have thus constructed a non-bounded sequence in the Furstenberg group which is therefore non-compact.

For the strong irreducibility, it is sufficient to use the transfer matrix which is a rotation to construct an infinite number of distinct directions by iteration from a well chosen initial direction. 

The discrete set of critical energies $M$ is obtained as the set of zeros of the reflection and transmission coefficients of the constructed Jost solutions, these coefficients being analytically dependent on $E$ and $f$ being assumed to be non zero.

The existence of a discrete set of energies for which the hypotheses of the Furstenberg theorem are not true will be found in all continuous models. Note that the study of the behavior of the Lyapunov exponents at these critical energies is a difficult question. In particular, in the absence of an integral representation for the Lyapunov exponents, due to the lack of reducibility of the Furstenberg group, one has to look directly at some properties of the Markov chain associated to the sequence of transfer matrices in order to prove continuity of the Lyapunov exponents near the critical energies. For this purpose one can try to adapt the ideas of \cite{GGG17} which have been extended in \cite{DuSi20}. Unfortunately, in these two references, there is an hypothesis of positivity on the random variables which is not satisfied in the Anderson model or in the study of a random Ising chain as studied by Chapman and Stolz\cite{CS15}.

\subsubsection{The one-dimensional Dirac model}

Consider the $\Z$-ergodic family $\{ D_{\omega} \}_{\omega \in \Omega}$ acting on $L^2(\R)\otimes \C^2$, 
$$D_{\omega} = \left( \begin{smallmatrix}
                      0 &  -\frac{\dd}{\dd x} \\
\frac{\dd}{\dd x} & 0
                      \end{smallmatrix}
 \right)+ V_{\mathrm{per}} +  V_{\omega}$$
where $V_{\mathrm{per}}$ is a $\Z$-periodic linear combination of the Pauli matrices $\sigma_1=\left(\begin{smallmatrix}                                                                                                              0 & 1 \\
1 & 0                                                                                                           \end{smallmatrix} \right)$, $\sigma_3=\left(\begin{smallmatrix}                                                                                                              1 & 0 \\
0 & -1                                                                                                           \end{smallmatrix} \right)$ and $I_2$, and $V_{\omega}$ is a random potential given by $V_{\omega}(x)=\displaystyle \sum_{n\in \Z} \omega^{(n)} f(x-n)$ where $f$ is a linear combination of $\sigma_1$ and $\sigma_3$ or a multiple of $I_2$.

\begin{thm}[\cite{Za21}]\label{thm_Z21}
 There exists a discrete set $M\subset \R$ such that for every $E\in \R \setminus M$,  $G_{\mu_E}$ is non compact and strongly irreducible. Hence, for every $E\in  \R \setminus M$, $\gamma(E)>0$ and on every interval $I\subset \R \setminus M$, $\{ D_{\omega} \}_{\omega\in \Omega}$ exhibits dynamical localization.
\end{thm}

The proof of this theorem is very similar to that of Theorem \ref{thm_DSS02} as far as the study of the Furstenberg group is concerned. It is also necessary to adapt to the case of the Dirac operator some parts of the proof of Theorem \ref{thm_localization}, in particular concerning the existence and the properties of the integrated density of states, including the Thouless formula. For this purpose, we can rely on the results of \cite{SSB09} and on their adaptation of Kotani's theory to the Dirac framework.

The Furstenberg theorem is sufficient to demonstrate the strict positivity of the dominant Lyapunov exponent for the discrete and continuous Anderson models or for this Dirac model. When we go to the quasi-one-dimensional models, we have to find new algebraic criteria.

\subsection{The Goldsheid-Margulis criterion}

When one wishes to verify the hypotheses of the Corollary \ref{cor_lyapmain} on concrete examples of quasi-one-dimensional Anderson operators, one comes up against a real difficulty. Even in the case of the discrete quasi-one-dimensional Anderson model presented by Goldsheid and Margulis in \cite{GM89}, whose transfer matrices are explicit and of a relatively simple form, the hypotheses of $p$-contractivity and especially of $L_p$-strong irreducibility are not easy to verify. This is why Goldsheid and Margulis used the following criterion.

\begin{thm}[Goldsheid and Margulis\cite{GM89}]\label{thm_GM}
Let $G$ be a subgroup of $\SpD$. If $G$ is Zariski-dense in $\SpD$, then $G$ is $p$-contracting and $L_{p}$-strongly irreducible for all $p\in \{ 1,\ldots ,N \}$.
\end{thm}

This criterion allows us to reduce the dynamical problem of separation and strict positivity of Lyapunov exponents to an algebraic problem of reconstruction of a Lie group. The key point is that the Zariski closure of a linear Lie group is still a linear Lie group. As we will see later on, the fact that we will work most of the time in the symplectic group which is connected, will allow us to bring most of the calculations to the level of Lie algebras. It simplifies the algebraic proofs since the algebraic calculations in a group become calculations of linear algebra and of Lie brackets.

\subsubsection{The discrete quasi-one-dimensional Anderson model}

We study the family of operators $\{ h_{\omega}^{(D)} \}_{\omega \in \Omega}$ defined by
\begin{equation}\label{eq_def_Anderson_discret_1d_N}
\forall \omega \in \Omega,\ h_{\omega}^{(D)}\ :\ \begin{array}{ccl}
                  \ell^2(\Z,\C^D) & \to & \ell^2(\Z,\C^D) \\
(u_n)_{n\in \Z} & \mapsto & (-(u_{n+1}+u_{n-1}) + V_{\omega^{(n)}} u_n )_{n\in \Z}
                  \end{array} 
\end{equation}
where $V_{\omega^{(n)}} = V_0 + \mathrm{diag}(\omega_{1}^{(n)}, \ldots, \omega_{D}^{(n)})$ with $V_0$ the tridiagonal matrix with zero diagonal terms and $1$ on the upper and lower diagonals.

Goldsheid and Margulis proved in \cite{GM89} that the Furstenberg group of  $\{h_{\omega}^{(D)} \}_{\omega \in \Omega}$ is Zariski-dense in $\SpD$. From this result, Klein, Martinelli and Speis proved in \cite{KLS90} the following result of localization.

\begin{thm}[\cite{GM89,KLS90}]\label{thm_loc_quasi1d_discret}
For every $E\in \R$, $G_{\mu_E}$ is Zariski-dense in $\SpD$. Hence, $\{h_{\omega}^{(D)} \}_{\omega \in \Omega}$ exhibits dynamical localization in every interval $I\subset \R$.  
\end{thm}

\noindent The transfer matrices associated with $\{ h_{\omega}^{(D)}\}_{\omega\in \Omega}$ are given by $T_{\omega^{(n)}}(E)=\left(\begin{smallmatrix}
                     V_{\omega^{(n)}} -E & -I_{\mathrm{D}}  \\
I_{\mathrm{D}} & 0
                    \end{smallmatrix} \right).$ These transfer matrices form an \emph{i.i.d. }sequence of matrices  in $\SpD$ of common law $\mu_E$ and one has 
$$G_{\mu_E}= \overline{\langle T_{\omega^{(0)}}(E)\ |\ \omega^{(0)} \in \supp (\nu_1 \otimes\cdots \otimes \nu_D)\rangle}.$$
According to the hypothesis made on the supports of the $\nu_i$'s and setting $a=0$ and $b=1$, 
$$G_{\{0,1\},E}:=\overline{\langle T_{\omega^{(0)}}(E)\ |\ \omega^{(0)}\in \{0,1\}^D \rangle} \subset G_{\mu_E}$$
and it is therefore sufficient to show that the subgroup generated by $2^D$ matrices, $G_{\{0,1\},E}$, is Zariski-dense in $\SpD$.  Let us denote $\mathrm{Cl}_{\mathrm{Z}}(G_{\{0,1\},E})$ its Zariski closure in $\SpD$ and $\mathfrak{g}(E)$ the Lie algebra of $\mathrm{Cl}_{\mathrm{Z}}(G_{\{0,1\},E})$. Recall that the Lie algebra of $\SpD$ is given by
$$\spD = \left\{ \left( \begin{smallmatrix}
A & B_{1} \\
B_{2} & -^{t}A
\end{smallmatrix} \right),\ A\in \MDR,\ b_{1}\ \mathrm{and}\ b_{2}\ \mbox{symmetric} \right\}.$$
For $i, j \in \{1,\ldots, D\}$, let $E_{ij}$ be the matrix in $\MDR$ with a coefficient $1$ at the intersection of the $i$-th row and the $j$-th column, and $0$ elsewhere. We also set
$$\forall i,j\in \{1,\ldots, D\},\ X_{ij}=\tfrac{1}{2} \left( \begin{smallmatrix}
0 & E_{ij}+E_{ji} \\
0 & 0
\end{smallmatrix} \right),\ Y_{ij}={^t}X_{ij},\ Z_{ij}= \left( \begin{smallmatrix}
E_{ij} & 0  \\
0 & -E_{ji}
\end{smallmatrix} \right).$$

An elementary calculation on Lie brackets of these three matrices allows to show that $\spD$ is generated by $\big\{ X_{ij},Y_{ij}\ |\ i,j\in \{1,\ldots, D\}, |i-j|\leq 1\big\}.$ From this, we are led to show that, for any $E\in \R$, $\mathfrak{g}(E)$ contains all the matrices $X_{ij}$ and $Y_{ij}$ for $i,j \in \{1,\ldots, D\}$, $|i-j|\leq 1$. We fix $E\in \R$ and proceed by successive steps. 

First, we prove that matrices of the form $\left( \begin{smallmatrix}
I & D \\
0 & I
\end{smallmatrix} \right)$ and $\left( \begin{smallmatrix}
I & 0 \\
D & I
\end{smallmatrix} \right)$ where $D$ is diagonal, are in $\mathrm{Cl}_{\mathrm{Z}}(G_{\{0,1\},E})$. For that we choose $T_{1}$ and $T_{2}$ in $G_{\{0,1\},E}$ associated to two realizations $V_1$ and $V_2$ of $V_{\omO}$. $\mathrm{Cl}_{\mathrm{Z}}(G_{\{0,1\},E})$ being a group, it is stable by inversion and product. So we have : 
$$B:=T_1 T_2^{-1} = \left( \begin{smallmatrix}
I & V_1-V_2 \\
0 & I 
\end{smallmatrix} \right)\in G_{\{0,1\},E}.$$ 
Let $i\in \{1,\ldots,D\}$. We can choose $V_{1}$ and $V_{2}$ so that  
$B = \left( \begin{smallmatrix}
I & E_{ii} \\
0 & I 
\end{smallmatrix} \right).$ 
Then, for every $n\in \Z$ : $B^{n} = \left( \begin{smallmatrix}
I & n E_{ii} \\
0 & I
\end{smallmatrix} \right).$
Let $P$ be a polynomial in $\R[X_{1,1},\ldots,X_{2D,2D}]$ such that : $\forall n\in \Z,\ P(B^{n})=0$. We fix $X_{j,j}=1$ for all $j \in \{1,\ldots, 2D\}$ and $X_{r,l}=0$ for $r \neq l$ except for $X_{i,D+i}$ and we consider : 
$$\widetilde{P} : X_{i,D+i} \mapsto P\left(  \left( \begin{smallmatrix} 
I & \left( \begin{smallmatrix} 
0 & | & 0 \\
- & X_{i,D+i} & - \\
0 & | & 0
\end{smallmatrix} \right)\\[1mm]
0 & I
\end{smallmatrix} \right) \right)$$

\noindent which is a polynomial in one variable with an infinite number of roots, the $n\in \Z$. So $\widetilde{P}$ is the null polynomial and: $\forall \alpha \in \R,\ \widetilde{P}(\alpha)=0$. This means that $P$ cancels on all matrices $\left( \begin{smallmatrix}
I & \alpha E_{ii} \\
0 & I
\end{smallmatrix} \right)$. By definition of the Zariski closure: 
$\forall \alpha \in \R,\ \left( \begin{smallmatrix}
I & \alpha E_{ii} \\
0 & I
\end{smallmatrix} \right) \in \mathrm{Cl}_{\mathrm{Z}}(G_{\{0,1\},E}).$
Since we have fixed an arbitrary $i$, we have : $\forall \alpha_{1},\ldots, \alpha_{D} \in \R,$
$$\left( \begin{smallmatrix}
I & \alpha_{1} E_{11} \\
0 & I
\end{smallmatrix} \right)\ldots \left( \begin{smallmatrix}
I & \alpha_{D} E_{DD} \\
0 & I
\end{smallmatrix} \right) =  \left( \begin{smallmatrix}
I & \alpha_{1} E_{11} +\ldots + \alpha_{D} E_{DD} \\
0 & I
\end{smallmatrix} \right) \in \mathrm{Cl}_{\mathrm{Z}}(G_{\{0,1\},E}).$$
This implies that $\left( \begin{smallmatrix}
I & D \\
0 & I
\end{smallmatrix} \right) \in \mathrm{Cl}_{\mathrm{Z}}(G_{\{0,1\},E})$ for every diagonal matrix $D$.

We have detailed a lot this first step because it is the one where the Zariski topology is involved. The other steps are a series of back and forth between $\mathrm{Cl}_{\mathrm{Z}}(G_{\{0,1\},E})$ and its Lie algebra: obtaining elements in $\mathrm{Cl}_{\mathrm{Z}}(G_{\{0,1\},E})$ allows to deduce elements contained in its Lie algebra and by linear combinations and Lie brackets we obtain new elements in the Lie algebra. By taking the exponential we recover elements in $\mathrm{Cl}_{\mathrm{Z}}(G_{\{0,1\},E})$ that we could not have constructed easily with only the product and the inverse. 

Finally we obtain that $\mathfrak{g}(E)=\spD$ and by connexity of $\SpD$ that $\mathrm{Cl}_{\mathrm{Z}}(G_{\{0,1\},E})=\SpD$. This allows us to deduce that the Furstenberg group associated to the family $\{ h_{\omega}^{(D)}\}_{\omega \in \Omega}$ is Zariski-dense in $\SpD$ which, applying Theorem \ref{thm_localization}, implies Theorem \ref{thm_loc_quasi1d_discret}.

\subsubsection{Point interactions}

We use the notations introduced in \eqref{def_modele_int_pt_formel_N} and \eqref{def_modele_int_pt_rig_N}. In this model, the random parameters intervene punctually in each integer, through interface conditions expressed on the eigenfunctions. This explains why we speak of point interactions. This model turns out to be very close to a discrete model like the quasi-one-dimensional discrete Anderson model. From the point of view of the randomness, we find the same dependence of the transfer matrices as in the discrete case, this one intervening only at integer points. On the other hand, considering a continuous Laplacian in dimension $1$ instead of a discrete Laplacian changes the dependence of the transfer matrices on the energy parameter, which leads to the existence of critical energies where the Lyapunov exponents could cancel. 

To determine the  transfer matrices associated to $\{ H_{P,\omega}\}_{\omega \in \Omega}$, we consider for $E\in \R$ the differential system $H_{P,\omega}u=Eu$ whose solutions are functions $u=(u_1,\ldots,u_D):\R \to \C^D$ satisfying $-u''+V_0 u=Eu$ on $\R \setminus \Z$ and such that each coordinate function $u_i$ satisfies the boundary conditions: 
$\forall i\in \{1,\ldots, D\},\ \forall n\in \Z,\ u_{i}'(n^+)=u_{i}'(n^-)+c_i \omega_{i}^{(n)} u_i (n).$
If $u$ is such a solution, the transfer matrix $T_{\omega^{(n)}}^{(n,n+1]}(E)$ from $n^+$ to $(n+1)^+$ is defined by 
$$ \forall n \in \Z,\ \left( \begin{smallmatrix}
u((n+1)^+) \\
u'((n+1)^+) 
\end{smallmatrix} \right) = T_{\omega^{(n)}}^{(n,n+1]}(E) \left( \begin{smallmatrix}
u(n^+) \\
u'(n^+) 
\end{smallmatrix} \right).$$

The sequence $( T_{\omega^{(n)}}^{(n,n+1]}(E))_{n\in \Z}$ is an  \emph{i.i.d.} sequence of matrices in $\SpD$ whose common law is $\mu_E$. As in the case of the quasi-one-dimensional discrete Anderson model, we compute explicitly the transfer matrices. To do so, we start by solving the free differential system on the open interval  $(n,n+1)$. By $1$-periodicity of $V_0$, it is sufficient to do it on the interval $(0,1)$. We then obtain the transfer matrix from $n^+$ to $(n+1)^-$ :
$$T_{(n,n+1)}(E)=T_{(0,1)}(E)=\exp \left( \begin{smallmatrix}
0 & I_{\mathrm{D}} \\
V_0-EI_{\mathrm{D}} & 0
\end{smallmatrix}\right).$$

We also set, for every $Q\in \MDR$, $M(Q)\in \mathcal{M}_{\mathrm{2D}}(\R)$ defined by $M(Q)=\left( \begin{smallmatrix}
I_{\mathrm{D}} & 0 \\
Q & I_{\mathrm{D}}
\end{smallmatrix}\right)$. Then, using the boundary conditions, the transfer matrix from $(n+1)^-$ to $(n+1)^+$ is non other than  $M(\mathrm{diag}(c_1\omega_1^{(n)},\ldots,c_D\omega_D^{(n)}))$. Finally, the transfer matrix from $n^+$ to $(n+1)^+$ is the product of the transfer matrices from $n^+$ to $(n+1)^-$ and from $(n+1)^-$ to $(n+1)^+$.
$$T_{\omega^{(n)}}^{(n,n+1]}(E)=M(\mathrm{diag}(c_1\omega_1^{(n)},\ldots,c_D\omega_D^{(n)})) \; T_{(0,1)}(E).$$
The first factor contains the random part and is independent of the energy $E$. The second factor is deterministic and depends only on $E$.

Finally, the \emph{i.i.d.} character of the transfer matrices implies the following internal description for the Furstenberg group: 
$$ G_{\mu_E}=\overline{<T_{\omega^{(0)}}^{(0,1]}(E) \ |\ \omega^{(0)} \in \supp \nu>}.$$
In \cite{BS07} and \cite{B09LMP} we proved the following result.

\begin{thm}\label{thm_sep_int_pt_N2}
There exists a discrete set $\mathcal{S}_{P,2} \subset \R$ (respectively  $\mathcal{S}_{P,3} \subset \R$), such that for every  
$E\in \R \setminus \mathcal{S}_{P,2}$ (respectively $E\in \R \setminus \mathcal{S}_{P,3}$) , $G_{\mu_E}$ is Zariski-dense in $\mathrm{Sp}_{2}(\R)$ (respectively $\mathrm{Sp}_{3}(\R)$). 

Hence, for $D=2$ and $D=3$, the almost-sure absolutely continuous spectrum of  $\{ H_{P,\omega}\}_{\omega \in \Omega}$ is empty.

Moreover, the integrated density of states associated to $\{ H_{P,\omega}\}_{\omega \in \Omega}$ is H\"older continuous on every interval included in $\R \setminus \mathcal{S}_{P,2}$ if $D=2$ and in $\R \setminus \mathcal{S}_{P,3}$ if $D=3$.
\end{thm}

Unfortunately, we were not able to go further in our analysis of  $\{ H_{P,\omega}\}_{\omega \in \Omega}$ and in particular we did not prove a weak Wegner estimate for this model. 

\begin{openq}
Prove the analog of Theorem \ref{thm_localization} in the framework of point interaction models.
\end{openq}

Up to our knowledge, there is very few results of localization for random point interactions models. For discrete point interaction models there is a proof of localization due to Delyon, Simon and Souillard in\cite{DSS85}. But this result requires randomness with an absolutely continuous component and does not cover the Bernoulli case. Also not covering the Bernoulli case are the results of Hislop, Kirsch and Krishna about localization\cite{HKK05} in dimension $d=1,2,3$ and eigenvalue statistics\cite{HKK20} again in dimension $d=1,2,3$. In the Bernoulli case, let us point out the recent result by Damanik, Fillman, Helman, Kesten and Sukhtaiev\cite{DFHKS21} and the result by Damanik, Fillman and Sukhtaiev\cite{DFS20} which obtained a localization result for point interactions on metric and discrete tree graphs.

To prove Theorem \ref{thm_sep_int_pt_N2},  the first thing to do is to compute explicitly the matrix exponential in $T_{(0,1)}(E)$. This leads us to separate the proof into several cases depending on the value of the energy $E$. Indeed, for $E>1$, if $U=\frac{1}{\sqrt{2}} \left( \begin{smallmatrix}
1 & 1 \\
1 & -1
\end{smallmatrix} \right)$
then 
$T_{(0,1)}(E)=\left( \begin{smallmatrix}
U & 0 \\
0 & U
\end{smallmatrix} \right) \; R_{\alpha,\beta} \; \left( \begin{smallmatrix}
U & 0 \\
0 & U
\end{smallmatrix} \right)\;,$
where $\alpha=\sqrt{E-1}$, $\beta=\sqrt{E+1}$, and
\[
R_{\alpha,\beta}=\left( \begin{smallmatrix}
\cos\alpha & 0 & \frac{1}{\alpha}\sin\alpha & 0 \\
0 & \cos\beta & 0 & \frac{1}{\beta}\sin\beta \\
-\alpha \sin\alpha & 0 & \cos\alpha & 0 \\
0 & -\beta \sin\beta & 0 & \cos\beta
\end{smallmatrix} \right).
\]

For $E\in (-1,1)$, we obtain the same expression by changing in $R_{\alpha,\beta}$ the cosines and sines of $\alpha$ by  cosines and hyperbolic sines. For $E<-1$ we still get the same expression by changing in $R_{\alpha,\beta}$ all cosines and sines by hyperbolic cosines and  hyperbolic sines.

With this explicit form of the transfer matrices, we proceed to a multi-step proof of the fact that the Zariski closure of $G_{\mu_E}$ is equal to $\mathrm{Sp}_{2}(\R)$ for all energies $E$ except those in a discrete set which we obtain in the course of the proof. First, by exploiting the definition of the Zariski closure and using the hypothesis made on the support of the $\nu_i$'s, we show that for any diagonal matrix $Q\in \mathcal{M}_2(\R)$, $\left( \begin{smallmatrix}
0 & 0 \\Q & 0 \end{smallmatrix} \right)$ is in the Lie algebra of $\mathrm{Cl}_{\textup{Z}}(G_{\mu_E})$. Thus $M(Q)$ is in $\mathrm{Cl}_{\mathrm{Z}}(G_{\mu_E})$ for any diagonal matrix $Q\in \mathcal{M}_2(\R)$. Then, we use the conjugation property in linear Lie groups to conjugate $ \left( \begin{smallmatrix}
0 & 0 \\Q & 0 \end{smallmatrix} \right)$ by powers of $R_{\alpha,\beta}$ and stay in the group $\mathrm{Cl}_{\textup{Z}}(G_{\mu_E})$. By choosing four such powers we obtain a family of four matrices in the Lie algebra of $\mathrm{Cl}_{\textup{Z}}(G_{\mu_E})$ which we show is free in $\mathcal{M}_2(\R)$ except for a discrete set of values of $E$. These values of $E$ are the ones for which the $4 \times 4$ determinant formed by the four columns, whose coefficients are the non-zero coefficients of the four matrices considered, vanish. Thus, we obtain a subspace of dimension 4 contained in the Lie algebra of $\mathrm{Cl}_{\textup{Z}}(G_{\mu_E})$ and by taking particular matrices in this space, we obtain new ones by Lie brackets. Then we consider a new family of 6 matrices whose non-zero coefficients are complementary to the non-zero coefficients of the first family of 4 matrices considered and which are still in the Lie algebra of $\mathrm{Cl}_{\textup{Z}}(G_{\mu_E})$. Again, we have to exclude a discrete set of values of $E$ for which a determinant $6\times 6$ cancels. We then obtain a free family of 10 matrices in the Lie algebra of $\mathrm{Cl}_{\textup{Z}}(G_{\mu_E})$. As the Lie algebra of $\mathrm{Sp}_{2}(\R)$ is of dimension $10$ we have well shown the equality of these two Lie algebras and by connexity of $\mathrm{Sp}_{2}(\R)$, the equality $\mathrm{Cl}_{\textup{Z}}(G_{\mu_E})=\mathrm{Sp}_{2}(\R)$. Hence the desired result using the  Goldsheid and Margulis criterion and versions of  Theorem \ref{thm_ishii_pastur_kotani} for the absence of absolutely continuous spectrum and of Theorem \ref{thm_holderids} for the regularity result of the integrated density of states which are adapted to the framework of point interactions models (see\cite{B09LMP}).

The proof in the case $D=3$ is very similar to the case $D=2$. From the construction, one remarks that we can choose $c_2=0$ and still obtain the Zariski denseness of $G_{\mu_E}$ in $\mathrm{Sp}_{3}(\R)$. This fact means that we are in the presence of a phenomenon of propagation of the randomness.  According to Kotani's theory, heuristically, if the second layer in our model becomes deterministic, we should not have separability of Lyapunov exponents. But, the first and third layers are coupled to the second through the deterministic potential $V_0$ and the randomness of the first and third layers is somehow transported to the second. To our knowledge there are only two other papers in which this phenomenon is observed\cite{Gl90,Go95}.

Due to the use of explicit determinants in the construction we were not able to obtain the Zariski-denseness in $\SpD$ of the Furstenberg group for arbitrary $D\geq 1$. 

\begin{openq}
For $D\geq 1$, prove that the  Furstenberg group associated to  $\{ H_{P,\omega}\}_{\omega \in \Omega}$ is Zariski-dense in $\SpD$.
\end{openq}

\subsubsection{A continuous quasi-one-dimensional Anderson model for $D=2$ (I)}

In \cite{BS07}, we have obtained a first result of Lyapunov exponents separability for the following quasi-one-dimensional continuous Anderson-Bernoulli model: 
\begin{equation}\label{def_modele_AB_N2_1}
\forall \omega \in \Omega,\ H_{\omega}^{(2)}=-\frac{\dd ^{2}}{\dd x^{2}}\otimes I_2+ \left(
\begin{smallmatrix}{}
0& 1 \\
1 &0
\end{smallmatrix} \right)    + \sum_{n\in \Z} \left(
\begin{smallmatrix}
 \omega_{1}^{(n)}  \mathds{1}_{[0,1]}(x-n)& 0 \\
0 & \omega_{2}^{(n)} \mathds{1}_{[0,1]}(x-n)
\end{smallmatrix} \right)  
\end{equation}
acting on $L^{2}(\R)\otimes \C^{2}$.

\begin{thm}[\cite{BS07}]\label{thm_sep_AB_N2_1}
There exists a countable set $\mathcal{C}$ such that, for any $E\in (2,+\infty)\setminus \mathcal{C}$, the Furstenberg group of $\{H_{\omega}^{(2)}\}_{\omega \in \Omega}$ is Zariski-dense in $\mathrm{Sp}_{2}(\R)$. Hence
\[
\forall E\in (2,+\infty)\setminus \mathcal{C},\ \gamma_{1}(E) > \gamma_{2}(E) >0
\] 
and there is no almost-sure absolutely continuous spectrum of  $\{H_{\omega}^{(2)}\}_{\omega \in \Omega}$  in $(2,+\infty)$. 
\end{thm}\vskip 3mm
\newpage

\noindent For $E\in \R$ fixed, the transfer matrix $T_{\omega^{(0)}}(E)$ is of the form $\exp\left(\begin{smallmatrix}
0 & I_{2} \\
V_{\omega}^{(0)}-E I_2& 0
\end{smallmatrix} \right)$ where $V_{\omega}^{(0)}=\left(\begin{smallmatrix}
\omega_1^{(0)} & 1 \\
1 & \omega_2^{(0)}
\end{smallmatrix} \right)$ and $I_2$ is the identity matrix of order 2. It is therefore necessary to compute an exponential matrix which this time depends on $\omega$, contrary to the case of the model of point interactions. This greatly complicates the calculations in the following. 

We notice that here we do not use the Zariski-denseness at the first step as in the discrete or point interactions cases, but an  argument of density of trajectories on the torus $\R^{2}/(2\pi \Z)^{2}$ hence the fact that we have to exclude a countable set of energies.

The fact of excluding only a countable set of energies is sufficient to apply Theorem \ref{thm_ishii_pastur_kotani} and get absence of absolutely continuous spectrum, but it is insufficient to apply Theorem \ref{thm_localization} which allows us to go from the separability of Lyapunov exponents to Anderson localization. Indeed, in our construction, nothing tells us that this countable set $\mathcal{C}$ is not dense in $(2,+\infty)$ and that we cannot therefore find a non-trivial interval of energies on which we have the Zariski-denseness of the Furstenberg group. We will see how to refine the algebraic techniques used in the proof of this theorem to obtain that in fact the Furstenberg group is dense in the sense of the usual topology in $\mathrm{Sp}_{2}(\R)$ for energies in $(2,+\infty)$ and outside a discrete set.

\subsection{The Breuillard-Gelander criterion}

Proving that a subgroup of $\SpD$ is Zariski-dense in $\SpD$ is a constructive problem that can be difficult to implement. In the previous Section, having transfer matrices given by matrix exponentials in which the random parameters and the energy parameter are mixed makes the explicit construction difficult and we can no longer follow the construction done in the discrete quasi-one-dimensional case or in the case of the point interaction model. In order to improve the result obtained in Theorem \ref{thm_sep_AB_N2_1}, it was therefore necessary to find a new algebraic criterion allowing to obtain the denseness and consequently the Zariski denseness of a subgroup in a Lie group having reasonable properties of semisimplicity and connectedness. We will therefore present in this Section a criterion allowing to reduce the question of whether a subgroup of a semisimple Lie group $G$ generated by a finite number of elements is dense, to a problem of reconstruction of the Lie algebra of $G$. 

\newpage
\begin{thm}[Breuillard and Gelander\cite{BG03}]\label{thm_BG03}
If $G$ is a real semisimple connected Lie group of Lie algebra $\mathfrak{g}$, then there exists a neighborhood $\mathcal{O} \subset G$ of the identity, on which $\log=\exp^{-1}$ is a well-defined diffeomorphism and such that $g_{1},\ldots,g_{m}\in \mathcal{O}$ generate a dense subgroup of $G$ whenever $\log(g_{1}),\ldots, \log(g_{m})$ generate $\mathfrak{g}$. 
\end{thm}

The term "generating" is used here in the sense of Lie algebras, thus taking into account both the linear combinations and the Lie bracket.

Theorem \ref{thm_BG03} gives us a clear plan to follow when we want to show that a subgroup of the symplectic group is dense in the symplectic group. We have to start by constructing elements of our subgroup (the Furstenberg group in our case) which are in the neighborhood $\mathcal{O}$ given by Theorem \ref{thm_BG03} when applied to the symplectic group, and then we compute the logarithms of these elements. We then consider the Lie algebra generated by these logarithms and show that it is equal to the Lie algebra of the symplectic group.

\subsubsection{A continuous quasi-one-dimensional Anderson model for $D=2$ (II)}

In \cite{B07}, I obtained the following result which improves Theorem \ref{thm_sep_AB_N2_1}.

\begin{thm}\label{thm_sep_AB_N2_2}
There exists a discrete set $\mathcal{S}\subset \R$ such that, for any $E\in (2,+\infty)\setminus \mathcal{S}$, the Furstenberg group of $\{H_{\omega}^{(2)}\}_{\omega \in \Omega}$ is dense (and therefore Zariski-dense) in $\mathrm{Sp}_{2}(\R)$. Thus $\{H_{\omega}^{(2)}\}_{\omega \in \Omega}$ exhibits dynamical localization in the intersection of every compact interval included in $(2,+\infty)\setminus \mathcal{S}$ with the almost-sure spectrum of $\{H_{\omega}^{(2)}\}_{\omega \in \Omega}$.
\end{thm}\vskip 3mm

First of all, we have to construct, from the transfer matrices, matrices in the neighborhood $\mathcal{O}$ of $I_4$ given by Theorem \ref{thm_BG03} applied to the group $\mathrm{Sp}_{2}(\R)$. This time we do not want to use an argument of density of trajectories on the torus because it would lead as before to a countable dense set of critical energies instead of a discrete set. We will instead use a simultaneous diophantine approximation argument which leads to the existence of 
$m_{\omega^{(0)}}(E)\in \N^{*}$ such that $1\leq m_{\omega^{(0)}}(E) \leq M$ and $(T_{\omega^{(0)}}(E))^{m_{\omega^{(0)}}(E)} \in \mathcal{O}$, for any $M\geq 1$ fixed.

The second step consists in calculating the logarithms of the matrices $(T_{\omega^{(0)}}(E))^{m_{\omega^{(0)}}(E)}$ which we have just shown to be in $\mathcal{O}$. Of course, even if these matrices are written as exponentials, nothing tells us that the arguments of these exponentials are in $\log \mathcal{O}$ and a priori 
$$\log (T_{\omega^{(0)}}(E))^{m_{\omega^{(0)}}(E)} \neq  m_{\omega^{(0)}}(E) \left(\begin{smallmatrix}
0 & I_{2} \\
V_{\omega}^{(0)}-E I_2& 0
\end{smallmatrix} \right).$$

Because of that, the logarithms have an expression which make difficult to prove that they generate the Lie algebra $\spDD$. We still manage to do very explicit computations and constructions which involves, like in the case of point interactions, many determinants from which we deduce the discrete set of critical energies to exclude. 

In view of the calculations carried out in this approach, it is clear that it is difficult to continue with this very explicit approach to tackle the case of quasi-one-dimensional operators whose matrix potentials are in $\MDR$ for any $D\geq 1$. To get around these difficulties and obtain a more general result, there will of course be a price to pay: we have to introduce a parameter of large disorder which was not present in our first results, just as it is not present in the results on the discrete quasi-one-dimensional Anderson model.

\subsubsection{A continuous quasi-one-dimensional Anderson model for $D\geq 1$}

 Again, let's take the notations introduced in \eqref{eq_def_espace_proba_tensor}. We introduce, for any $\omega \in \Omega$ and any real number $\ell >0$, the operator

\begin{equation}\label{def_modele_AB_N}
H_{\omega,\ell}^{(D)} = -\frac{\dd^2}{\dd x^2}\otimes I_{\mathrm{D}} + V_0 + \sum_{n\in \Z}  \left(
\begin{smallmatrix}
c_1 \omega_{1}^{(n)} \mathbf{1}_{[0,\ell]}(x-\ell n) & & 0\\ 
 & \ddots &  \\
0 & & c_D \omega_{D}^{(n)} \mathbf{1}_{[0,\ell]}(x-\ell n)\\ 
\end{smallmatrix}\right),
\end{equation}
acting on $L^2(\R)\otimes \C^D$. The numbers $c_1,\ldots,c_D$ are non-zero real numbers, $V_0$ denotes the multiplication operator by the tridiagonal matrix whose diagonal coefficients are zero and those on the sub and upper diagonals are $1$, $ \mathbf{1}_{[0,\ell]}$ is the characteristic function of the interval $[0,\ell]$.

The real $\ell >0$ can be seen as a parameter measuring the intensity of the disorder. Indeed, in an interval of fixed length, the smaller $\ell$ is, the more this fixed interval will contain random variables of the sequence $(\omega^{(n)})_{n\in \Z}$, these being found in each point of the $\ell \Z$ lattice. For example the interval $[0,1]$ will contain $\mathsf{E}(\frac{1}{\ell})$ random variables. So when $\ell$ decreases towards $0$, the randomness fills more and more the bounded intervals of $\R$.

Let $\mathcal{O}$ be the neighborhood of $I_{2\mathrm{D}}$ in $\SpD$ given by the Breuillard and Gelander theorem applied to the symplectic group of order $D$. We set
$$\dlO=\max\{ R>0\ |\ B(0,R)\subset \log\,\mathcal{O} \},$$
where $B(0,R)$ is the open ball centered in $0$ and of radius $R>0$ for the topology induced on the Lie algebra $\spD$ of $\SpD$ by the matrix norm associated to the Euclidean norm on $\R^{2D}$.

Let $M_{\omO} = V_0 + \mathrm{diag}(c_1 \omega_1^{(0)},\ldots,c_D \omega_D^{(0)}).$ The matrix $M_{\omO}$ is symmetric, so its eigenvalues $\lambda_1^{\omO},\ldots ,\lambda_D^{\omO}$ are real. We set:
\begin{equation}\label{eq_lambda_minmax}
\lambda_{\mathrm{min}}=\min_{\omO\in \{ 0,1\}^D} \min_{1\leq i\leq D} \lambda_i^{\omO},\qquad \lambda_{\mathrm{max}}=\max_{\omO\in \{ 0,1\}^D} \max_{1\leq i\leq D} \lambda_i^{\omO}
\end{equation}
and $\lambda_0=\frac{\lambda_{\mathrm{max}}-\lambda_{\mathrm{min}}}{2}$. We also set $\ell_C:=\ell_C(D)=\min \left( 1, \tfrac{\dlO}{\lambda_0}\right)$ and for every $\ell<\ell_C$,
\begin{equation}\label{eq_def_IlN}
I(\ell,D)=\left[ \lambda_{\mathrm{max}} - \frac{\dlO}{\ell},\lambda_{\mathrm{min}} + \frac{\dlO}{\ell} \right]\subset \R.
\end{equation}
\vskip 3mm

With all these notations, we state the localization result obtained in\cite{B09MPAG}.

\begin{thm}[\cite{B09MPAG}]\label{thm_sep_AB_N}
For every $\ell \in (0,\ell_C)$ and every $E\in I(\ell,D)$, the Furstenberg group of $\{H_{\omega,\ell}^{(D)}\}_{\omega \in \Omega}$ is equal to $\SpD$. Therefore, the family $\{H_{\omega,\ell}^{(D)}\}_{\omega \in \Omega}$ exhibits dynamical localization on every compact interval $I\subset I(\ell,D)\cap \Sigma$.
\end{thm}

To prove this theorem it suffices to prove the first point and to use Theorem \ref{thm_localization}.

For $E\in \R$, $n\in \Z$ and $\omega^{(n)}\in \tilde{\Omega}^{\otimes D}$, let $M_{\omega^{(n)}}(E)=V_0 + \mathrm{diag}(c_1 \omega_1^{(n)},\ldots,c_N \omega_D^{(n)})-EI_{\mathrm{D}}.$
So, if we also define  
\begin{equation}\label{eq_def_mat_X_AB_N}
X_{\omega^{(n)}}(E)=\left( \begin{smallmatrix}
0 & I_{\mathrm{D}} \\
M_{\omega^{(n)}}(E) & 0
\end{smallmatrix}\right) \in \mathcal{M}_{\mathrm{2D}}(\R),
\end{equation}
by solving the differential system with constant coefficients  $H_{\omega,\ell}^{(D)}u = Eu$ on $[\ell n,\ell (n+1)]$, 
\begin{equation}\label{eq_expr_mat_T_AB_N}
\forall \ell >0,\ \forall n\in \Z,\ \forall E\in \R,\ T_{\omega^{(n)}}(E)=\exp\left(\ell X_{\omega^{(n)}}(E)\right).
\end{equation}

Being able to write the transfer matrices $T_{\omega ^{(n)}}(E)$ as a matrix exponential is very important in the sequel, in particular to be able to apply Theorem \ref{thm_BG03}. 

We have the internal description of the Furstenberg group $G_{\mu_E}$ associated to the family  $\{H_{\omega,\ell}^{(D)}\}_{\omega \in \Omega}$ :
$$\forall E\in \R,\ G_{\mu_E}=\overline{<T_{\omO} (E)\ |\ \omO \in \supp \nu>}.$$
Since $\{0,1\}^D \subset \supp \nu$, $\overline{<T_{\omO} (E)\ |\ \omO \in \{0,1\}^N >}\subset G_{\mu_E}.$

Theorem \ref{thm_BG03} gives us the plan of our demonstration.

\begin{enumerate}[1.]
\item We construct $\ell_C$ and $I(\ell,D)$ so that for all $\ell \in (0,\ell_C)$ and all $E\in I(\ell,D)$, $T_{\omO}(E)\in \mathcal{O}$, for any $\omO \in \{0,1\}^D$, where $\mathcal{O}$ is the neighborhood of $I_{2D}$ given by Theorem \ref{thm_BG03} applied to $\SpD$.
\item For $\ell <\ell_C$, we compute $\log T_{\omO}(E)$.
\item We finally show that $\mathrm{Lie}\{ \log T_{\omO}(E)\ |\ \omO \in \{0,1\}^D \}=\spD.$
\end{enumerate}

The last point will be a consequence of the following algebraic lemma which we do not prove here but which is shown in detail in \cite{B09MPAG}.

\begin{lem}\label{lem_a_spN}
Let $D\geq 1$ and $E\in \R$.

\noindent The Lie algebra generated by the set $\{ X_{\omO}(E) \ |\ \omO \in \{ 0,1\}^D \}$ is equal to $\spD$.
\end{lem}

Let's go back to the first step of the demonstration of Theorem \ref{thm_sep_AB_N}. First of all, the singular values of $X_{\omO}(E)$ are $1$, $(\lambda_1^{\omO}-E)^2$, $\ldots$, $(\lambda_D^{\omO}-E)^2$, hence :
$$||X_{\omO}(E)||=\max \left(1, \max_{1\leq i \leq D} |\lambda_i^{\omO}-E|\right),$$
where $||\ ||$ is the matrix norm associated with the Euclidean norm on $\R^{2D}$.

Since the neighborhood $\mathcal{O}$ depends only on $\SpD$, thus only on $D$, we construct an interval of values of $E$ such that, for $\ell$ small enough, 
\begin{equation}\label{eq_prop_GE_1_AB_N}
\forall \omO\in \{ 0,1\}^D,\ 0<\ell ||X_{\omO}(E)|| < \dlO,
\end{equation}
or equivalently
\begin{equation}\label{eq_prop_GE_2_AB_N}
0<\ell \max \left(1,\max_{\omO\in \{ 0,1\}^D} \max_{1\leq i\leq D} |\lambda_i^{\omO}-E|\right) < \dlO.
\end{equation}
Let us assume that $\ell \leq \dlO$ and let $r_{\ell}=\frac{1}{\ell}\dlO \geq 1$. Then, because $r_{\ell}\geq 1$, the set :
\begin{equation}\label{eq_prop_GE_3_AB_N}
I(\ell,D)=\left\{E\in \R\ \bigg|\  \max \left(1,\max_{\omO\in \{ 0,1\}^D} \max_{1\leq i\leq D} |\lambda_i^{\omO}-E|\right) \leq r_{\ell} \right\}
\end{equation}
can be written as the following intersection,
\begin{equation}\label{eq_prop_GE_4_AB_N}
I(\ell,D)=\bigcap_{\omO\in \{ 0,1\}^D} \bigcap_{1\leq i\leq D} [\lambda_i^{\omO}-r_{\ell}, \lambda_i^{\omO}+r_{\ell}].
\end{equation}
With the definitions of $\lambda_{\mathrm{min}}$, $\lambda_{\mathrm{max}}$ and $\lambda_{0}$ given in (\ref{eq_lambda_minmax}), if $\lambda_0 <r_{\ell}$, $I(\ell,D) \neq \emptyset$ and more precisely, $I(\ell,D)=[\lambda_{\mathrm{max}}-r_{\ell},\lambda_{\mathrm{min}}+r_{\ell}]$. 
This interval is centered in $\frac{\lambda_{\mathrm{min}} + \lambda_{\mathrm{max}}}{2}$ and has length $2r_{\ell}-2\lambda_0 >0$ which tends to $+\infty$ when $\ell$ tends to $0^+$.

Moreover, $\lambda_{\mathrm{min}}$, $\lambda_{\mathrm{max}}$ and $\dlO$ depending only on $D$, $I(\ell,D)$ depends only on $\ell$ and $D$ and the condition $\lambda_0 <r_{\ell}$ is equivalent to $\ell <\tfrac{\dlO}{\lambda_0}=\ell_C.$ We have just constructed $\ell_C$ and $I(\ell,D)$ so that  
\begin{equation}\label{eq_prop_GE_5_AB_N}
\forall \ell\in (0,\ell_C),\ \forall E\in I(\ell,D),\ 0<\ell ||X_{\omO}(E)|| \leq \dlO.
\end{equation}
Now, let us recall that from the definition of $\mathcal{O}$ in Theorem \ref{thm_BG03}, \emph{exp} is a diffeomorphism from $\log \mathcal{O}$ onto $\mathcal{O}$. Therefore, for any $E\in I(\ell,D)$, $\log T_{\omO}(E)=\ell X_{\omO}(E)$, which brings us immediately to the third step of our demonstration. 

For the third step, it is enough to apply Lemma \ref{lem_a_spN} to obtain :
\begin{equation}\label{eq_prop_GE_6_AB_N}
\forall \ell >0,\ \forall E\in \R,\ \mathrm{Lie} \{ \ell X_{\omO}(E) \ |\ \omO \in \{0,1\}^D \}=\spD,
\end{equation}
which finishes the proof, applying Theorem \ref{thm_BG03}.

With the help of the parameter $\ell$, we avoid using the simultaneous diophantine approximation to obtain elements in $\mathcal{O}$. There is also no difficulty in computing the logarithms because this time we can choose $\ell$ so that the computed logarithms are well within $\log \mathcal{O}$. Finally, the algebraic construction of the Lie algebra can be done for any $D\geq 1$ because the logarithms have a simple enough algebraic expression. So we don't have to construct by hand families of linearly independent matrices whose linear independance is shown by means of determinants. Moreover, this absence of determinants in our construction means that unlike in Theorem \ref{thm_sep_int_pt_N2} and Theorem \ref{thm_sep_AB_N2_2}, there is no discrete set of critical energies to exclude here. On the other hand the interval on which we have localization depends on the parameter $\ell$, but it has the good taste to tend to the whole real line when $\ell$ tends to $0^+$.

\subsubsection{Extension to a generic interaction potential}

The use of the Breuillard and Gelander theorem to obtain the Lyapunov exponents separability leads us to show an algebraic property on a Lie algebra generated by a finite number of matrices. As explained in \cite{BPec06}, this is an open condition and better, the $n$-uplets of $\spD$ elements which do not generate a dense subgroup are contained in a closed analytic subvariety. This allows to perturb the interaction potential $V_0$ in \eqref{def_modele_AB_N} while preserving the fact that the Furstenberg group is equal to $\SpD$ for any energy in $I(\ell,D)$ for $\ell \in (0,\ell_C)$. 

In \eqref{def_modele_AB_N} we replace the matrix $V_0$ by any real symmetric matrix $V$ of size $D\geq 1$. We denote by $\SD$ the space of real symmetric matrices of size $D\times D$.

\begin{thm}[\cite{B13}]
For almost every $V\in \SD$, there exist a finite set $\mathcal{S}_V\subset \R$ and $\ell_C=\ell_C(D,V) >0$ such that, for every $\ell\in (0,\ell_C)$, there exists a compact interval $I(\ell,D,V)\subset \R$ such that :
\begin{itemize}
 \item[1.] $\forall E\in I(\ell,D,V) \setminus \mathcal{S}_V,\quad G_{\mu_E}=\SpD.$
\item[2.] If $I\subset I(\ell,D,V)\setminus \mathcal{S}_V$ is an open interval with $\Sigma \cap I \neq \emptyset$, then $H_{\omega,\ell}^{(D)}$ exhibits dynamical localization in $\Sigma \cap I$.
\end{itemize}
\end{thm}

In this theorem, the genericity is understood in the sense of the Lebesgue measure on $\SD$ identified with the Lebesgue measure on $\R^{\frac{D(D+1)}{2}}$. This is the statement such as presented in \cite{B13}. However, the proof of this theorem implies a stronger result which is the Zariski genericity, in the sense that one can choose $V$ in a dense Zariski open set of $\SD$.

It is the algebraic nature of the objects involved that allows us to demonstrate a generic result in $V$ and the finiteness of the set of critical energies. We simply summarize the ideas used in the proof by recalling that the set of zeros of a non-zero one-variable polynomial is finite and that more generally, the set of zeros of a non-zero multivariable polynomial has zero Lebesgue measure. For a more precise proof we refer to \cite{B13}.

\section{The unitary case}\label{sec_unitary}

As we will see for four different examples, the role played by the symplectic group for the quasi-one-dimensional models of Schr\"odinger type will be played by the so-called Lorentz group. 

The Lorentz group $\mbox{U}(D,D)$ of signature $(D,D)$ is defined as the set of matrices of size $2D \times 2D$ which preserve the form $\Ll
= \left(\begin{smallmatrix}
 \one & 0 \\
0 & -\one
\end{smallmatrix} \right)$ in the sense that $T$ is in $\mbox{U}(D,D)$ if and only if $T^* \Ll T = \Ll$. 

To pass from the results on Lyapunov exponents in the symplectic framework to the Lorentz group framework, one uses the Cayley transform. By the Cayley transform, the group $\mathrm{U}(D,D)$ is unitarily equivalent to the complex symplectic group. More precisely, if $C=\frac{1}{\sqrt{2}} \left( \begin{smallmatrix}
\one & -i\one \\
\one & i\one
\end{smallmatrix}  \right) \in \mathcal{M}_{2D}(\C)$ and if $J= \left( \begin{smallmatrix}
0 & -\one \\
\one & 0
\end{smallmatrix}  \right),$
then
$\mathrm{U}(D,D) = C \mathrm{Sp}_{\mathrm{D}}(\C) C^*,$
where 
$$\mathrm{Sp}_{\mathrm{D}}(\C) = \{ M\in \mathcal{M}_{2D}(\C) | M^* J M = J \}.$$

In order to apply directly the results of\cite{BL85}, we have to pass from the complex symplectic group to the real symplectic group. For that we follow \cite{ABJ10} and introduce the application which separates the real and imaginary parts of a matrix with complex coefficients and place them in blocks: 
$$\pi\ :\ \begin{array}{cll}
           \mathcal{M}_{2D}(\C) & \to & \mathcal{M}_{4D}(\R) \\[2mm]
	    A+iB & \mapsto & \left( \begin{smallmatrix}
				      A & -B \\
				      B & A
				      \end{smallmatrix} \right) .
				\end{array}$$
Finally, $\pi(C^*\cdot \mathrm{U}(D,D)\cdot C) \subset \mathrm{Sp}_{\mathrm{2D}}(\R)$ which allows to use the results on Lyapunov exponents in the symplectic group to study the Lyapunov exponents in the unitary setting.

\subsection{The unitary Anderson model}

The first example of unitary model in dimension one for which we present a localization result is the unitary analog of the Anderson model. It was studied by Hamza, Joye and Stolz in \cite{HJS09}. We present their result and use their notations.

First, consider two $2\times 2$ unitary matrices $B_1=\left( \begin{smallmatrix}
r & t \\
-t & r \end{smallmatrix} \right)$ and $B_2=\left( \begin{smallmatrix}
r & -t \\
t & r \end{smallmatrix} \right)$ with $(r,t)\in \R^2$ satisfying $r^2+t^2=1$. These real numbers correspond to reflection and transition coefficients. Then, let $U_e$ the unitary matrix operator in $\ell^2(\Z)$ defined as the direct sum of identical $B_1$-blocks with blocks starting at even indices. Construct also $U_o$ the unitary matrix operator in $\ell^2(\Z)$ defined as the direct sum of identical $B_2$-blocks with blocks starting at odd indices. Let $S=U_eU_o$, unitary operator on $\ell^2(\Z)$ with band structure.

Then, introduce the probability space $(\Omega, \mathcal{F}, \mathbb{P})$ where $\Omega =\T^{\Z}$ (with $\T=\R/(2\pi \Z)$), $\mathcal{F}$ is the $\sigma$-algebra generated by cylinders of Borel sets and $\mathbb{P}=\bigotimes_{k\in \Z} \mu$ where $\mu$ is a non-trivial probability measure on $\T$. Assume that $\mu$ is absolutely continuous with bounded density. Then, define a sequence of random variables $(\theta_k)_{k\in \Z}$ on  $(\Omega, \mathcal{F}, \mathbb{P})$ which are $\T$-valued and \emph{i.i.d.} with common law $\mu$. With these random variables one defines the diagonal random operator $D_{\omega}$ acting on $\ell^2(\Z)$ and defined by :
$$\forall \omega \in \Omega,\ \forall k\in \Z, D_{\omega} e_k = \ee^{-i\theta_k(\omega)} e_k$$
where $(e_k)_{k\in \Z}$ is the canonical basis of $\ell^2(\Z)$.

The \textbf{unitary Anderson model} is the family $\{U_{\omega} \}_{\omega \in \Omega}$ of unitary operators acting on $\ell^2(\Z)$ where for every $\omega \in \Omega$,  $U_{\omega}=D_{\omega}S$. 

The family $\{U_{\omega} \}_{\omega \in \Omega}$ is ergodic with respect to the $2$-shift in $\Omega$ and one can show that its almost-sure spectrum is equal to: $\Sigma = \{ \ee^{ia} \ |\ a \in [-\lambda_0,\lambda_0]-\supp \mu \} \subset \mathbb{S}^1$
where $\lambda_0=\arccos(r^2-t^2)$\cite{HJS09}. 

For $\omega\in \Omega$ and $z\in \C \setminus \mathbb{S}^1$ let $G_{\omega}(z)=(U_{\omega}-z)^{-1}$ and for $k,l\in \Z$, let $G_{\omega}(k,l,z)=\langle e_k | G_{\omega}(z) e_l\rangle$ be the Green function of $U_{\omega}$.

Using the Fractional Moments Method\cite{AW15}, Hamza, Joye and Stolz proved the following result:

\begin{thm}[\cite{HJS09}]\label{thm_HJS09}
For every $t<1$ there exists $s>0$, $C<\infty$ and $\alpha>0$ such that 
$$\E(|G_{\omega}(k,l,z)|^s)\leq C \ee^{-\alpha |k-l|}$$
for all $z\in \C$ such that $0 < ||z|-1|<\frac12$ and all $k,l\in \Z$.
Therefore,  $\{U_{\omega} \}_{\omega \in \Omega}$ exhibits dynamical localization throughout $\Sigma$.
\end{thm}

The proof of Theorem \ref{thm_HJS09} relies on the formalism of transfer matrices and on the use of the Furstenberg theorem to get positivity of the Lyapunov exponent which allows to prove results of exponential decay of some products of transfer matrices. 

Let $\omega \in \Omega$. Consider the equation $U_{\omega}\psi = z\psi$ for $z\in \C \setminus \{ 0\}$ and with $\psi$ not necessarily in $\ell^2(\Z)$. For all $(\theta,\eta)\in \T^2$, let
$$T_z(\theta,\eta)=\left(\begin{smallmatrix}
-\frac{1}{z} \ee^{-i\eta} & \frac{r}{t} \left( \ee^{i(\theta-\eta}) - \frac{1}{z} \ee^{-i\eta} \right) \\
\frac{r}{t} \left(1 - \frac{1}{z} \ee^{-i\eta} \right) & -\frac{z}{t^2} \ee^{i\theta} + \frac{r^2}{t^2}  \left(1+ \ee^{i(\theta-\eta}) - \frac{1}{z} \ee^{-i\eta} \right)  \end{smallmatrix} \right).$$
Then, one has 
$$\forall k \in \Z,\ \left(\begin{smallmatrix}
 \psi_{2k+1} \\ \psi_{2k+2} \end{smallmatrix}\right) = T_z(\theta_{2k}(\omega),\theta_{2k+1}(\omega)) \left(\begin{smallmatrix}
\psi_{2k-1} \\ \psi_{2k} \end{smallmatrix} \right). $$
Let $\gamma(z)$ be the Lyapunov exponent associated to the sequence of \emph{i.i.d.} matrices in $\SLD$, $(T_z(\theta_{2k}(\omega),\theta_{2k+1}(\omega)))_{k\in \Z}$. Using Furstenberg theorem in a very similar way as for the discrete scalar-valued one-dimensional Anderson model, it was proven in\cite{HS07} that
\begin{itemize}
 \item[(i)] if $\supp \mu = \{a ,b \}$ with $|a-b|=\pi$ then $\gamma(-a)=\gamma(-b)=0$ and for all $z\in \T\setminus \{-a,-b\}$, $\gamma(z)>0$.
 \item[(ii)] If $\{a ,b \} \subset \supp \mu$ with $|a-b|\notin \{0 ,\pi\}$ then for every $z\in \T$, $\gamma(z) >0$. 
\end{itemize}
This result of positivity of Lyapunov exponents implies that for every compact $K\subset \C$ there exist $\alpha>0$, $\delta\in (0,1)$ and $C<\infty$ such that: $\forall z\in K,\ \forall n \in \N, \forall v\in \C^2,\ ||v||=1,$
$$ \E(||  T_z(\theta_{2(n-1)}(\cdot),\theta_{2(n-1)+1}(\cdot))\cdots T_z(\theta_{0}(\cdot),\theta_{1}(\cdot))v||^{-\delta}) \leq C \ee^{-\alpha n}.$$
This estimate is the core of the proof of Theorem \ref{thm_HJS09}.

\subsection{CMV matrices}

The unitary Anderson model is a particular case of more general one-dimensional unitary random operators, the so-called CMV matrices, named after the authors of\cite{CMV03} who popularized them.

Let $(\alpha_n)_{n\in \N}$ a sequence of complex numbers such that for every $n\in \N$, $|\alpha_n|<1$. If $\rho_n=(1-|\alpha_n|^2)^{\frac{1}{2}}$ then the CMV matrix associated to the Verblunsky sequence $(\alpha_n)_{n\in \N}$ is the operator acting on $\ell^2(\N,\C)$ given by the semi-infinite matrix,
\begin{equation}\label{eq_def_CMV_matrix}
\mathcal{C}=\left(\begin{smallmatrix}
\overline{\alpha_0} & \overline{\alpha_1} \rho_0 & \rho_1 \rho_0 & & & & \\
\rho_0 & -\overline{\alpha_1} \alpha_0 & -\rho_1 \alpha0 & & & & \\
 & \overline{\alpha_2} \rho_1 & -\overline{\alpha_2} \alpha_1 & \overline{\alpha_3} \rho_2 & \rho_3 \rho_2 & & \\
 & \rho_2 \rho_1 & -\rho_2\alpha_1 & -\overline{\alpha_3} \alpha_2 & -\rho_3 \alpha_2 & & \\
 & & & \overline{\alpha_4} \rho_3 &  -\overline{\alpha_4} \alpha_3 & \overline{\alpha_5} \rho_4 & \rho_5 \rho_4 \\
 & & &  \rho_4 \rho_3 & -\rho_4\alpha_3 & -\overline{\alpha_5} \alpha_4 & -\rho_5 \alpha_4 \\
 & & & & \ddots & \ddots &\ddots
      \end{smallmatrix}
 \right) 
\end{equation}

One also define the extended CMV matrix associated to $(\alpha_n)_{n\in \Z}$, a sequence of complex numbers such that for every $n\in \Z$, $|\alpha_n|<1$, as the operator acting on $\ell^2(\Z,\C)$ given by the infinite matrix,
\begin{equation}\label{eq_def_extCMV_matrix}
\mathcal{E}=\left(\begin{smallmatrix}
\ddots & \ddots & \ddots & & & & & \\
\overline{\alpha_0}\rho_{-1}  & -\overline{\alpha_0} \alpha_{-1} & \overline{\alpha_1} \rho_0 & \rho_1 \rho_0 & & & & \\
\rho_0 \rho_{-1} & -\rho_0 \alpha_{-1} & -\overline{\alpha_1} \alpha_0 & -\rho_1 \alpha0 & & & & \\
& & \overline{\alpha_2} \rho_1 & -\overline{\alpha_2} \alpha_1 & \overline{\alpha_3} \rho_2 & \rho_3 \rho_2 & & \\
& & \rho_2 \rho_1 & -\rho_2\alpha_1 & -\overline{\alpha_3} \alpha_2 & -\rho_3 \alpha_2 & & \\
& & & & \overline{\alpha_4} \rho_3 &  -\overline{\alpha_4} \alpha_3 & \overline{\alpha_5} \rho_4 & \rho_5 \rho_4 \\
& & & &  \rho_4 \rho_3 & -\rho_4\alpha_3 & -\overline{\alpha_5} \alpha_4 & -\rho_5 \alpha_4 \\
& & & & & \ddots & \ddots &\ddots
      \end{smallmatrix}
 \right) 
\end{equation}

The CMV matrices are the unitary analog of Jacobi matrices and were originally introduced in the study of orthogonal polynomials on the unit circle. Indeed they arise in the representation of the map $f\mapsto zf$ on $L^2(\SM^1,\dd \mu)$ in the basis given by the orthonormalization of the set of Laurent polynomials $\{1, z, z^{-1}, z^2, z^{-2},\ldots \}$ according to the usual scalar product on $L^2(\SM^1,\dd \mu)$. Here $\SM^1$ denote the unit circle in $\C$ and $\mu$ denotes a probability measure on $\SM^1$ that does not admit a finite support.   See \cite{Sim05_1, Sim05_2} for a comprehensive review of this vast subject. 

In order to introduce randomness in extended CMV matrices, let $\nu$ be a Borel probability measure supported on a compact set $S$ of the open unit disk in $\C$ which contains at least two points. Let $\Omega = S^{\Z}$ and consider $(\alpha_n(\omega))_{n\in \Z}=(\omega_n)_{n\in \Z}\in \Omega $ a sequence of \emph{i.i.d.} random variables of common law $\nu$. In particular, the $\omega_n$ could be Bernoulli variables.

The random Verblunsky sequence $(\alpha_n(\omega))_{n\in \Z}$ defines a random extended CMV matrix $\mathcal{E}_{\omega}$. The family $\{ \mathcal{E}_{\omega} \}_{\omega \in \Omega}$ is $\Z$-ergodic. For such an ergodic family, there is no multi-scale analysis or Kunz-Souillard approach available and the first localization result in the Bernoulli case is found in\cite[Theorems 7.1-7.2]{BDFGVW19} :

\begin{thm}
There exists a set $\mathcal{D}\subset \SM^1$ which contains at most three points such that, for every compact interval $I\subset \SM^1\setminus \mathcal{D}$, the family $\{ \mathcal{E}_{\omega} \}_{\omega \in \Omega}$ exhibits Anderson localization and dynamical localization on $I$ .
\end{thm}

This result was also obtained by Zhu in \cite[Theorems 1-2]{Z21} with slightly different techniques. In both cases the proofs are based upon the Furstenberg theorem and large deviations estimates.

Note also that there is deep connection between CMV matrices and  coined quantum walks as presented in\cite{ASW11,W13}. Indeed, to each coined quantum walk which is obtained by a single coin and a single shift, one associates a random CMV operator as explained in\cite{BDFGVW19,W13}. This make possible to deduce spectral and dynamical properties of such a coined quantum walk from the study of the associated random CMV operator.

\subsection{The quasi-1d Chalker-Coddington model}

In this Section we review a localization result for the quasi-one-dimensional version of the Chalker-Coddington model as presented in\cite{ABJ10}. This model is introduced to help to understand the delocalization transition of the quantum Hall effect.

Like in the unitary Anderson model, we start by introducing two parameters $r,t\in [0,1]$ such that $r^2+t^2=1$, corresponding to reflection and transition coefficients. If $\T=\R/(2\pi \Z)$, let $q=(q_1,q_2,q_3)\in \T^3$ and set 
$S(q)=\left( \begin{smallmatrix}
q_1q_2 & 0 \\ 0 & q_1\overline{q_2}  \end{smallmatrix}  \right) \left( \begin{smallmatrix}
t & -r \\ r & t \end{smallmatrix}  \right)
\left(\begin{smallmatrix}
q_3 & 0 \\ 0 & \overline{q_3}\end{smallmatrix}  \right).$

Let $(\Omega, \mathcal{F}, \mathbb{P})$ where $\Omega =(\T^6)^{(2\Z)^2}$, $\mathcal{F}$ is the $\sigma$-algebra generated by cylinders of Borel sets and $\mathbb{P}=\bigotimes_{(2\Z)^2} \lambda^{\otimes 6}$ where $\lambda$ is the Haar measure on $\T$. 

Set, for $p\in \Omega$ and every $j,k\in \Z$, $p(2j,2k)=(p_e(2j,2k),p_o(2j+1,2k+1))$ where $p_e(2j,2k)=(p_1(2j,2k),p_2(2j,2k),p_3(2j,2k))$ and $p_o(2j,2k)=(p_4(2j,2k),p_5(2j,2k),p_6(2j,2k))$.

With these notations, one introduces the family of unitary operators acting on $\ell^2(\Z^2)$, $\{ \hat{U}(p) \}_{p\in \Omega}$ where for every $p\in \Omega$, $\hat{U}(p)$ is defined by its matrix elements, $\hat{U}(p)_{\mu,\nu}=\langle e_{\mu}|\hat{U}(p)e_{\nu} \rangle$ for $\mu,\nu\in \Z^2$ and $(e_\mu)_{\mu\in \Z^2}$ the canonical basis of $\ell^2(\Z^2)$. One sets $\hat{U}(p)_{\mu,\nu}=0$ except for the blocks
$$\left( \begin{matrix}
  \hat{U}(p)_{(2j+1,2k),(2j,2k)} & \hat{U}(p)_{(2j+1,2k),(2j+1,2k+1)} \\
  \hat{U}(p)_{(2j,2k+1),(2j,2k)} & \hat{U}(p)_{(2j,2k+1),(2j+1,2k+1)}
         \end{matrix} \right):= S(p_e(2j,2k)), \forall j,k\in \Z$$
and 
$$\left( \begin{matrix}
 \hat{U}(p)_{(2j+2,2k+2),(2j+2,2k+1)} & \hat{U}(p)_{(2j+2,2k+2),(2j+1,2k+2)} \\
 \hat{U}(p)_{(2j+1,2k+1),(2j+2,2k+1)} & \hat{U}(p)_{(2j+1,2k+1),(2j+1,2k+2)}
         \end{matrix} \right):= S(p_o(2j+1,2k+1)), \forall j,k\in \Z.$$
For a complete physical interpretation of this family of unitary operators, which is called the Chalker-Coddington model, we refer to \cite{ABJ10}. These operators acting on $\ell^2(\Z^2)$ are not quasi-one-dimensional random operators. In order to enter into the framework of quasi-one-dimensional random operators, one restrict in one direction the action of $\hat{U}(p)$. Let $M\in \N$. Set $\Z_{2M} = \Z / (2M\Z)$ for the discrete circle of perimeter $2M$. Let $p\in\Omega$ and consider the restriction of $\hat{U}(p)$ to $\ell^2(\Z\times \Z_{2M})$, denoted by $U(p) : \ell^2(\Z\times \Z_{2M}) \to \ell^2(\Z\times \Z_{2M})$ and defined as
$$\forall (\mu,\nu) \in (\Z^2)^2,\ U(p)_{\mu,\nu} = \hat{U}(p)_{(\mu_1,\mu_2 \mbox{ mod } 2M),(\nu_1,\nu_2 \mbox{ mod } 2M)}.$$
The family $\{ U(p) \}_{p\in \Omega}$ is a $2\Z$-ergodic family of quasi-one-dimensional unitary operators which is called the Chalker-Coddington model on the cylinder. Rewriting this family to a unitary equivalent family of operators, one obtains that the asymptotic behavior of the generalized eigenfunctions is given by a sequence of \emph{i.i.d.}  transfer matrices which are in the Lorentz group $\mathrm{U}(M,M)$. One associates to this sequence of transfer matrices a Furstenberg group and $2M$ Lyapunov exponents which are paired as opposite real numbers. Hence one only considers the positive exponents, $\lambda_1 \geq \cdots \geq \lambda_M\geq 0$.

It is proven in \cite[Theorem 6.1]{ABJ10} that these Lyapunov exponents are all distincts and strictly positive. This result is obtained by proving that the Furstenberg group is equal to the whole Lorentz group $\mathrm{U}(M,M)$. By connectedness, it suffices to show that the Lie algebras are equal. This reconstruction of the Lie algebra of the Lorentz group from the transfer matrices is very close to the one done in the next Section on the random scattering zipper model and for which we give more details. One must precise that the construction of \cite{ABJ10} was done before and strongly inspired the one done in \cite{BM15JMP}.

Using the result of strict positivity of the Lyapunov exponents, the Fractional Moments Method in a similar way as in \cite{HJS09} and the use of spectral averaging (which is one of the reasons why the randomness is given through the Haar measure and not a singular measure), the following result of localization is obtained in \cite{ABJ12}:

\begin{thm}
Let $M\in \N$ and assume $rt\neq 0$. Let $\varphi$ be the angle such that $(t,r)=(\cos(\varphi), \sin(\varphi))$. Then, there exists $\varphi_0 >0$ such that if $|\varphi \mbox{ mod } \frac{\pi}{2}|<\varphi_0$, the Chalker-Coddington model on the cylinder, $\{ U(p) \}_{p\in \Omega}$, exhibits dynamical localization throughout its almost-sure spectrum.
\end{thm}

\subsection{The random scattering zipper}

A scattering zipper is a system obtained by concatenating scattering events each having a fixed even number of outgoing and incoming channels. The number of outgoing channels is equal to the number of incoming channels for each scattering element and for all the elements taken separately. More precisely, a scattering zipper is described by a sequence $(S_n)_{n \in \Z}$ of scattering matrices in the subset $\mbox{U}(2D)_{\inv}$ of the unitary group $\mbox{U}(2D)$ defined by
$$\mbox{\rm U}(2D)_\inv
\; =\;
\left\{\left. S(\alpha,U,V) \in \mbox{\rm U}(2D) \;\right|\; \alpha^*\alpha<\one \;\mbox{ et }U,V\in\mbox{\rm U}(D) 
\right\}
\;,$$%
where $S(\alpha,U,V)\; =\; 
\left( \begin{smallmatrix} \alpha & \rho(\alpha) U 
\\ 
V\widetilde{\rho}(\alpha) & -V\alpha^*U 
\end{smallmatrix} \right) \mbox{ and }  \rho(\alpha) \;=\; (\one-\alpha\alpha^*)^{\frac{1}{2}}, \;\; \widetilde{\rho}(\alpha) \;=\; (\one-\alpha^*\alpha)^{\frac{1}{2}}\;. $

The scattering zipper operator $\CMV$ associated to the sequence $(S_n)_{n \in \Z}$ is the operator acting on $\ell^2( \Z,\C^D)$ and defined by : 
\begin{equation}\label{eq_def_cmv_intro}
\CMV\;=\;\CMVL\,\CMVR\;, 
\end{equation}

\noindent where the two unitary operators $\CMVL$ and $\CMVR$ act on $\ell^2( \Z,\C^D)$ and are given by  
\begin{equation}\label{eq_def_VW_intro}
\CMVL \;=\;  \left( \begin{smallmatrix}
\ddots & & & \\
 & S_0 & &  \\
& & S_{2} &  \\
& & & \ddots 
\end{smallmatrix} \right) \circ s_g^{D}\;, \quad \CMVR \;=\; \left( \begin{smallmatrix}
\ddots & & &  \\
& S_{-1}  & &  \\
 & & S_{1}  &  \\
& & & \ddots   
\end{smallmatrix}\right)\; , 
\end{equation}
\noindent where $s_g$ is the shift operator defined by 
\begin{equation}\label{def_shift_scatt_zip}
s_g : \begin{array}{ccl}
\ell^2( \Z,\C) & \to & \ell^2( \Z,\C) \\
(v_n)_{n\in \Z} & \mapsto & (v_{n+1})_{n\in \Z}.
\end{array}
\end{equation}

Considering the factorization of CMV matrices proved in \cite{CMV03}, the scattering zipper model can be seen as a version of the CMV matrices with matrix-valued coefficients.

We now introduce as in \cite{BM15JMP} a random version of the scattering zipper. Let $\Omega_0 = \UL \times \UL$, $\mathcal{B}_0$ the Borel $\sigma$-algebra on  $\UL \times \UL$ endowed with its usual Lie group topology and  $\mathbb{P}_0 \;=\; \nu_D \otimes \nu_D$ where $\nu_D$ is the Haar measure on $\UL$. We then define the probability space:
$$(\Omega,\mathcal{B},\mathbb{P})=\left( \bigotimes_{n\in \Z} \Omega_0,\; \bigotimes_{n\in \Z} \mathcal{B}_0,\; \bigotimes_{n\in \Z} \mathbb{P}_0\right)$$
For $\omega \in \Omega$ and $n\in \Z$, we denote $\omega_n = (U_n(\omega),V_n(\omega))\in \Omega_0.$

Let $\omega \in \Omega$ and let $(\alpha_n)_{n\in \Z}$ be a sequence of matrices in $\mathcal{M}_{D}(\C)$ independent of $\omega$ such that, for any $n\in \Z$, $\alpha_n^* \alpha_n <\one$. Then, for all $n\in \Z$, we set
$$S_n(\omega) = S(\alpha_n,U_n(\omega),V_n(\omega)) \in  \mbox{U}(2L)_\inv$$
where the sequence $((U_n(\omega),V_n(\omega)))_{n\in \Z}$ is a sequence of \textit{i.i.d.} random variables on the probability space $(\Omega_0,\mathcal{B}_0,\mathbb{P}_0)$. 
Once such a sequence of independent random matrices in $\mbox{U}(2D)_\inv$ is defined, we define, for any $\omega \in \Omega$, the operators $\mathbb{V}_{\omega}$, $\mathbb{W}_{\omega}$ and $\mathbb{U}_{\omega}=\mathbb{V}_{\omega}\; \mathbb{W}_{\omega}$ as in \eqref{eq_def_cmv_intro} and \eqref{eq_def_VW_intro}. 

One calls \emph{random scattering zipper} associated to the sequence $(S_n(\omega))_{n\in \Z}$, the family of operators $\{ \mathbb{U}_{\omega} \}_{\omega \in \Omega}$.

In order to obtain a property of $2\Z$-ergodicity for the family  $\{ \mathbb{U}_{\omega} \}_{\omega \in \Omega}$, we assume that \textbf{the sequence of Verblunsky coefficients $(\alpha_n)_{n\in \N}$ is constant}, equal to a matrix $\alpha \neq 0$ satisfying $\alpha^* \alpha < \one$. This hypothesis also ensures that the independent matrices $S_n(\omega)$ are identically distributed. Comparing with the quasi-one-dimensional Anderson models, the constancy hypothesis of the Verblunsky sequence can be understood as the simple site hypothesis for the Anderson potential: a deterministic potential with constant compact support that is translated along $\Z$ and only the random variables that multiply it differ at each point of the lattice $\Z$. Here $\alpha$ is the analogue of the deterministic potential which is constant along $\Z$ and only the phases which multiply it in the expression of $S_n(\omega)$ vary when $n$ travels along $\Z$. It is therefore not surprising that this hypothesis ensures the ergodicity of  $\{ \mathbb{U}_{\omega} \}_{\omega \in \Omega}$, just as the simple site hypothesis ensures the ergodicity in the case of the quasi-one-dimensional Anderson model.

Using the transfer matrix formalism, we reduce the study of the asymptotic behavior of a solution $\phi$  of 
\begin{equation}\label{eq_vp_scatzip}
\mathbb{U}_{\omega}\phi=z\phi,\quad \mbox{for}\ z\in \SM^1,  
\end{equation}
to the asymptotic behavior of a product of random matrices. Instead of looking at the input-output relations of the scattering matrix $S_n(\omega)$, we look for a new matrix which allows to express $\left( \begin{smallmatrix}                                                                                                                                                                                                                                                                                                                                                                                                              \phi_{n+1} \\
\psi_{n+1}                                                                                                                                                                                                                                                                                                                                                                                                                                \end{smallmatrix}\right)$ in terms of $\left( \begin{smallmatrix}                                                                                                                                                                                                                                                                                                                                                                                                              \phi_n \\
\psi_n                                                                                                                                                                                                                                                                                                                                                                                                                                \end{smallmatrix}
 \right)$ for $\phi$ a solution of \eqref{eq_vp_scatzip} and $\psi= \CMVR \phi$. This is done by transforming the scattering matrices $S_n(\omega)$ belonging to $\mbox{\rm U}(2D)_\inv$ into elements of the Lorentz group via the bijection:
$$\varphi\ :\ \begin{array}{ccl}
            \mbox{\rm U}(2D)_\inv & \to & \mathrm{U}(D,D) \\[2mm]
	    \left( \begin{smallmatrix}
	    \alpha & \beta \\
	    \gamma & \delta
	    \end{smallmatrix} \right) & \mapsto & \left(\begin{smallmatrix}
	    \gamma-\delta \beta^{-1} \alpha & \delta \beta^{-1} \\
	    -\beta^{-1} \alpha & \beta^{-1}
	    \end{smallmatrix} \right)
            \end{array}$$
We have the following relations, proven in \cite{MSB13}:
\begin{equation}\label{eq_trans_mat_odd_even}
\forall n\in \Z,\  \left( \begin{smallmatrix}                                                                                                                                                                                                                                                                                                                                                                                                              \phi_{2n} \\
\psi_{2n}                                                                                                                                                                                                                                                                                                                                                                                                                                \end{smallmatrix}
 \right) = \varphi(z^{-1} S_{2n}(\omega))  \left( \begin{smallmatrix}                                                                                                                                                                                                                                                                                                                                                                                                              \psi_{2n-1} \\
\phi_{2n-1}                                                                                                                                                                                                                                                                                                                                                                                                                                \end{smallmatrix}
 \right) \quad \mbox{and} \quad \left( \begin{smallmatrix}                                                                                                                                                                                                                                                                                                                                                                                                              \psi_{2n+1} \\
\phi_{2n+1}                                                                                                                                                                                                                                                                                                                                                                                                                                \end{smallmatrix}
 \right) = \varphi( S_{2n+1}(\omega))  \left( \begin{smallmatrix}                                                                                                                                                                                                                                                                                                                                                                                                              \phi_{2n} \\
\psi_{2n}                                                                                                                                                                                                                                                                                                                                                                                                                                \end{smallmatrix} \right).
\end{equation}

These relations lead to introduce the application $T(z,\cdot ): \Omega \to \mathrm{U}(D,D)$, 
\begin{equation}\label{eq_def_trans_mat}
 \forall \omega \in \Omega,\ T(z,\omega) =  \left(\begin{smallmatrix} V_{0}(\omega)  & 0 \\ 0 & (U_{0}(\omega))^* \end{smallmatrix} \right)
\hat{T}_0(z)
\left( \begin{smallmatrix} V_{1}(\omega)  & 0 \\ 0 & (U_{1}(\omega))^* \end{smallmatrix} \right)
\hat{T}_1
\end{equation}
with
$$
\hat{T}_0(z) = \left( \begin{smallmatrix}
z^{-1} (\widetilde{\rho}(\alpha))^{-1} & (\widetilde{\rho}(\alpha))^{-1}\alpha^* \\
\alpha (\widetilde{\rho}(\alpha))^{-1} & z (\rho(\alpha))^{-1}
\end{smallmatrix} \right) \quad \mbox{ and }\quad \hat{T}_1 = \left( \begin{smallmatrix}
 (\widetilde{\rho}(\alpha))^{-1} & (\widetilde{\rho}(\alpha))^{-1}\alpha^* \\
\alpha (\widetilde{\rho}(\alpha))^{-1} &  (\rho(\alpha))^{-1}
\end{smallmatrix} \right).
$$
If $\tau : \Omega \to \Omega$ is defined by: $\forall \omega\in \Omega,\ \forall n\in \Z,\ (\tau(\omega))_n \; =\; \omega_{n+2},$ one has: 
\begin{equation}\label{eq_link_trans_scat}
 \forall \omega \in \Omega,\ \forall z\in \SM^1,\ \forall n\in \Z,\ T(z,\tau^{n}(\omega))= \varphi(z^{-1}S_{2n}(\omega))\cdot \varphi(S_{2n-1}(\omega)).
\end{equation}
The matrix $T(z,\tau^{n}(\omega))$ is the n-th transfer matrix associated to $\mathbb{U}_{\omega}$. Then $(T(z,\tau^{n}(\omega)))_{n\in \Z}$ is a sequence  of \emph{i.i.d.} random matrices in $\mathrm{U}(D,D)$ because of the \emph{i.i.d.} character of the sequence $((U_n(\omega),V_n(\omega)))_{n\in \Z}$ in $\Omega$ and because of the constancy of the Verblunsky sequence.

The transfer matrices $T(z,\cdot)$ generate a cocycle $\Phi(z,\cdot,\cdot): \Omega \times \Z \to \mathrm{U}(D,D) $ on the ergodic dynamical system $(\Omega, \mathcal{B}, \mathbb{P}, (\tau^n)_{n \in \Z} )$ defined by   
$$\forall \omega \in\Omega,\ \forall n\in \Z,\ \Phi(z,\omega,n) \;=\; \left\lbrace 
\begin{array}{lcl}
T(z,\tau^{n-1}(\omega)) \dots T(z,\omega) & \mbox{ if } & n >0 \\
I_{2L} & \mbox{ if } & n=0 \\
(T(z,\tau^{n}(\omega)))^{-1} \dots (T(z,\tau^{-1}(\omega)))^{-1} & \mbox{ if } & n <0.
\end{array} \right.
$$

From this cocycle we define the Lyapunov exponents associated to the ergodic family $\{ \mathbb{U}_{\omega} \}_{\omega \in \Omega}$. Let $z\in \SM^1$. For $\mathbb{P}$-almost every $\omega \in \Omega$, the following limits exists and are equal:
\begin{equation}\label{eq_lim_Lyap}
 \Psi(z,\omega)  :=  \lim_{n\to +\infty} ((\Phi(z,\omega,n))^*\Phi(z,\omega,n))^{1/2n}   =  \lim_{n\to -\infty} ((\Phi(z,\omega,n))^*\Phi(z,\omega,n))^{1/2|n|}.
\end{equation}
For every $k\in \{ 1,\ldots, 2D\}$, let $\lambda_k(z,\omega)$ the $k$-th  eigenvalue of $\Psi(z,\omega)$, the eigenvalues being ordered in increasing order. There are then real numbers $\lambda_k(z) \geq 0$ such that, for $\mathbb{P}$-almost every $\omega \in \Omega$, $\lambda_k(z,\omega)=\lambda_k(z)$. We then define the Lyapunov exponents associated to the ergodic family $\{ \mathbb{U}_{\omega} \}_{\omega \in \Omega}$ as being the real numbers $\gamma_k(z)$ defined by : 
$$\forall z\in \SM^1,\ \forall \ k\in \{ 1,\ldots,2D\},\ \gamma_k(z) := \log( \lambda_k(z) ).$$
The fact that the transfer matrices belong to the Lorentz group implies a symmetry relation on the Lyapunov exponents which is the same as in the case where the transfer matrices are in the symplectic group:
$\forall k\ \in \{0,\ldots D \},\ \gamma_{2D-k+1}(z)=-\gamma_{k}(z).$

In\cite{BM15JMP}, we proved the following result.

\begin{thm}[\cite{BM15JMP}]\label{thm_pos_Lyap_scatzip}
If the Verblusnky sequence is constant, for every $z\in \mathbb{S}^1$, 
$$ \gamma_1(z)>  \gamma_2(z)> \cdots >  \gamma_D(z)> 0$$
and the almost-sure absolutely continuous spectrum of $\{\mathbb{U}_{\omega}\}_{\omega \in \Omega}$ is empty.
\end{thm}

The idea is to prove directly that the Furstenberg group $G_{\mu_z}$ associated with the family $\{\mathbb{U}_{\omega}\}_{\omega \in \Omega}$ satisfy, for every $z\in \SM^1$, $G_{\mu_z}  \;=\; \mathrm{U}(D,D)$.
 
Using the \emph{i.i.d.} character of the transfer matrices and the fact that the Haar measure is supported by the unitary group $\mathrm{U}(D)$: \begin{equation}\label{eq_def_G_scatzip}
 G_{\mu_z} \; = \; \overline{  \left\langle \left\{ \left(\begin{smallmatrix} V_{0} & 0 \\ 0 & (U_{0})^* \end{smallmatrix} \right)
\hat{T}_0(z)
 \left(\begin{smallmatrix} V_{1}  & 0 \\ 0 & (U_{1})^* \end{smallmatrix}\right)
\hat{T}_1  \Big| (U_0,V_0,U_1,V_1) \in \mathrm{U}(D)^4 \right\} \right\rangle  }.
\end{equation}

Then, by connectedness of the Lorentz group $\mathrm{U}(D,D)$, to show that $G_{\mu_z}  \;=\; \mathrm{U}(D,D)$, it is sufficient to show the equality of the Lie algebras of these two Lie groups.
The Lie algebra of $\mathrm{U}(D,D)$ is given by: $\mathfrak{u}(D,D) =   \{ T \in \mathcal{M}_{2D} (\C) \;|\; T^*\Ll + \Ll T =0 \},$
or more explicitely,
$$\mathfrak{u}(D,D) = \left\{ \left. \left( \begin{smallmatrix} A & B \\ B^* & C \end{smallmatrix} \right) \in \mathcal{M}_{2D} (\C) \;\right|\; A^*= -A, C^* = -C, (A,B,D) \in  \mathcal{M}_{D} (\C) \right\}.$$

Let $\mathfrak{g}(z):=\mathrm{Lie}(G_{\mu_z})$. For $U_0 = V_0 = U_1 = V_1 = \one$, $\hat{T}_0(z)\hat{T}_1 \in G_{\mu_z}$. Hence, $(\hat{T}_0(z)\hat{T}_1)^{-1}$ is also in $G_{\mu_z}$. Taking this time $ U_1 = V_1 = \one$, letting $U_0, V_0$ take any value in $\mathrm{U}(D)$ and multiplying the right side by $(\hat{T}_0(z)\hat{T}_1)^{-1}$ : $\forall (U_0,V_0)\in \mathrm{U}(D)^2,\ \left( \begin{smallmatrix} V_0 & 0 \\ 0 & U_0^* \end{smallmatrix} \right) \in G_{\mu_z}.$

From the first step, we deduce that $\mathrm{U}(D) \oplus \mathrm{U}(D) \subset G_{\mu_z}$ which implies that $\text{Lie} (\mathrm{U}(D)) \oplus \text{Lie} (\mathrm{U}(D)) \subset \mathfrak{g}(z)$ and since $\text{Lie} (\mathrm{U}(D)) = \{ A \in \mathcal{M}_D (\C) | A^* = -A \},$
$$\mathfrak{a}_1 = \left\{ \left. \left(\begin{smallmatrix} A & 0 \\ 0 & C \end{smallmatrix}\right) \right| A^* = -A, C^*= -C, (A,D) \in \mathcal{M}_D (\C)^2 \right\} \subset \mathfrak{g}(z).$$

The construction of the diagonal blocks of the matrices in $\mathfrak{g}(z)$ is thus obtained very quickly because we have assumed a very regular randomness for the unitary phases $U$ and $V$. The next step is to construct the non-diagonal blocks.

For that, we start by taking $V_0 = U_0 = \one$ and for $j \in \{1,\ldots ,D\}$ and $t \in \R$, $V_1=\mbox{diag}(1,\ldots,1,e^{\ii t},1,\ldots,1)$ with $e^{\ii t}$  at the $j$-th place and $U_1=\one$. We derive at $t=0$ to get : 
$\forall j \in \{1,\ldots ,D\},\ \ii \hat{T}_1^{-1} \left( \begin{smallmatrix} E_{jj} & 0 \\ 0 & 0 \end{smallmatrix} \right) \hat{T}_1 \in \mathfrak{g}(z),$
where $E_{kl}\in \mathcal{M}_{D}(\C)$ is the elementary matrix whose coefficients are all $0$ except the coefficient $(k,l)$ which is $1$. Summing over $j$ it comes, 
$ \ii \hat{T}_1^{-1}  \left(\begin{smallmatrix} \one & 0 \\ 0 & 0 \end{smallmatrix} \right) \hat{T}_1 \in \mathfrak{g}(z)$
or more explicitly, $\ii \left( \begin{smallmatrix} 0 & (\widetilde{\rho}(\alpha))^{-2} \alpha^* \\ - \alpha (\widetilde{\rho}(\alpha))^{-2} & 0 \end{smallmatrix} \right) \in \mathfrak{g}(z).$
So we have matrices in $\mathfrak{g}(z)$ with blocks off the diagonal.  Using the fact that $\alpha \neq 0$ and the inversibility of $(\widetilde{\rho}(\alpha))^{-2}$, using Lie brackets, we get the existence of a couple of indices $(j_0,k_0) \in \{1,\ldots ,D\}^2$ and of $c \in \C$, with $c\neq 0 $, such that  $\ii \left( \begin{smallmatrix} 0 & c E_{k_0j_0} \\ -\overline{c} E_{j_0k_0} & 0 \end{smallmatrix} \right)  \in \mathfrak{g}(z).$ By using good combinations of Lie brackets we then show successively that  
\begin{equation}\label{eq_step5_3}
\forall (j,k) \in \{1,\ldots ,D\}^2 \setminus \{ (j_0,k_0) \},\  \left( \begin{smallmatrix} 0 & E_{kj} \\ E_{jk} & 0  \end{smallmatrix} \right) \in \mathfrak{g}(z)\quad \mbox{and}\quad \left( \begin{smallmatrix} 0 & \ii E_{kj} \\ -\ii E_{jk} & 0  \end{smallmatrix} \right) \in \mathfrak{g}(z),
\end{equation}
then
\begin{equation}\label{eq_step6_2}
\left( \begin{smallmatrix} 0 & E_{k_0j_0} \\ E_{j_0k_0} & 0  \end{smallmatrix} \right) \in \mathfrak{g}(z) \quad \mbox{and finally}\quad \left( \begin{smallmatrix}
0 & \ii E_{k_0j_0} \\ -\ii E_{j_0k_0} & 0  \end{smallmatrix}\right) \in \mathfrak{g}(z).                                                                                                                               
\end{equation}
If we set $\mathfrak{a}_2 = \left\{ \left. \left(  \begin{smallmatrix} 0 & B \\ B^* & 0 \end{smallmatrix} \right)\right| B \in \mathcal{M}_D (\C) \right\},$ as the elements constructed in \eqref{eq_step5_3} and \eqref{eq_step6_2} form a basis of $\mathfrak{a}_2$, we get $\mathfrak{a}_2  \subset \mathfrak{g}(z).$

Finally, since $\mathfrak{u}(D,D)= \mathfrak{a}_1 \oplus \mathfrak{a}_2 \subset \mathfrak{g}(z)$ we get $\mathfrak{u}(D,D)=\mathfrak{g}(z)$ which shows the first point of Theorem \ref{thm_pos_Lyap_scatzip}.

To prove the absence of absolutely continuous spectrum for the family $\{ \mathbb{U}_{\omega} \}_{\omega \in \Omega}$, we have adapted the results of Kotani theory to the framework of quasi-one-dimensional random models of unitary type\cite{BM15JMP}. 

Unfortunately, we do not have yet a localization result for the random scattering zipper model. This could be done by using Fractional Moments Method instead of multi-scale analysis, as in \cite{HJS09,ABJ10}.

\begin{openq}
Prove the analog of Theorem \ref{thm_localization} in the framework of the random scattering zipper.
\end{openq}

It would also be interesting to relate the study of the scattering zippers to some coined quantum walk. For this purpose one should first find some generalized coined quantum walk to which a given scattering zipper is associated in a similar one associates to a random CMV operator a coined quantum walk. The computations should be analog to the one done in\cite{CGMV10,CGMV12,W13} and from the properties of the scattering zippers one could retrieve properties of the underlying generalized coined quantum walk.

\newpage
\begin{acknowledgments}
The author would like to thank Jake Fillman for several suggestions which helped to improve a lot bibliographical aspects of this review article.  He is also thankful to an anonymous referee for many useful comments and suggestions which helped to improve the overall quality of the present paper.

The author is supported by ANR JCJC Project RAW. 
\end{acknowledgments}

\bibliography{HDRBoumaza.bib}

\begin{thebibliography}{83}%
\makeatletter
\providecommand \@ifxundefined [1]{%
 \@ifx{#1\undefined}
}%
\providecommand \@ifnum [1]{%
 \ifnum #1\expandafter \@firstoftwo
 \else \expandafter \@secondoftwo
 \fi
}%
\providecommand \@ifx [1]{%
 \ifx #1\expandafter \@firstoftwo
 \else \expandafter \@secondoftwo
 \fi
}%
\providecommand \natexlab [1]{#1}%
\providecommand \enquote  [1]{``#1''}%
\providecommand \bibnamefont  [1]{#1}%
\providecommand \bibfnamefont [1]{#1}%
\providecommand \citenamefont [1]{#1}%
\providecommand \href@noop [0]{\@secondoftwo}%
\providecommand \href [0]{\begingroup \@sanitize@url \@href}%
\providecommand \@href[1]{\@@startlink{#1}\@@href}%
\providecommand \@@href[1]{\endgroup#1\@@endlink}%
\providecommand \@sanitize@url [0]{\catcode `\\12\catcode `\$12\catcode
  `\&12\catcode `\#12\catcode `\^12\catcode `\_12\catcode `\%12\relax}%
\providecommand \@@startlink[1]{}%
\providecommand \@@endlink[0]{}%
\providecommand \url  [0]{\begingroup\@sanitize@url \@url }%
\providecommand \@url [1]{\endgroup\@href {#1}{\urlprefix }}%
\providecommand \urlprefix  [0]{URL }%
\providecommand \Eprint [0]{\href }%
\providecommand \doibase [0]{https://doi.org/}%
\providecommand \selectlanguage [0]{\@gobble}%
\providecommand \bibinfo  [0]{\@secondoftwo}%
\providecommand \bibfield  [0]{\@secondoftwo}%
\providecommand \translation [1]{[#1]}%
\providecommand \BibitemOpen [0]{}%
\providecommand \bibitemStop [0]{}%
\providecommand \bibitemNoStop [0]{.\EOS\space}%
\providecommand \EOS [0]{\spacefactor3000\relax}%
\providecommand \BibitemShut  [1]{\csname bibitem#1\endcsname}%
\let\auto@bib@innerbib\@empty
\bibitem [{\citenamefont {Ahlbrecht}, \citenamefont {Scholz},\ and\
  \citenamefont {Werner}(2011)}]{ASW11}%
  \BibitemOpen
  \bibfield  {author} {\bibinfo {author} {\bibnamefont {Ahlbrecht},
  \bibfnamefont {A.}}, \bibinfo {author} {\bibnamefont {Scholz}, \bibfnamefont
  {V.~B.}}, and\ \bibinfo {author} {\bibnamefont {Werner}, \bibfnamefont
  {A.~H.}},\ }\bibfield  {title} {{\selectlanguage {English}\enquote {\bibinfo
  {title} {Disordered quantum walks in one lattice dimension},}\ }}\href
  {https://doi.org/10.1063/1.3643768} {\bibfield  {journal} {\bibinfo
  {journal} {J. Math. Phys.}\ }\textbf {\bibinfo {volume} {52}},\ \bibinfo
  {pages} {102201, 48} (\bibinfo {year} {2011})}\BibitemShut {NoStop}%
\bibitem [{\citenamefont {{Aizenman}}\ and\ \citenamefont
  {{Warzel}}(2015)}]{AW15}%
  \BibitemOpen
  \bibfield  {author} {\bibinfo {author} {\bibnamefont {{Aizenman}},
  \bibfnamefont {M.}}and\ \bibinfo {author} {\bibnamefont {{Warzel}},
  \bibfnamefont {S.}},\ }\href@noop {} {{\selectlanguage {English}\emph
  {\bibinfo {title} {{Random operators. Disorder effects on quantum spectra and
  dynamics}}}}},\ Vol.\ \bibinfo {volume} {168}\ (\bibinfo  {publisher}
  {Providence, RI: American Mathematical Society (AMS)},\ \bibinfo {year}
  {2015})\ pp.\ \bibinfo {pages} {xiv + 326}\BibitemShut {NoStop}%
\bibitem [{\citenamefont {{Albeverio}}\ \emph {et~al.}(2005)\citenamefont
  {{Albeverio}}, \citenamefont {{Gesztesy}}, \citenamefont {{H{\o}egh-Krohn}},\
  and\ \citenamefont {{Holden}}}]{AGHKH05}%
  \BibitemOpen
  \bibfield  {author} {\bibinfo {author} {\bibnamefont {{Albeverio}},
  \bibfnamefont {S.}}, \bibinfo {author} {\bibnamefont {{Gesztesy}},
  \bibfnamefont {F.}}, \bibinfo {author} {\bibnamefont {{H{\o}egh-Krohn}},
  \bibfnamefont {R.}}, and\ \bibinfo {author} {\bibnamefont {{Holden}},
  \bibfnamefont {H.}},\ }\href@noop {} {{\selectlanguage {English}\emph
  {\bibinfo {title} {{Solvable models in quantum mechanics. With an appendix by
  Pavel Exner}}}}}\ (\bibinfo  {publisher} {Providence, RI: AMS Chelsea
  Publishing},\ \bibinfo {year} {2005})\ pp.\ \bibinfo {pages} {xiv +
  488}\BibitemShut {NoStop}%
\bibitem [{\citenamefont {Anderson}(2018)}]{And18}%
  \BibitemOpen
  \bibfield  {author} {\bibinfo {author} {\bibnamefont {Anderson},
  \bibfnamefont {P.}},\ }\href {https://books.google.fr/books?id=tF0PEAAAQBAJ}
  {\emph {\bibinfo {title} {Basic Notions Of Condensed Matter Physics}}}\
  (\bibinfo  {publisher} {CRC Press},\ \bibinfo {year} {2018})\BibitemShut
  {NoStop}%
\bibitem [{\citenamefont {Anderson}(1958)}]{A58}%
  \BibitemOpen
  \bibfield  {author} {\bibinfo {author} {\bibnamefont {Anderson},
  \bibfnamefont {P.~W.}},\ }\bibfield  {title} {\enquote {\bibinfo {title}
  {Absence of diffusion in certain random lattices},}\ }\href
  {https://doi.org/10.1103/PhysRev.109.1492} {\bibfield  {journal} {\bibinfo
  {journal} {Phys. Rev.}\ }\textbf {\bibinfo {volume} {109}},\ \bibinfo {pages}
  {1492--1505} (\bibinfo {year} {1958})}\BibitemShut {NoStop}%
\bibitem [{\citenamefont {{Arnold}}(1998)}]{Ar98}%
  \BibitemOpen
  \bibfield  {author} {\bibinfo {author} {\bibnamefont {{Arnold}},
  \bibfnamefont {L.}},\ }\href@noop {} {{\selectlanguage {English}\emph
  {\bibinfo {title} {{Random dynamical systems}}}}}\ (\bibinfo  {publisher}
  {Berlin: Springer},\ \bibinfo {year} {1998})\ pp.\ \bibinfo {pages} {xi +
  586}\BibitemShut {NoStop}%
\bibitem [{\citenamefont {Asch}, \citenamefont {Bourget},\ and\ \citenamefont
  {Joye}(2010)}]{ABJ10}%
  \BibitemOpen
  \bibfield  {author} {\bibinfo {author} {\bibnamefont {Asch}, \bibfnamefont
  {J.}}, \bibinfo {author} {\bibnamefont {Bourget}, \bibfnamefont {O.}}, and\
  \bibinfo {author} {\bibnamefont {Joye}, \bibfnamefont {A.}},\ }\bibfield
  {title} {{\selectlanguage {English}\enquote {\bibinfo {title} {Localization
  properties of the {Chalker}-{Coddington} model},}\ }}\href
  {https://doi.org/10.1007/s00023-010-0056-1} {\bibfield  {journal} {\bibinfo
  {journal} {Ann. Henri Poincar{\'e}}\ }\textbf {\bibinfo {volume} {11}},\
  \bibinfo {pages} {1341--1373} (\bibinfo {year} {2010})}\BibitemShut {NoStop}%
\bibitem [{\citenamefont {Asch}, \citenamefont {Bourget},\ and\ \citenamefont
  {Joye}(2012)}]{ABJ12}%
  \BibitemOpen
  \bibfield  {author} {\bibinfo {author} {\bibnamefont {Asch}, \bibfnamefont
  {J.}}, \bibinfo {author} {\bibnamefont {Bourget}, \bibfnamefont {O.}}, and\
  \bibinfo {author} {\bibnamefont {Joye}, \bibfnamefont {A.}},\ }\bibfield
  {title} {{\selectlanguage {English}\enquote {\bibinfo {title} {Dynamical
  localization of the {Chalker}-{Coddington} model far from transition},}\
  }}\href {https://doi.org/10.1007/s10955-012-0477-y} {\bibfield  {journal}
  {\bibinfo  {journal} {J. Stat. Phys.}\ }\textbf {\bibinfo {volume} {147}},\
  \bibinfo {pages} {194--205} (\bibinfo {year} {2012})}\BibitemShut {NoStop}%
\bibitem [{\citenamefont {Basko}, \citenamefont {Aleiner},\ and\ \citenamefont
  {Altshuler}(2006)}]{BAA06}%
  \BibitemOpen
  \bibfield  {author} {\bibinfo {author} {\bibnamefont {Basko}, \bibfnamefont
  {D.~M.}}, \bibinfo {author} {\bibnamefont {Aleiner}, \bibfnamefont {I.~L.}},
  and\ \bibinfo {author} {\bibnamefont {Altshuler}, \bibfnamefont {B.~L.}},\
  }\bibfield  {title} {{\selectlanguage {English}\enquote {\bibinfo {title}
  {Metal--insulator transition in a weakly interacting many-electron system
  with localized single-particle states},}\ }}\href
  {https://doi.org/10.1016/j.aop.2005.11.014} {\bibfield  {journal} {\bibinfo
  {journal} {Ann. Phys.}\ }\textbf {\bibinfo {volume} {321}},\ \bibinfo {pages}
  {1126--1205} (\bibinfo {year} {2006})}\BibitemShut {NoStop}%
\bibitem [{\citenamefont {Berezin}\ and\ \citenamefont {Faddeev}(1961)}]{BF61}%
  \BibitemOpen
  \bibfield  {author} {\bibinfo {author} {\bibnamefont {Berezin}, \bibfnamefont
  {F.~A.}}and\ \bibinfo {author} {\bibnamefont {Faddeev}, \bibfnamefont
  {L.~D.}},\ }\bibfield  {title} {{\selectlanguage {English}\enquote {\bibinfo
  {title} {A remark on {Schr{\"o}dinger}'s equation with a singular
  potential},}\ }}\href@noop {} {\bibfield  {journal} {\bibinfo  {journal}
  {Sov. Math., Dokl.}\ }\textbf {\bibinfo {volume} {2}},\ \bibinfo {pages}
  {372--375} (\bibinfo {year} {1961})}\BibitemShut {NoStop}%
\bibitem [{\citenamefont {{Bougerol}}\ and\ \citenamefont
  {{Lacroix}}(1985)}]{BL85}%
  \BibitemOpen
  \bibfield  {author} {\bibinfo {author} {\bibnamefont {{Bougerol}},
  \bibfnamefont {P.}}and\ \bibinfo {author} {\bibnamefont {{Lacroix}},
  \bibfnamefont {J.}},\ }\href@noop {} {{\selectlanguage {English}\emph
  {\bibinfo {title} {{Products of random matrices with applications to
  Schr\"odinger operators}}}}},\ Vol.~\bibinfo {volume} {8}\ (\bibinfo
  {publisher} {Birkh\"auser, Boston, MA},\ \bibinfo {year} {1985})\BibitemShut
  {NoStop}%
\bibitem [{\citenamefont {{Boumaza}}(2007)}]{B07}%
  \BibitemOpen
  \bibfield  {author} {\bibinfo {author} {\bibnamefont {{Boumaza}},
  \bibfnamefont {H.}},\ }\bibfield  {title} {{\selectlanguage {English}\enquote
  {\bibinfo {title} {{Positivity of Lyapunov exponents for a continuous
  matrix-valued Anderson model}},}\ }}\href
  {https://doi.org/10.1007/s11040-007-9023-6} {\bibfield  {journal} {\bibinfo
  {journal} {{Math. Phys. Anal. Geom.}}\ }\textbf {\bibinfo {volume} {10}},\
  \bibinfo {pages} {97--122} (\bibinfo {year} {2007})}\BibitemShut {NoStop}%
\bibitem [{\citenamefont {{Boumaza}}(2008)}]{B08RMP}%
  \BibitemOpen
  \bibfield  {author} {\bibinfo {author} {\bibnamefont {{Boumaza}},
  \bibfnamefont {H.}},\ }\bibfield  {title} {{\selectlanguage {English}\enquote
  {\bibinfo {title} {{H\"older continuity of the integrated density of states
  for matrix-valued Anderson models}},}\ }}\href
  {https://doi.org/10.1142/S0129055X08003456} {\bibfield  {journal} {\bibinfo
  {journal} {{Rev. Math. Phys.}}\ }\textbf {\bibinfo {volume} {20}},\ \bibinfo
  {pages} {873--900} (\bibinfo {year} {2008})}\BibitemShut {NoStop}%
\bibitem [{\citenamefont {{Boumaza}}(2009{\natexlab{a}})}]{B09LMP}%
  \BibitemOpen
  \bibfield  {author} {\bibinfo {author} {\bibnamefont {{Boumaza}},
  \bibfnamefont {H.}},\ }\bibfield  {title} {{\selectlanguage {English}\enquote
  {\bibinfo {title} {{A matrix-valued point interactions model}},}\ }}\href
  {https://doi.org/10.1007/s11005-008-0289-9} {\bibfield  {journal} {\bibinfo
  {journal} {{Lett. Math. Phys.}}\ }\textbf {\bibinfo {volume} {87}},\ \bibinfo
  {pages} {81--97} (\bibinfo {year} {2009}{\natexlab{a}})}\BibitemShut
  {NoStop}%
\bibitem [{\citenamefont {{Boumaza}}(2009{\natexlab{b}})}]{B09MPAG}%
  \BibitemOpen
  \bibfield  {author} {\bibinfo {author} {\bibnamefont {{Boumaza}},
  \bibfnamefont {H.}},\ }\bibfield  {title} {{\selectlanguage {English}\enquote
  {\bibinfo {title} {{Localization for a matrix-valued Anderson model}},}\
  }}\href {https://doi.org/10.1007/s11040-009-9061-3} {\bibfield  {journal}
  {\bibinfo  {journal} {{Math. Phys. Anal. Geom.}}\ }\textbf {\bibinfo {volume}
  {12}},\ \bibinfo {pages} {255--286} (\bibinfo {year}
  {2009}{\natexlab{b}})}\BibitemShut {NoStop}%
\bibitem [{\citenamefont {{Boumaza}}(2013)}]{B13}%
  \BibitemOpen
  \bibfield  {author} {\bibinfo {author} {\bibnamefont {{Boumaza}},
  \bibfnamefont {H.}},\ }\bibfield  {title} {{\selectlanguage {English}\enquote
  {\bibinfo {title} {{Localization for an Anderson-Bernoulli model with generic
  interaction potential}},}\ }}\href {https://doi.org/10.2748/tmj/1365452625}
  {\bibfield  {journal} {\bibinfo  {journal} {{Tohoku Math. J. (2)}}\ }\textbf
  {\bibinfo {volume} {65}},\ \bibinfo {pages} {57--74} (\bibinfo {year}
  {2013})}\BibitemShut {NoStop}%
\bibitem [{\citenamefont {{Boumaza}}\ and\ \citenamefont
  {{Marin}}(2015)}]{BM15JMP}%
  \BibitemOpen
  \bibfield  {author} {\bibinfo {author} {\bibnamefont {{Boumaza}},
  \bibfnamefont {H.}}and\ \bibinfo {author} {\bibnamefont {{Marin}},
  \bibfnamefont {L.}},\ }\bibfield  {title} {{\selectlanguage {English}\enquote
  {\bibinfo {title} {{Absence of absolutely continuous spectrum for random
  scattering zippers}},}\ }}\href {https://doi.org/10.1063/1.4906809}
  {\bibfield  {journal} {\bibinfo  {journal} {{J. Math. Phys.}}\ }\textbf
  {\bibinfo {volume} {56}},\ \bibinfo {pages} {022701, 13} (\bibinfo {year}
  {2015})}\BibitemShut {NoStop}%
\bibitem [{\citenamefont {{Boumaza}}\ and\ \citenamefont
  {{Najar}}(2015)}]{BN15JSP}%
  \BibitemOpen
  \bibfield  {author} {\bibinfo {author} {\bibnamefont {{Boumaza}},
  \bibfnamefont {H.}}and\ \bibinfo {author} {\bibnamefont {{Najar}},
  \bibfnamefont {H.}},\ }\bibfield  {title} {{\selectlanguage {English}\enquote
  {\bibinfo {title} {{Lifshitz tails for continuous matrix-valued Anderson
  models}},}\ }}\href {https://doi.org/10.1007/s10955-015-1255-4} {\bibfield
  {journal} {\bibinfo  {journal} {{J. Stat. Phys.}}\ }\textbf {\bibinfo
  {volume} {160}},\ \bibinfo {pages} {371--396} (\bibinfo {year}
  {2015})}\BibitemShut {NoStop}%
\bibitem [{\citenamefont {{Boumaza}}\ and\ \citenamefont
  {{Stolz}}(2007)}]{BS07}%
  \BibitemOpen
  \bibfield  {author} {\bibinfo {author} {\bibnamefont {{Boumaza}},
  \bibfnamefont {H.}}and\ \bibinfo {author} {\bibnamefont {{Stolz}},
  \bibfnamefont {G.}},\ }\bibfield  {title} {{\selectlanguage {English}\enquote
  {\bibinfo {title} {{Positivity of Lyapunov exponents for Anderson-type models
  on two coupled strings}},}\ }}\href@noop {} {\bibfield  {journal} {\bibinfo
  {journal} {{Electron. J. Differ. Equ.}}\ }\textbf {\bibinfo {volume}
  {2007}},\ \bibinfo {pages} {18} (\bibinfo {year} {2007})},\ \bibinfo {note}
  {id/No 47}\BibitemShut {NoStop}%
\bibitem [{\citenamefont {Bourgain}\ and\ \citenamefont {Kenig}(2005)}]{BK05}%
  \BibitemOpen
  \bibfield  {author} {\bibinfo {author} {\bibnamefont {Bourgain},
  \bibfnamefont {J.}}and\ \bibinfo {author} {\bibnamefont {Kenig},
  \bibfnamefont {C.~E.}},\ }\bibfield  {title} {{\selectlanguage
  {English}\enquote {\bibinfo {title} {On localization in the continuous
  {Anderson}-{Bernoulli} model in higher dimension},}\ }}\href
  {https://doi.org/10.1007/s00222-004-0435-7} {\bibfield  {journal} {\bibinfo
  {journal} {Invent. Math.}\ }\textbf {\bibinfo {volume} {161}},\ \bibinfo
  {pages} {389--426} (\bibinfo {year} {2005})}\BibitemShut {NoStop}%
\bibitem [{\citenamefont {Brandes}\ and\ \citenamefont
  {Kettemann}(2003)}]{BrKe03}%
  \BibitemOpen
  \bibinfo {editor} {\bibnamefont {Brandes}, \bibfnamefont {T.}}and\ \bibinfo
  {editor} {\bibnamefont {Kettemann}, \bibfnamefont {S.}},\ eds.,\ \href
  {https://doi.org/10.1007/b13139} {{\selectlanguage {English}\emph {\bibinfo
  {title} {Anderson localization and its ramification. {Disorder}, phase
  coherence, and electron correlations.}}}},\ \bibinfo {series} {Lect. Notes
  Phys.}, Vol.\ \bibinfo {volume} {630}\ (\bibinfo  {publisher} {Berlin:
  Springer},\ \bibinfo {year} {2003})\BibitemShut {NoStop}%
\bibitem [{\citenamefont {{Breuillard}}(2006)}]{BPec06}%
  \BibitemOpen
  \bibfield  {author} {\bibinfo {author} {\bibnamefont {{Breuillard}},
  \bibfnamefont {E.}},\ }\href
  {https://www.imo.universite-paris-saclay.fr/~breuilla/Peccot4.pdf} {\enquote
  {\bibinfo {title} {{Cours Peccot 2006, Propri\'et\'es qualitatives des
  groupes discrets}},}\ } (\bibinfo {year} {2006})\BibitemShut {NoStop}%
\bibitem [{\citenamefont {{Breuillard}}\ and\ \citenamefont
  {{Gelander}}(2003)}]{BG03}%
  \BibitemOpen
  \bibfield  {author} {\bibinfo {author} {\bibnamefont {{Breuillard}},
  \bibfnamefont {E.}}and\ \bibinfo {author} {\bibnamefont {{Gelander}},
  \bibfnamefont {T.}},\ }\bibfield  {title} {{\selectlanguage {English}\enquote
  {\bibinfo {title} {{On dense free subgroups of Lie groups}},}\ }}\href
  {https://doi.org/10.1016/S0021-8693(02)00675-0} {\bibfield  {journal}
  {\bibinfo  {journal} {{J. Algebra}}\ }\textbf {\bibinfo {volume} {261}},\
  \bibinfo {pages} {448--467} (\bibinfo {year} {2003})}\BibitemShut {NoStop}%
\bibitem [{\citenamefont {Bucaj}\ \emph
  {et~al.}(2019{\natexlab{a}})\citenamefont {Bucaj}, \citenamefont {Damanik},
  \citenamefont {Fillman}, \citenamefont {Gerbuz}, \citenamefont {Vandenboom},
  \citenamefont {Wang},\ and\ \citenamefont {Zhang}}]{BDFGVW19}%
  \BibitemOpen
  \bibfield  {author} {\bibinfo {author} {\bibnamefont {Bucaj}, \bibfnamefont
  {V.}}, \bibinfo {author} {\bibnamefont {Damanik}, \bibfnamefont {D.}},
  \bibinfo {author} {\bibnamefont {Fillman}, \bibfnamefont {J.}}, \bibinfo
  {author} {\bibnamefont {Gerbuz}, \bibfnamefont {V.}}, \bibinfo {author}
  {\bibnamefont {Vandenboom}, \bibfnamefont {T.}}, \bibinfo {author}
  {\bibnamefont {Wang}, \bibfnamefont {F.}}, and\ \bibinfo {author}
  {\bibnamefont {Zhang}, \bibfnamefont {Z.}},\ }\bibfield  {title}
  {{\selectlanguage {English}\enquote {\bibinfo {title} {Localization for the
  one-dimensional {Anderson} model via positivity and large deviations for the
  {Lyapunov} exponent},}\ }}\href {https://doi.org/10.1090/tran/7832}
  {\bibfield  {journal} {\bibinfo  {journal} {Trans. Am. Math. Soc.}\ }\textbf
  {\bibinfo {volume} {372}},\ \bibinfo {pages} {3619--3667} (\bibinfo {year}
  {2019}{\natexlab{a}})}\BibitemShut {NoStop}%
\bibitem [{\citenamefont {Bucaj}\ \emph
  {et~al.}(2019{\natexlab{b}})\citenamefont {Bucaj}, \citenamefont {Damanik},
  \citenamefont {Fillman}, \citenamefont {Gerbuz}, \citenamefont {VandenBoom},
  \citenamefont {Wang},\ and\ \citenamefont {Zhang}}]{BDFGVW19b}%
  \BibitemOpen
  \bibfield  {author} {\bibinfo {author} {\bibnamefont {Bucaj}, \bibfnamefont
  {V.}}, \bibinfo {author} {\bibnamefont {Damanik}, \bibfnamefont {D.}},
  \bibinfo {author} {\bibnamefont {Fillman}, \bibfnamefont {J.}}, \bibinfo
  {author} {\bibnamefont {Gerbuz}, \bibfnamefont {V.}}, \bibinfo {author}
  {\bibnamefont {VandenBoom}, \bibfnamefont {T.}}, \bibinfo {author}
  {\bibnamefont {Wang}, \bibfnamefont {F.}}, and\ \bibinfo {author}
  {\bibnamefont {Zhang}, \bibfnamefont {Z.}},\ }\bibfield  {title}
  {{\selectlanguage {English}\enquote {\bibinfo {title} {Positive {Lyapunov}
  exponents and a large deviation theorem for continuum {Anderson} models,
  briefly},}\ }}\href {https://doi.org/10.1016/j.jfa.2019.05.028} {\bibfield
  {journal} {\bibinfo  {journal} {J. Funct. Anal.}\ }\textbf {\bibinfo {volume}
  {277}},\ \bibinfo {pages} {3179--3186} (\bibinfo {year}
  {2019}{\natexlab{b}})}\BibitemShut {NoStop}%
\bibitem [{\citenamefont {Cantero}\ \emph {et~al.}(2012)\citenamefont
  {Cantero}, \citenamefont {Gr{\"u}nbaum}, \citenamefont {Moral},\ and\
  \citenamefont {Vel{\'a}zquez}}]{CGMV12}%
  \BibitemOpen
  \bibfield  {author} {\bibinfo {author} {\bibnamefont {Cantero}, \bibfnamefont
  {M.~J.}}, \bibinfo {author} {\bibnamefont {Gr{\"u}nbaum}, \bibfnamefont
  {F.~A.}}, \bibinfo {author} {\bibnamefont {Moral}, \bibfnamefont {L.}}, and\
  \bibinfo {author} {\bibnamefont {Vel{\'a}zquez}, \bibfnamefont {L.}},\
  }\bibfield  {title} {{\selectlanguage {English}\enquote {\bibinfo {title}
  {One-dimensional quantum walks with one defect},}\ }}\href
  {https://doi.org/10.1142/S0129055X1250002X} {\bibfield  {journal} {\bibinfo
  {journal} {Rev. Math. Phys.}\ }\textbf {\bibinfo {volume} {24}},\ \bibinfo
  {pages} {1250002, 52} (\bibinfo {year} {2012})}\BibitemShut {NoStop}%
\bibitem [{\citenamefont {Cantero}\ \emph {et~al.}(2010)\citenamefont
  {Cantero}, \citenamefont {Moral}, \citenamefont {Gr{\"u}nbaum},\ and\
  \citenamefont {Vel{\'a}zquez}}]{CGMV10}%
  \BibitemOpen
  \bibfield  {author} {\bibinfo {author} {\bibnamefont {Cantero}, \bibfnamefont
  {M.~J.}}, \bibinfo {author} {\bibnamefont {Moral}, \bibfnamefont {L.}},
  \bibinfo {author} {\bibnamefont {Gr{\"u}nbaum}, \bibfnamefont {F.~A.}}, and\
  \bibinfo {author} {\bibnamefont {Vel{\'a}zquez}, \bibfnamefont {L.}},\
  }\bibfield  {title} {{\selectlanguage {English}\enquote {\bibinfo {title}
  {Matrix-valued {Szeg{\H{o}}} polynomials and quantum random walks},}\ }}\href
  {https://doi.org/10.1002/cpa.20312} {\bibfield  {journal} {\bibinfo
  {journal} {Commun. Pure Appl. Math.}\ }\textbf {\bibinfo {volume} {63}},\
  \bibinfo {pages} {464--507} (\bibinfo {year} {2010})}\BibitemShut {NoStop}%
\bibitem [{\citenamefont {Cantero}, \citenamefont {Moral},\ and\ \citenamefont
  {Vel{\'a}zquez}(2003)}]{CMV03}%
  \BibitemOpen
  \bibfield  {author} {\bibinfo {author} {\bibnamefont {Cantero}, \bibfnamefont
  {M.~J.}}, \bibinfo {author} {\bibnamefont {Moral}, \bibfnamefont {L.}}, and\
  \bibinfo {author} {\bibnamefont {Vel{\'a}zquez}, \bibfnamefont {L.}},\
  }\bibfield  {title} {{\selectlanguage {English}\enquote {\bibinfo {title}
  {Five-diagonal matrices and zeros of orthogonal polynomials on the unit
  circle},}\ }}\href {https://doi.org/10.1016/S0024-3795(02)00457-3} {\bibfield
   {journal} {\bibinfo  {journal} {Linear Algebra Appl.}\ }\textbf {\bibinfo
  {volume} {362}},\ \bibinfo {pages} {29--56} (\bibinfo {year}
  {2003})}\BibitemShut {NoStop}%
\bibitem [{\citenamefont {Carmona}, \citenamefont {Klein},\ and\ \citenamefont
  {Martinelli}(1987)}]{CKM87}%
  \BibitemOpen
  \bibfield  {author} {\bibinfo {author} {\bibnamefont {Carmona}, \bibfnamefont
  {R.}}, \bibinfo {author} {\bibnamefont {Klein}, \bibfnamefont {A.}}, and\
  \bibinfo {author} {\bibnamefont {Martinelli}, \bibfnamefont {F.}},\
  }\bibfield  {title} {{\selectlanguage {English}\enquote {\bibinfo {title}
  {Anderson localization for {Bernoulli} and other singular potentials},}\
  }}\href {https://doi.org/10.1007/BF01210702} {\bibfield  {journal} {\bibinfo
  {journal} {Commun. Math. Phys.}\ }\textbf {\bibinfo {volume} {108}},\
  \bibinfo {pages} {41--66} (\bibinfo {year} {1987})}\BibitemShut {NoStop}%
\bibitem [{\citenamefont {{Carmona}}\ and\ \citenamefont
  {{Lacroix}}(1990)}]{CL90}%
  \BibitemOpen
  \bibfield  {author} {\bibinfo {author} {\bibnamefont {{Carmona}},
  \bibfnamefont {R.}}and\ \bibinfo {author} {\bibnamefont {{Lacroix}},
  \bibfnamefont {J.}},\ }\href@noop {} {{\selectlanguage {English}\emph
  {\bibinfo {title} {{Spectral theory of random Schr\"odinger operators}}}}}\
  (\bibinfo  {publisher} {Basel etc.: Birkh\"auser Verlag},\ \bibinfo {year}
  {1990})\ pp.\ \bibinfo {pages} {xxvi + 587}\BibitemShut {NoStop}%
\bibitem [{\citenamefont {Chapman}\ and\ \citenamefont {Stolz}(2015)}]{CS15}%
  \BibitemOpen
  \bibfield  {author} {\bibinfo {author} {\bibnamefont {Chapman}, \bibfnamefont
  {J.}}and\ \bibinfo {author} {\bibnamefont {Stolz}, \bibfnamefont {G.}},\
  }\bibfield  {title} {{\selectlanguage {English}\enquote {\bibinfo {title}
  {Localization for random block operators related to the {{\(XY\)}} spin
  chain},}\ }}\href {https://doi.org/10.1007/s00023-014-0328-2} {\bibfield
  {journal} {\bibinfo  {journal} {Ann. Henri Poincar{\'e}}\ }\textbf {\bibinfo
  {volume} {16}},\ \bibinfo {pages} {405--435} (\bibinfo {year}
  {2015})}\BibitemShut {NoStop}%
\bibitem [{\citenamefont {Damanik}\ \emph {et~al.}(2021)\citenamefont
  {Damanik}, \citenamefont {Fillman}, \citenamefont {Helman}, \citenamefont
  {Kesten},\ and\ \citenamefont {Sukhtaiev}}]{DFHKS21}%
  \BibitemOpen
  \bibfield  {author} {\bibinfo {author} {\bibnamefont {Damanik}, \bibfnamefont
  {D.}}, \bibinfo {author} {\bibnamefont {Fillman}, \bibfnamefont {J.}},
  \bibinfo {author} {\bibnamefont {Helman}, \bibfnamefont {M.}}, \bibinfo
  {author} {\bibnamefont {Kesten}, \bibfnamefont {J.}}, and\ \bibinfo {author}
  {\bibnamefont {Sukhtaiev}, \bibfnamefont {S.}},\ }\bibfield  {title}
  {{\selectlanguage {English}\enquote {\bibinfo {title} {Random {Hamiltonians}
  with arbitrary point interactions in one dimension},}\ }}\href
  {https://doi.org/10.1016/j.jde.2021.01.044} {\bibfield  {journal} {\bibinfo
  {journal} {J. Differ. Equations}\ }\textbf {\bibinfo {volume} {282}},\
  \bibinfo {pages} {104--126} (\bibinfo {year} {2021})}\BibitemShut {NoStop}%
\bibitem [{\citenamefont {Damanik}, \citenamefont {Fillman},\ and\
  \citenamefont {Sukhtaiev}(2020)}]{DFS20}%
  \BibitemOpen
  \bibfield  {author} {\bibinfo {author} {\bibnamefont {Damanik}, \bibfnamefont
  {D.}}, \bibinfo {author} {\bibnamefont {Fillman}, \bibfnamefont {J.}}, and\
  \bibinfo {author} {\bibnamefont {Sukhtaiev}, \bibfnamefont {S.}},\ }\bibfield
   {title} {{\selectlanguage {English}\enquote {\bibinfo {title} {Localization
  for {Anderson} models on metric and discrete tree graphs},}\ }}\href
  {https://doi.org/10.1007/s00208-019-01912-6} {\bibfield  {journal} {\bibinfo
  {journal} {Math. Ann.}\ }\textbf {\bibinfo {volume} {376}},\ \bibinfo {pages}
  {1337--1393} (\bibinfo {year} {2020})}\BibitemShut {NoStop}%
\bibitem [{\citenamefont {{Damanik}}, \citenamefont {{Sims}},\ and\
  \citenamefont {{Stolz}}(2002)}]{DSS02}%
  \BibitemOpen
  \bibfield  {author} {\bibinfo {author} {\bibnamefont {{Damanik}},
  \bibfnamefont {D.}}, \bibinfo {author} {\bibnamefont {{Sims}}, \bibfnamefont
  {R.}}, and\ \bibinfo {author} {\bibnamefont {{Stolz}}, \bibfnamefont {G.}},\
  }\bibfield  {title} {{\selectlanguage {English}\enquote {\bibinfo {title}
  {{Localization for one-dimensional, continuum, Bernoulli-Anderson models.}}}\
  }}\href {https://doi.org/10.1215/S0012-7094-02-11414-8} {\bibfield  {journal}
  {\bibinfo  {journal} {{Duke Math. J.}}\ }\textbf {\bibinfo {volume} {114}},\
  \bibinfo {pages} {59--100} (\bibinfo {year} {2002})}\BibitemShut {NoStop}%
\bibitem [{\citenamefont {Damanik}\ and\ \citenamefont
  {Stollmann}(2001)}]{DS01}%
  \BibitemOpen
  \bibfield  {author} {\bibinfo {author} {\bibnamefont {Damanik}, \bibfnamefont
  {D.}}and\ \bibinfo {author} {\bibnamefont {Stollmann}, \bibfnamefont {P.}},\
  }\bibfield  {title} {{\selectlanguage {English}\enquote {\bibinfo {title}
  {Multi-scale analysis implies strong dynamical localization},}\ }}\href
  {https://doi.org/10.1007/PL00001666} {\bibfield  {journal} {\bibinfo
  {journal} {Geom. Funct. Anal.}\ }\textbf {\bibinfo {volume} {11}},\ \bibinfo
  {pages} {11--29} (\bibinfo {year} {2001})}\BibitemShut {NoStop}%
\bibitem [{\citenamefont {Delyon}, \citenamefont {Simon},\ and\ \citenamefont
  {Souillard}(1985)}]{DSS85}%
  \BibitemOpen
  \bibfield  {author} {\bibinfo {author} {\bibnamefont {Delyon}, \bibfnamefont
  {F.}}, \bibinfo {author} {\bibnamefont {Simon}, \bibfnamefont {B.}}, and\
  \bibinfo {author} {\bibnamefont {Souillard}, \bibfnamefont {B.}},\ }\bibfield
   {title} {{\selectlanguage {English}\enquote {\bibinfo {title} {From power
  pure point to continuous spectrum in disordered systems},}\ }}\href@noop {}
  {\bibfield  {journal} {\bibinfo  {journal} {Ann. Inst. Henri Poincar{\'e},
  Phys. Th{\'e}or.}\ }\textbf {\bibinfo {volume} {42}},\ \bibinfo {pages}
  {283--309} (\bibinfo {year} {1985})}\BibitemShut {NoStop}%
\bibitem [{\citenamefont {Ding}\ and\ \citenamefont {Smart}(2020)}]{DS20}%
  \BibitemOpen
  \bibfield  {author} {\bibinfo {author} {\bibnamefont {Ding}, \bibfnamefont
  {J.}}and\ \bibinfo {author} {\bibnamefont {Smart}, \bibfnamefont {C.~K.}},\
  }\bibfield  {title} {{\selectlanguage {English}\enquote {\bibinfo {title}
  {Localization near the edge for the {Anderson} {Bernoulli} model on the two
  dimensional lattice},}\ }}\href {https://doi.org/10.1007/s00222-019-00910-4}
  {\bibfield  {journal} {\bibinfo  {journal} {Invent. Math.}\ }\textbf
  {\bibinfo {volume} {219}},\ \bibinfo {pages} {467--506} (\bibinfo {year}
  {2020})}\BibitemShut {NoStop}%
\bibitem [{\citenamefont {{Disertori}}\ \emph {et~al.}(2008)\citenamefont
  {{Disertori}}, \citenamefont {{Kirsch}}, \citenamefont {{Klein}},
  \citenamefont {{Klopp}},\ and\ \citenamefont {{Rivasseau}}}]{Ki08}%
  \BibitemOpen
  \bibfield  {author} {\bibinfo {author} {\bibnamefont {{Disertori}},
  \bibfnamefont {M.}}, \bibinfo {author} {\bibnamefont {{Kirsch}},
  \bibfnamefont {W.}}, \bibinfo {author} {\bibnamefont {{Klein}}, \bibfnamefont
  {A.}}, \bibinfo {author} {\bibnamefont {{Klopp}}, \bibfnamefont {F.}}, and\
  \bibinfo {author} {\bibnamefont {{Rivasseau}}, \bibfnamefont {V.}},\
  }\href@noop {} {{\selectlanguage {English}\emph {\bibinfo {title} {{Random
  Schr\"odinger operators}}}}},\ Vol.~\bibinfo {volume} {25}\ (\bibinfo
  {publisher} {Paris: Soci\'et\'e Math\'ematique de France (SMF)},\ \bibinfo
  {year} {2008})\ pp.\ \bibinfo {pages} {xiv + 213}\BibitemShut {NoStop}%
\bibitem [{\citenamefont {Duarte}\ and\ \citenamefont {Klein}(2020)}]{DuSi20}%
  \BibitemOpen
  \bibfield  {author} {\bibinfo {author} {\bibnamefont {Duarte}, \bibfnamefont
  {P.}}and\ \bibinfo {author} {\bibnamefont {Klein}, \bibfnamefont {S.}},\
  }\bibfield  {title} {{\selectlanguage {English}\enquote {\bibinfo {title}
  {Large deviations for products of random two dimensional matrices},}\ }}\href
  {https://doi.org/10.1007/s00220-019-03586-2} {\bibfield  {journal} {\bibinfo
  {journal} {Commun. Math. Phys.}\ }\textbf {\bibinfo {volume} {375}},\
  \bibinfo {pages} {2191--2257} (\bibinfo {year} {2020})}\BibitemShut {NoStop}%
\bibitem [{\citenamefont {{Furstenberg}}(1963)}]{F63}%
  \BibitemOpen
  \bibfield  {author} {\bibinfo {author} {\bibnamefont {{Furstenberg}},
  \bibfnamefont {H.}},\ }\bibfield  {title} {{\selectlanguage {English}\enquote
  {\bibinfo {title} {{Noncommuting random products}},}\ }}\href
  {https://doi.org/10.2307/1993589} {\bibfield  {journal} {\bibinfo  {journal}
  {{Trans. Am. Math. Soc.}}\ }\textbf {\bibinfo {volume} {108}},\ \bibinfo
  {pages} {377--428} (\bibinfo {year} {1963})}\BibitemShut {NoStop}%
\bibitem [{\citenamefont {{Furstenberg}}\ and\ \citenamefont
  {{Kesten}}(1960)}]{FK60}%
  \BibitemOpen
  \bibfield  {author} {\bibinfo {author} {\bibnamefont {{Furstenberg}},
  \bibfnamefont {H.}}and\ \bibinfo {author} {\bibnamefont {{Kesten}},
  \bibfnamefont {H.}},\ }\bibfield  {title} {{\selectlanguage {English}\enquote
  {\bibinfo {title} {{Products of random matrices}},}\ }}\href
  {https://doi.org/10.1214/aoms/1177705909} {\bibfield  {journal} {\bibinfo
  {journal} {{Ann. Math. Stat.}}\ }\textbf {\bibinfo {volume} {31}},\ \bibinfo
  {pages} {457--469} (\bibinfo {year} {1960})}\BibitemShut {NoStop}%
\bibitem [{\citenamefont {Ge}\ and\ \citenamefont {Zhao}(2020)}]{LX20}%
  \BibitemOpen
  \bibfield  {author} {\bibinfo {author} {\bibnamefont {Ge}, \bibfnamefont
  {L.}}and\ \bibinfo {author} {\bibnamefont {Zhao}, \bibfnamefont {X.}},\
  }\bibfield  {title} {{\selectlanguage {English}\enquote {\bibinfo {title}
  {Exponential dynamical localization in expectation for the one dimensional
  {Anderson} model},}\ }}\href {https://doi.org/10.4171/JST/315} {\bibfield
  {journal} {\bibinfo  {journal} {J. Spectr. Theory}\ }\textbf {\bibinfo
  {volume} {10}},\ \bibinfo {pages} {887--904} (\bibinfo {year}
  {2020})}\BibitemShut {NoStop}%
\bibitem [{\citenamefont {Genovese}, \citenamefont {Giacomin},\ and\
  \citenamefont {Greenblatt}(2017)}]{GGG17}%
  \BibitemOpen
  \bibfield  {author} {\bibinfo {author} {\bibnamefont {Genovese},
  \bibfnamefont {G.}}, \bibinfo {author} {\bibnamefont {Giacomin},
  \bibfnamefont {G.}}, and\ \bibinfo {author} {\bibnamefont {Greenblatt},
  \bibfnamefont {R.~L.}},\ }\bibfield  {title} {{\selectlanguage
  {English}\enquote {\bibinfo {title} {Singular behavior of the leading
  {Lyapunov} exponent of a product of random {{\({2 \times 2}\)}} matrices},}\
  }}\href {https://doi.org/10.1007/s00220-017-2855-4} {\bibfield  {journal}
  {\bibinfo  {journal} {Commun. Math. Phys.}\ }\textbf {\bibinfo {volume}
  {351}},\ \bibinfo {pages} {923--958} (\bibinfo {year} {2017})}\BibitemShut
  {NoStop}%
\bibitem [{\citenamefont {Germinet}\ and\ \citenamefont {Klein}(2001)}]{GK01}%
  \BibitemOpen
  \bibfield  {author} {\bibinfo {author} {\bibnamefont {Germinet},
  \bibfnamefont {F.}}and\ \bibinfo {author} {\bibnamefont {Klein},
  \bibfnamefont {A.}},\ }\bibfield  {title} {{\selectlanguage {English}\enquote
  {\bibinfo {title} {Bootstrap multiscale analysis and localization in random
  media},}\ }}\href {https://doi.org/10.1007/s002200100518} {\bibfield
  {journal} {\bibinfo  {journal} {Commun. Math. Phys.}\ }\textbf {\bibinfo
  {volume} {222}},\ \bibinfo {pages} {415--448} (\bibinfo {year}
  {2001})}\BibitemShut {NoStop}%
\bibitem [{\citenamefont {Germinet}, \citenamefont {Klein},\ and\ \citenamefont
  {Schenker}(2007)}]{GKS07}%
  \BibitemOpen
  \bibfield  {author} {\bibinfo {author} {\bibnamefont {Germinet},
  \bibfnamefont {F.}}, \bibinfo {author} {\bibnamefont {Klein}, \bibfnamefont
  {A.}}, and\ \bibinfo {author} {\bibnamefont {Schenker}, \bibfnamefont
  {J.~H.}},\ }\bibfield  {title} {{\selectlanguage {English}\enquote {\bibinfo
  {title} {Dynamical delocalization in random {Landau} {Hamiltonians}},}\
  }}\href {https://doi.org/10.4007/annals.2007.166.215} {\bibfield  {journal}
  {\bibinfo  {journal} {Ann. Math. (2)}\ }\textbf {\bibinfo {volume} {166}},\
  \bibinfo {pages} {215--244} (\bibinfo {year} {2007})}\BibitemShut {NoStop}%
\bibitem [{\citenamefont {Girvin}\ and\ \citenamefont {Yang}(2019)}]{GY19}%
  \BibitemOpen
  \bibfield  {author} {\bibinfo {author} {\bibnamefont {Girvin}, \bibfnamefont
  {S.}}and\ \bibinfo {author} {\bibnamefont {Yang}, \bibfnamefont {K.}},\
  }\href {https://books.google.fr/books?id=2ESIDwAAQBAJ} {\emph {\bibinfo
  {title} {Modern Condensed Matter Physics}}}\ (\bibinfo  {publisher}
  {Cambridge University Press},\ \bibinfo {year} {2019})\BibitemShut {NoStop}%
\bibitem [{\citenamefont {{Glaffig}}(1990)}]{Gl90}%
  \BibitemOpen
  \bibfield  {author} {\bibinfo {author} {\bibnamefont {{Glaffig}},
  \bibfnamefont {C.}},\ }\bibfield  {title} {{\selectlanguage {English}\enquote
  {\bibinfo {title} {{Smoothness of the integrated density of states on
  strips}},}\ }}\href {https://doi.org/10.1016/0022-1236(90)90061-O} {\bibfield
   {journal} {\bibinfo  {journal} {{J. Funct. Anal.}}\ }\textbf {\bibinfo
  {volume} {92}},\ \bibinfo {pages} {509--534} (\bibinfo {year}
  {1990})}\BibitemShut {NoStop}%
\bibitem [{\citenamefont {{Goldsheid}}(1995)}]{Go95}%
  \BibitemOpen
  \bibfield  {author} {\bibinfo {author} {\bibnamefont {{Goldsheid}},
  \bibfnamefont {I.~Y.}},\ }\bibfield  {title} {{\selectlanguage
  {English}\enquote {\bibinfo {title} {{Zariski closure of subgroups of the
  symplectic group and Lyapunov exponents of the Schr\"odinger operator on the
  strip}},}\ }}\href {https://doi.org/10.1007/BF02099606} {\bibfield  {journal}
  {\bibinfo  {journal} {{Commun. Math. Phys.}}\ }\textbf {\bibinfo {volume}
  {174}},\ \bibinfo {pages} {347--365} (\bibinfo {year} {1995})}\BibitemShut
  {NoStop}%
\bibitem [{\citenamefont {{Goldsheid}}\ and\ \citenamefont
  {{Margulis}}(1989)}]{GM89}%
  \BibitemOpen
  \bibfield  {author} {\bibinfo {author} {\bibnamefont {{Goldsheid}},
  \bibfnamefont {I.~Y.}}and\ \bibinfo {author} {\bibnamefont {{Margulis}},
  \bibfnamefont {G.~A.}},\ }\bibfield  {title} {{\selectlanguage
  {English}\enquote {\bibinfo {title} {{Lyapunov indices of a product of random
  matrices}},}\ }}\href {https://doi.org/10.1070/RM1989v044n05ABEH002214}
  {\bibfield  {journal} {\bibinfo  {journal} {{Russ. Math. Surv.}}\ }\textbf
  {\bibinfo {volume} {44}},\ \bibinfo {pages} {11--71} (\bibinfo {year}
  {1989})}\BibitemShut {NoStop}%
\bibitem [{\citenamefont {Gorodetski}\ and\ \citenamefont
  {Kleptsyn}(2021)}]{GoKl21}%
  \BibitemOpen
  \bibfield  {author} {\bibinfo {author} {\bibnamefont {Gorodetski},
  \bibfnamefont {A.}}and\ \bibinfo {author} {\bibnamefont {Kleptsyn},
  \bibfnamefont {V.}},\ }\bibfield  {title} {{\selectlanguage {English}\enquote
  {\bibinfo {title} {Parametric {Furstenberg} theorem on random products of
  {{\(\mathrm{SL}(2, \mathbb{R})\)}} matrices},}\ }}\href
  {https://doi.org/10.1016/j.aim.2020.107522} {\bibfield  {journal} {\bibinfo
  {journal} {Adv. Math.}\ }\textbf {\bibinfo {volume} {378}},\ \bibinfo {pages}
  {82} (\bibinfo {year} {2021})},\ \bibinfo {note} {id/No 107522}\BibitemShut
  {NoStop}%
\bibitem [{\citenamefont {{Guivarc'h}}\ and\ \citenamefont
  {{Raugi}}(1985)}]{GR85}%
  \BibitemOpen
  \bibfield  {author} {\bibinfo {author} {\bibnamefont {{Guivarc'h}},
  \bibfnamefont {Y.}}and\ \bibinfo {author} {\bibnamefont {{Raugi}},
  \bibfnamefont {A.}},\ }\bibfield  {title} {{\selectlanguage {French}\enquote
  {\bibinfo {title} {{Fronti\`ere de Furstenberg, propri\'et\'es de contraction
  et th\'eor\`emes de convergence}},}\ }}\href
  {https://doi.org/10.1007/BF02450281} {\bibfield  {journal} {\bibinfo
  {journal} {{Z. Wahrscheinlichkeitstheor. Verw. Geb.}}\ }\textbf {\bibinfo
  {volume} {69}},\ \bibinfo {pages} {187--242} (\bibinfo {year}
  {1985})}\BibitemShut {NoStop}%
\bibitem [{\citenamefont {Hamza}, \citenamefont {Joye},\ and\ \citenamefont
  {Stolz}(2009)}]{HJS09}%
  \BibitemOpen
  \bibfield  {author} {\bibinfo {author} {\bibnamefont {Hamza}, \bibfnamefont
  {E.}}, \bibinfo {author} {\bibnamefont {Joye}, \bibfnamefont {A.}}, and\
  \bibinfo {author} {\bibnamefont {Stolz}, \bibfnamefont {G.}},\ }\bibfield
  {title} {{\selectlanguage {English}\enquote {\bibinfo {title} {Dynamical
  localization for unitary {Anderson} models},}\ }}\href
  {https://doi.org/10.1007/s11040-009-9068-9} {\bibfield  {journal} {\bibinfo
  {journal} {Math. Phys. Anal. Geom.}\ }\textbf {\bibinfo {volume} {12}},\
  \bibinfo {pages} {381--444} (\bibinfo {year} {2009})}\BibitemShut {NoStop}%
\bibitem [{\citenamefont {Hamza}\ and\ \citenamefont {Stolz}(2007)}]{HS07}%
  \BibitemOpen
  \bibfield  {author} {\bibinfo {author} {\bibnamefont {Hamza}, \bibfnamefont
  {E.}}and\ \bibinfo {author} {\bibnamefont {Stolz}, \bibfnamefont {G.}},\
  }\bibfield  {title} {{\selectlanguage {English}\enquote {\bibinfo {title}
  {Lyapunov exponents for unitary {Anderson} models},}\ }}\href
  {https://doi.org/10.1063/1.2713996} {\bibfield  {journal} {\bibinfo
  {journal} {J. Math. Phys.}\ }\textbf {\bibinfo {volume} {48}},\ \bibinfo
  {pages} {043301, 16} (\bibinfo {year} {2007})}\BibitemShut {NoStop}%
\bibitem [{\citenamefont {Hislop}(2008)}]{His08}%
  \BibitemOpen
  \bibfield  {author} {\bibinfo {author} {\bibnamefont {Hislop}, \bibfnamefont
  {P.~D.}},\ }\bibfield  {title} {{\selectlanguage {English}\enquote {\bibinfo
  {title} {Lectures on random {Schr{\"o}dinger} operators},}\ }}in\ \href@noop
  {} {{\selectlanguage {English}\emph {\bibinfo {booktitle} {Fourth summer
  school in analysis and mathematical physics. Topics in spectral theory and
  quantum mechanics, Cuernavaca, M\'exico, May 2005}}}}\ (\bibinfo  {publisher}
  {Providence, RI: American Mathematical Society (AMS)},\ \bibinfo {year}
  {2008})\ pp.\ \bibinfo {pages} {41--131}\BibitemShut {NoStop}%
\bibitem [{\citenamefont {Hislop}, \citenamefont {Kirsch},\ and\ \citenamefont
  {Krishna}(2005)}]{HKK05}%
  \BibitemOpen
  \bibfield  {author} {\bibinfo {author} {\bibnamefont {Hislop}, \bibfnamefont
  {P.~D.}}, \bibinfo {author} {\bibnamefont {Kirsch}, \bibfnamefont {W.}}, and\
  \bibinfo {author} {\bibnamefont {Krishna}, \bibfnamefont {M.}},\ }\bibfield
  {title} {{\selectlanguage {English}\enquote {\bibinfo {title} {Spectral and
  dynamical properties of random models with nonlocal and singular
  interactions},}\ }}\href {https://doi.org/10.1002/mana.200310261} {\bibfield
  {journal} {\bibinfo  {journal} {Math. Nachr.}\ }\textbf {\bibinfo {volume}
  {278}},\ \bibinfo {pages} {627--664} (\bibinfo {year} {2005})}\BibitemShut
  {NoStop}%
\bibitem [{\citenamefont {Hislop}, \citenamefont {Kirsch},\ and\ \citenamefont
  {Krishna}(2020)}]{HKK20}%
  \BibitemOpen
  \bibfield  {author} {\bibinfo {author} {\bibnamefont {Hislop}, \bibfnamefont
  {P.~D.}}, \bibinfo {author} {\bibnamefont {Kirsch}, \bibfnamefont {W.}}, and\
  \bibinfo {author} {\bibnamefont {Krishna}, \bibfnamefont {M.}},\ }\bibfield
  {title} {{\selectlanguage {English}\enquote {\bibinfo {title} {Eigenvalue
  statistics for {Schr{\"o}dinger} operators with random point interactions on
  {{\(\mathbb{R}^d, d = 1, 2, 3\)}}},}\ }}\href
  {https://doi.org/10.1063/5.0002885} {\bibfield  {journal} {\bibinfo
  {journal} {J. Math. Phys.}\ }\textbf {\bibinfo {volume} {61}},\ \bibinfo
  {pages} {092103, 24} (\bibinfo {year} {2020})}\BibitemShut {NoStop}%
\bibitem [{\citenamefont {Hurt}(2000)}]{H00}%
  \BibitemOpen
  \bibfield  {author} {\bibinfo {author} {\bibnamefont {Hurt}, \bibfnamefont
  {N.~E.}},\ }\href@noop {} {{\selectlanguage {English}\emph {\bibinfo {title}
  {Mathematical physics of quantum wires and devices. {From} spectral
  resonances to {Anderson} localization}}}},\ \bibinfo {series} {Math. Appl.,
  Dordr.}, Vol.\ \bibinfo {volume} {506}\ (\bibinfo  {publisher} {Dordrecht:
  Kluwer Academic Publishers},\ \bibinfo {year} {2000})\BibitemShut {NoStop}%
\bibitem [{\citenamefont {Jitomirskaya}\ and\ \citenamefont
  {Zhu}(2019)}]{JZ19}%
  \BibitemOpen
  \bibfield  {author} {\bibinfo {author} {\bibnamefont {Jitomirskaya},
  \bibfnamefont {S.}}and\ \bibinfo {author} {\bibnamefont {Zhu}, \bibfnamefont
  {X.}},\ }\bibfield  {title} {{\selectlanguage {English}\enquote {\bibinfo
  {title} {Large deviations of the {Lyapunov} exponent and localization for the
  1d {Anderson} model},}\ }}\href {https://doi.org/10.1007/s00220-019-03502-8}
  {\bibfield  {journal} {\bibinfo  {journal} {Commun. Math. Phys.}\ }\textbf
  {\bibinfo {volume} {370}},\ \bibinfo {pages} {311--324} (\bibinfo {year}
  {2019})}\BibitemShut {NoStop}%
\bibitem [{\citenamefont {{Kingman}}(1973)}]{Kg73}%
  \BibitemOpen
  \bibfield  {author} {\bibinfo {author} {\bibnamefont {{Kingman}},
  \bibfnamefont {J.~F.~C.}},\ }\bibfield  {title} {{\selectlanguage
  {English}\enquote {\bibinfo {title} {{Subadditive ergodic theory}},}\ }}\href
  {https://doi.org/10.1214/aop/1176996798} {\bibfield  {journal} {\bibinfo
  {journal} {{Ann. Probab.}}\ }\textbf {\bibinfo {volume} {1}},\ \bibinfo
  {pages} {883--909} (\bibinfo {year} {1973})}\BibitemShut {NoStop}%
\bibitem [{\citenamefont {Kirsch}\ and\ \citenamefont
  {Martinelli}(1982)}]{KM82}%
  \BibitemOpen
  \bibfield  {author} {\bibinfo {author} {\bibnamefont {Kirsch}, \bibfnamefont
  {W.}}and\ \bibinfo {author} {\bibnamefont {Martinelli}, \bibfnamefont {F.}},\
  }\bibfield  {title} {{\selectlanguage {English}\enquote {\bibinfo {title} {On
  the ergodic properties of the spectrum of general random operators},}\
  }}\href@noop {} {\bibfield  {journal} {\bibinfo  {journal} {J. Reine Angew.
  Math.}\ }\textbf {\bibinfo {volume} {334}},\ \bibinfo {pages} {141--156}
  (\bibinfo {year} {1982})}\BibitemShut {NoStop}%
\bibitem [{\citenamefont {Klein}(2008)}]{Kl08}%
  \BibitemOpen
  \bibfield  {author} {\bibinfo {author} {\bibnamefont {Klein}, \bibfnamefont
  {A.}},\ }\bibfield  {title} {{\selectlanguage {English}\enquote {\bibinfo
  {title} {Multiscale analysis and localization of random operators},}\ }}in\
  \href@noop {} {{\selectlanguage {English}\emph {\bibinfo {booktitle} {Random
  Schr\"odinger operators}}}}\ (\bibinfo  {publisher} {Paris: Soci{\'e}t{\'e}
  Math{\'e}matique de France (SMF)},\ \bibinfo {year} {2008})\ pp.\ \bibinfo
  {pages} {221--259}\BibitemShut {NoStop}%
\bibitem [{\citenamefont {{Klein}}, \citenamefont {{Lacroix}},\ and\
  \citenamefont {{Speis}}(1990)}]{KLS90}%
  \BibitemOpen
  \bibfield  {author} {\bibinfo {author} {\bibnamefont {{Klein}}, \bibfnamefont
  {A.}}, \bibinfo {author} {\bibnamefont {{Lacroix}}, \bibfnamefont {J.}}, and\
  \bibinfo {author} {\bibnamefont {{Speis}}, \bibfnamefont {A.}},\ }\bibfield
  {title} {{\selectlanguage {English}\enquote {\bibinfo {title} {{Localization
  for the Anderson model on a strip with singular potentials}},}\ }}\href
  {https://doi.org/10.1016/0022-1236(90)90031-F} {\bibfield  {journal}
  {\bibinfo  {journal} {{J. Funct. Anal.}}\ }\textbf {\bibinfo {volume} {94}},\
  \bibinfo {pages} {135--155} (\bibinfo {year} {1990})}\BibitemShut {NoStop}%
\bibitem [{\citenamefont {Kotani}(1984)}]{Ko84}%
  \BibitemOpen
  \bibfield  {author} {\bibinfo {author} {\bibnamefont {Kotani}, \bibfnamefont
  {S.}},\ }\href@noop {} {{\selectlanguage {English}\enquote {\bibinfo {title}
  {Ljapunov indices determine absolutely continuous spectra of stationary
  random one-dimensional {Schr{\"o}dinger} operators},}\ }}\bibinfo
  {howpublished} {Stochastic analysis, {Proc}. {Taniguchi} {Int}. {Symp}.,
  {Katata} \& {Kyoto}/{Jap}. 1982, {North}-{Holland} {Math}. {Libr}. 32,
  225-247 (1984).} (\bibinfo {year} {1984})\BibitemShut {NoStop}%
\bibitem [{\citenamefont {{Kotani}}\ and\ \citenamefont
  {{Simon}}(1988)}]{KS88}%
  \BibitemOpen
  \bibfield  {author} {\bibinfo {author} {\bibnamefont {{Kotani}},
  \bibfnamefont {S.}}and\ \bibinfo {author} {\bibnamefont {{Simon}},
  \bibfnamefont {B.}},\ }\bibfield  {title} {{\selectlanguage {English}\enquote
  {\bibinfo {title} {{Stochastic Schr\"odinger operators and Jacobi matrices on
  the strip}},}\ }}\href {https://doi.org/10.1007/BF01218080} {\bibfield
  {journal} {\bibinfo  {journal} {{Commun. Math. Phys.}}\ }\textbf {\bibinfo
  {volume} {119}},\ \bibinfo {pages} {403--429} (\bibinfo {year}
  {1988})}\BibitemShut {NoStop}%
\bibitem [{\citenamefont {Kunz}\ and\ \citenamefont {Souillard}(1980)}]{KS80}%
  \BibitemOpen
  \bibfield  {author} {\bibinfo {author} {\bibnamefont {Kunz}, \bibfnamefont
  {H.}}and\ \bibinfo {author} {\bibnamefont {Souillard}, \bibfnamefont {B.}},\
  }\bibfield  {title} {{\selectlanguage {French}\enquote {\bibinfo {title} {Sur
  le spectre des op{\'e}rateurs aux diff{\'e}rences finies al{\'e}atoires},}\
  }}\href {https://doi.org/10.1007/BF01942371} {\bibfield  {journal} {\bibinfo
  {journal} {Commun. Math. Phys.}\ }\textbf {\bibinfo {volume} {78}},\ \bibinfo
  {pages} {201--246} (\bibinfo {year} {1980})}\BibitemShut {NoStop}%
\bibitem [{\citenamefont {Li}\ and\ \citenamefont {Zhang}(2022)}]{LZ22}%
  \BibitemOpen
  \bibfield  {author} {\bibinfo {author} {\bibnamefont {Li}, \bibfnamefont
  {L.}}and\ \bibinfo {author} {\bibnamefont {Zhang}, \bibfnamefont {L.}},\
  }\bibfield  {title} {{\selectlanguage {English}\enquote {\bibinfo {title}
  {Anderson-{Bernoulli} localization on the three-dimensional lattice and
  discrete unique continuation principle},}\ }}\href
  {https://doi.org/10.1215/00127094-2021-0038} {\bibfield  {journal} {\bibinfo
  {journal} {Duke Math. J.}\ }\textbf {\bibinfo {volume} {171}},\ \bibinfo
  {pages} {327--415} (\bibinfo {year} {2022})}\BibitemShut {NoStop}%
\bibitem [{\citenamefont {Macera}\ and\ \citenamefont {Sodin}(2022)}]{MS21}%
  \BibitemOpen
  \bibfield  {author} {\bibinfo {author} {\bibnamefont {Macera}, \bibfnamefont
  {D.}}and\ \bibinfo {author} {\bibnamefont {Sodin}, \bibfnamefont {S.}},\
  }\bibfield  {title} {{\selectlanguage {English}\enquote {\bibinfo {title}
  {Anderson localisation for quasi-one-dimensional random operators},}\ }}\href
  {https://doi.org/10.1007/s00023-022-01191-z} {\bibfield  {journal} {\bibinfo
  {journal} {Ann. Henri Poincar{\'e}}\ }\textbf {\bibinfo {volume} {23}},\
  \bibinfo {pages} {4227--4247} (\bibinfo {year} {2022})}\BibitemShut {NoStop}%
\bibitem [{\citenamefont {Marin}\ and\ \citenamefont
  {Schulz-Baldes}(2013)}]{MSB13}%
  \BibitemOpen
  \bibfield  {author} {\bibinfo {author} {\bibnamefont {Marin}, \bibfnamefont
  {L.}}and\ \bibinfo {author} {\bibnamefont {Schulz-Baldes}, \bibfnamefont
  {H.}},\ }\bibfield  {title} {{\selectlanguage {English}\enquote {\bibinfo
  {title} {Scattering zippers and their spectral theory},}\ }}\href
  {https://doi.org/10.4171/JST/37} {\bibfield  {journal} {\bibinfo  {journal}
  {J. Spectr. Theory}\ }\textbf {\bibinfo {volume} {3}},\ \bibinfo {pages}
  {47--82} (\bibinfo {year} {2013})}\BibitemShut {NoStop}%
\bibitem [{\citenamefont {{Oseledets}}(1968)}]{Os68}%
  \BibitemOpen
  \bibfield  {author} {\bibinfo {author} {\bibnamefont {{Oseledets}},
  \bibfnamefont {V.~I.}},\ }\bibfield  {title} {{\selectlanguage
  {English}\enquote {\bibinfo {title} {{A multiplicative ergodic theorem.
  Lyapunov characteristic numbers for dynamical systems}},}\ }}\href@noop {}
  {\bibfield  {journal} {\bibinfo  {journal} {{Trans. Mosc. Math. Soc.}}\
  }\textbf {\bibinfo {volume} {19}},\ \bibinfo {pages} {197--231} (\bibinfo
  {year} {1968})}\BibitemShut {NoStop}%
\bibitem [{\citenamefont {Pastur}(1980)}]{P80}%
  \BibitemOpen
  \bibfield  {author} {\bibinfo {author} {\bibnamefont {Pastur}, \bibfnamefont
  {L.~A.}},\ }\bibfield  {title} {{\selectlanguage {English}\enquote {\bibinfo
  {title} {Spectral properties of disordered systems in the one-body
  approximation},}\ }}\href {https://doi.org/10.1007/BF01222516} {\bibfield
  {journal} {\bibinfo  {journal} {Commun. Math. Phys.}\ }\textbf {\bibinfo
  {volume} {75}},\ \bibinfo {pages} {179--196} (\bibinfo {year}
  {1980})}\BibitemShut {NoStop}%
\bibitem [{\citenamefont {Rangamani}(2019)}]{R19}%
  \BibitemOpen
  \bibfield  {author} {\bibinfo {author} {\bibnamefont {Rangamani},
  \bibfnamefont {N.}},\ }\bibfield  {title} {{\selectlanguage {English}\enquote
  {\bibinfo {title} {Singular-unbounded random {Jacobi} matrices},}\ }}\href
  {https://doi.org/10.1063/1.5085027} {\bibfield  {journal} {\bibinfo
  {journal} {J. Math. Phys.}\ }\textbf {\bibinfo {volume} {60}},\ \bibinfo
  {pages} {081904, 11} (\bibinfo {year} {2019})}\BibitemShut {NoStop}%
\bibitem [{\citenamefont {Rangamani}(2022)}]{R22}%
  \BibitemOpen
  \bibfield  {author} {\bibinfo {author} {\bibnamefont {Rangamani},
  \bibfnamefont {N.}},\ }\bibfield  {title} {{\selectlanguage {English}\enquote
  {\bibinfo {title} {Exponential dynamical localization for random word
  models},}\ }}\href {https://doi.org/10.1007/s00023-022-01190-0} {\bibfield
  {journal} {\bibinfo  {journal} {Ann. Henri Poincar{\'e}}\ }\textbf {\bibinfo
  {volume} {23}},\ \bibinfo {pages} {4171--4193} (\bibinfo {year}
  {2022})}\BibitemShut {NoStop}%
\bibitem [{\citenamefont {{Reed}}\ and\ \citenamefont {{Simon}}(1978)}]{RS4}%
  \BibitemOpen
  \bibfield  {author} {\bibinfo {author} {\bibnamefont {{Reed}}, \bibfnamefont
  {M.}}and\ \bibinfo {author} {\bibnamefont {{Simon}}, \bibfnamefont {B.}},\
  }\href@noop {} {{\selectlanguage {English}\enquote {\bibinfo {title}
  {{Methods of modern mathematical physics. IV: Analysis of operators}},}\
  }}\bibinfo {howpublished} {{New York - San Francisco - London: Academic
  Press. XV, 396 p. (1978).}} (\bibinfo {year} {1978})\BibitemShut {NoStop}%
\bibitem [{\citenamefont {{Reed}}\ and\ \citenamefont {{Simon}}(1979)}]{RS3}%
  \BibitemOpen
  \bibfield  {author} {\bibinfo {author} {\bibnamefont {{Reed}}, \bibfnamefont
  {M.}}and\ \bibinfo {author} {\bibnamefont {{Simon}}, \bibfnamefont {B.}},\
  }\href@noop {} {{\selectlanguage {English}\enquote {\bibinfo {title}
  {{Methods of modern mathematical physics. III: Scattering theory}},}\
  }}\bibinfo {howpublished} {{New York, San Francisco, London: Academic Press.
  XV, 463 p. (1979).}} (\bibinfo {year} {1979})\BibitemShut {NoStop}%
\bibitem [{\citenamefont {Ruelle}(1979)}]{R79}%
  \BibitemOpen
  \bibfield  {author} {\bibinfo {author} {\bibnamefont {Ruelle}, \bibfnamefont
  {D.}},\ }\bibfield  {title} {{\selectlanguage {English}\enquote {\bibinfo
  {title} {Ergodic theory of differentiable dynamical systems},}\ }}\href
  {https://doi.org/10.1007/BF02684768} {\bibfield  {journal} {\bibinfo
  {journal} {Publ. Math., Inst. Hautes {\'E}tud. Sci.}\ }\textbf {\bibinfo
  {volume} {50}},\ \bibinfo {pages} {27--58} (\bibinfo {year}
  {1979})}\BibitemShut {NoStop}%
\bibitem [{\citenamefont {Sadel}\ and\ \citenamefont
  {Schulz-Baldes}(2010)}]{SSB09}%
  \BibitemOpen
  \bibfield  {author} {\bibinfo {author} {\bibnamefont {Sadel}, \bibfnamefont
  {C.}}and\ \bibinfo {author} {\bibnamefont {Schulz-Baldes}, \bibfnamefont
  {H.}},\ }\bibfield  {title} {{\selectlanguage {English}\enquote {\bibinfo
  {title} {Random {Dirac} operators with time reversal symmetry},}\ }}\href
  {https://doi.org/10.1007/s00220-009-0956-4} {\bibfield  {journal} {\bibinfo
  {journal} {Commun. Math. Phys.}\ }\textbf {\bibinfo {volume} {295}},\
  \bibinfo {pages} {209--242} (\bibinfo {year} {2010})}\BibitemShut {NoStop}%
\bibitem [{\citenamefont {Shubin}, \citenamefont {Vakilian},\ and\
  \citenamefont {Wolff}(1998)}]{SVW98}%
  \BibitemOpen
  \bibfield  {author} {\bibinfo {author} {\bibnamefont {Shubin}, \bibfnamefont
  {C.}}, \bibinfo {author} {\bibnamefont {Vakilian}, \bibfnamefont {R.}}, and\
  \bibinfo {author} {\bibnamefont {Wolff}, \bibfnamefont {T.}},\ }\bibfield
  {title} {{\selectlanguage {English}\enquote {\bibinfo {title} {Some harmonic
  analysis questions suggested by {Anderson}-{Bernoulli} models. {Appendix} by
  {T}.{H}.{Wolff}},}\ }}\href {https://doi.org/10.1007/s000390050078}
  {\bibfield  {journal} {\bibinfo  {journal} {Geom. Funct. Anal.}\ }\textbf
  {\bibinfo {volume} {8}},\ \bibinfo {pages} {932--964} (\bibinfo {year}
  {1998})}\BibitemShut {NoStop}%
\bibitem [{\citenamefont {Simon}(2005{\natexlab{a}})}]{Sim05_1}%
  \BibitemOpen
  \bibfield  {author} {\bibinfo {author} {\bibnamefont {Simon}, \bibfnamefont
  {B.}},\ }\href {www.ams.org/books/coll/054.1/} {{\selectlanguage
  {English}\emph {\bibinfo {title} {Orthogonal polynomials on the unit circle.
  {Part} 1: {Classical} theory}}}},\ \bibinfo {series} {Colloq. Publ., Am.
  Math. Soc.}, Vol.~\bibinfo {volume} {54}\ (\bibinfo  {publisher} {Providence,
  RI: American Mathematical Society (AMS)},\ \bibinfo {year}
  {2005})\BibitemShut {NoStop}%
\bibitem [{\citenamefont {Simon}(2005{\natexlab{b}})}]{Sim05_2}%
  \BibitemOpen
  \bibfield  {author} {\bibinfo {author} {\bibnamefont {Simon}, \bibfnamefont
  {B.}},\ }\href {www.ams.org/books/coll/054.2/} {{\selectlanguage
  {English}\emph {\bibinfo {title} {Orthogonal polynomials on the unit circle.
  {Part} 2: {Spectral} theory}}}},\ \bibinfo {series} {Colloq. Publ., Am. Math.
  Soc.}, Vol.~\bibinfo {volume} {54}\ (\bibinfo  {publisher} {Providence, RI:
  American Mathematical Society},\ \bibinfo {year} {2005})\BibitemShut
  {NoStop}%
\bibitem [{\citenamefont {{Stollmann}}(2001)}]{S01}%
  \BibitemOpen
  \bibfield  {author} {\bibinfo {author} {\bibnamefont {{Stollmann}},
  \bibfnamefont {P.}},\ }\href@noop {} {{\selectlanguage {English}\emph
  {\bibinfo {title} {{Caught by disorder. Bound states in random media}}}}},\
  Vol.~\bibinfo {volume} {20}\ (\bibinfo  {publisher} {Boston: Birkh\"auser},\
  \bibinfo {year} {2001})\ pp.\ \bibinfo {pages} {xvi + 166}\BibitemShut
  {NoStop}%
\bibitem [{\citenamefont {Veseli{\'c}}(2008)}]{Ves08}%
  \BibitemOpen
  \bibfield  {author} {\bibinfo {author} {\bibnamefont {Veseli{\'c}},
  \bibfnamefont {I.}},\ }\href {https://doi.org/10.1007/978-3-540-72691-3}
  {{\selectlanguage {English}\emph {\bibinfo {title} {Existence and regularity
  properties of the integrated density of states of random {Schr{\"o}dinger}
  operators}}}},\ \bibinfo {series} {Lect. Notes Math.}, Vol.\ \bibinfo
  {volume} {1917}\ (\bibinfo  {publisher} {Berlin: Springer},\ \bibinfo {year}
  {2008})\BibitemShut {NoStop}%
\bibitem [{\citenamefont {Werner}(2013)}]{W13}%
  \BibitemOpen
  \bibfield  {author} {\bibinfo {author} {\bibnamefont {Werner}, \bibfnamefont
  {A.~H.}},\ }\emph {\bibinfo {title} {Localization and Recurrence in Quantum
  Walks}},\ \href@noop {} {Ph.D. thesis},\ \bibinfo  {school} {Gottfried
  Wilhelm Leibniz Universit\"at Hannover} (\bibinfo {year} {2013})\BibitemShut
  {NoStop}%
\bibitem [{\citenamefont {Zalczer}(2023)}]{Za21}%
  \BibitemOpen
  \bibfield  {author} {\bibinfo {author} {\bibnamefont {Zalczer}, \bibfnamefont
  {S.}},\ }\bibfield  {title} {{\selectlanguage {English}\enquote {\bibinfo
  {title} {Localization for one-dimensional {Anderson}-{Dirac} models},}\
  }}\href {https://doi.org/10.1007/s00023-022-01203-y} {\bibfield  {journal}
  {\bibinfo  {journal} {Ann. Henri Poincar{\'e}}\ }\textbf {\bibinfo {volume}
  {24}},\ \bibinfo {pages} {37--72} (\bibinfo {year} {2023})}\BibitemShut
  {NoStop}%
\end{thebibliography}%




%
%
%
%
%
%
%
%
%
%
%


\end{document}